\documentclass[twocolumn,twocolappendix]{aastex63}

\usepackage{amsmath,amsopn,amsxtra,txfonts}
\usepackage{comment}  

\usepackage{threeparttablex}
\usepackage{natbib}
\usepackage{hyperref}

\newcommand{\nsightlines}{170}

\newcommand{\ha}{H$\alpha$}

\newcommand{\hi}{H\textsc{~i}}

\newcommand{\kms}{ \ifmmode{\rm km\thinspace s^{-1}}\else km\thinspace s$^{-1}$\fi}
\newcommand{\K}{\ensuremath{ \, \mathrm{K}}}

\newcommand{\kpc}{\ensuremath{\, \mathrm{kpc}}}
\newcommand{\cm}{\ensuremath{ \, \mathrm{cm}}}

\newcommand{\s}{\ensuremath{ \, \mathrm{s}}}
\newcommand{\yr}{\ensuremath{ \, \mathrm{yr}}}

\newcommand{\R}{\ensuremath{ \, \mathrm{R}}}

\newcommand{\vlsr}{\ifmmode{{v}_{\rm{LSR}}}\else ${v}_{\rm{LSR}}$\fi}

\newcommand{\dg}{\ifmmode{^{\circ}}\else $^{\circ}$\fi}

\begin{document}

\author[0000-0003-0536-3081]{Suraj Poudel}
\affiliation{Department of Physics \& Astronomy, Texas Christian University, Fort Worth, TX 76129, USA}

\author[0009-0005-3076-1104]{April Horton}
\affiliation{Department of Physics \& Astronomy, Texas Christian University, Fort Worth, TX 76129, USA}

\author[0000-0001-5817-0932]{Kathleen A. Barger}
\affiliation{Department of Physics \& Astronomy, Texas Christian University, Fort Worth, TX 76129, USA}

\author[0000-0003-4237-3553]{Frances H. Cashman}
\affiliation{Department of Physics, 
Presbyterian College, 
Clinton, SC 29325, USA}

\author[0000-0003-2308-8351]{Jo Vazquez}
\affiliation{Department of Physics \& Astronomy, Texas Christian University, Fort Worth, TX 76129, USA}

\author[0000-0003-0724-4115]{Andrew J. Fox}
\affil{AURA for ESA, Space Telescope Science Institute, 3700 San Martin Drive, Baltimore, MD, 21218, USA}

\author[0000-0001-9158-0829]{Nicolas Lehner}
\affiliation{Department of Physics and Astronomy, University of Notre Dame, Notre Dame, IN 46556, USA}

\author[0000-0002-2591-3792]{J. Christopher Howk}
\affiliation{Department of Physics and Astronomy, University of Notre Dame, Notre Dame, IN 46556, USA}

\author[0000-0001-9982-0241]{Scott Lucchini}
\affiliation{Center for Astrophysics $|$ Harvard \& Smithsonian, 60 Garden Street, Cambridge, MA 02138, USA}

\author[0000-0002-7955-7359]{Dhanesh Krishnarao}
\affiliation{Department of Physics, Colorado College, Colorado Springs, CO 80903, USA}

\author{Sabra Catalano}\affiliation{Department of Physics, Colorado College, Colorado Springs, CO 80903, USA}

\author{Anders Ripley}\affiliation{Department of Physics, Colorado College, Colorado Springs, CO 80903, USA}

\author{Wanyan Yuan}\affiliation{Department of Physics, Colorado College, Colorado Springs, CO 80903, USA}

\author{Natalie Van Tol}\affiliation{Department of Physics, Colorado College, Colorado Springs, CO 80903, USA}

\author{Citlali Alcala}
\affiliation{Department of Physics \& Astronomy, Texas Christian University, Fort Worth, TX 76129, USA}

\author{Jordan Elliott}
\affiliation{Department of Physics \& Astronomy, Texas Christian University, Fort Worth, TX 76129, USA}

\author[0000-0003-2730-957X]{N.\ M.\ McClure-Griffiths}
\affiliation{Research School of Astronomy \& Astrophysics, Australian National University, Canberra, Australia}

\author[0000-0003-2676-8344]{Elena D'Onghia}
\affiliation{Department of Astronomy, University of Wisconsin, Madison, WI 53706, USA}

\author[0000-0002-7982-412X]{Jason Tumlinson}\affiliation{Space Telescope Science Institute, 
3700 San Martin Drive,
Baltimore, MD 21218, USA}

\author[0000-0001-6398-5466]{Ananya Goon Tuli}
\affiliation{Department of Physics and Astronomy, University of Notre Dame, Notre Dame, IN 46556, USA}

\author[0000-0002-1272-3017]{Jacco Th. van Loon}
\affiliation{Lennard-Jones Laboratories, Keele University, ST5 5BG, UK}

\author[0000-0001-6326-7069]{Julia Roman-Duval}
\affiliation{Space Telescope Science Institute, 3700 San Martin Drive, Baltimore, MD 21218, USA}

\author[0000-0003-0742-2006]{Yik Ki Ma}
\affiliation{Max-Planck-Institut f\"ur Radioastronomie, Auf dem H\"ugel 69, 53121 Bonn, Germany}

\author[0000-0001-6846-5347]{Callum Lynn}
\affiliation{Research School of Astronomy \& Astrophysics, Australian National University, Canberra, Australia}

\title{Tracing Winds through the Fog: A Comprehensive Survey of LMC's Galactic Outflows with ULLYSES\footnote{Based on observations made with the NASA/ESA \textit{Hubble Space Telescope}, obtained at the Space Telescope Science Institute, which is operated by the Association of Universities for Research in Astronomy, Inc. under NASA contract No. NAS5-26555.}}

\begin{abstract}
We investigate nearside stellar-driven outflows from the Large Magellanic Cloud (LMC) using UV absorption-line spectroscopy of 170~OB stars, complemented by \hi\, 21-cm emission and \ha\, emission. Using Voigt-profile fitting and apparent-optical-depth analysis of Si\textsc{~ii}, O\textsc{~i}, and S\textsc{~ii} transitions, we map the velocity-dependent structure of foreground gas. The Si\textsc{~ii} column densities for the Voigt-fitted components decline smoothly from the LMC disk and reach a minimum at $v_{\rm LMCSR} \approx -100$ to $-200~{\rm km~s^{-1}}$, marking the transition from a denser, slower wind to more diffuse high-velocity material. Comparisons with local star-formation rate surface densities reveal a modest positive correlation for the  slower wind component, linking it to recent massive-stellar feedback. Photoionization modeling of 13 absorbers in the $+100\lesssim v_{\rm LSR} \lesssim +150~{\rm km~s^{-1}}$ range reveals wide diversity in metallicity, dust depletion, and ionization conditions, consistent with multiple possible origins. While a substantial fraction of this predominately ionized gas is consistent with the high-velocity extension of the LMC wind, part of it may arise from contamination by Milky Way (MW) high-velocity clouds (HVCs) and Magellanic circumgalactic medium (CGM). We estimate a nearside cool-gas outflow mass of ${\sim}1.8\times10^{7}\,M_{\odot}$, implying $\dot{M}_{\rm out}\approx0.15$--$0.33\,M_{\odot}\,\mathrm{yr^{-1}}$ and a mass-loading factor of $\eta\approx0.6$--1.3. On regional scales, 30~Doradus contributes $\sim10\%$ and N11 contributes $\sim3\%$ of the total outflow mass, while the trailing side contains more wind material than the leading side, consistent with ram-pressure stripping. These results provide the comprehensive kinematic and physical characterization of how stellar feedback, galactic environment, and foreground contamination shape the multiphase wind emerging from the LMC.
\end{abstract}



\section{Introduction}
Galactic outflows are driven by a combination of feedback from stars via radiation pressure and supernova explosions and from active galactic nuclei (AGN); together, these processes inject energy and momentum into the interstellar medium (ISM) that can propel gas out of the host galaxy \citep{2005ARAA..43..769V, 2008MNRAS.387..577O}. These outflows can either escape the galaxy’s gravitational potential---enriching the intergalactic medium (IGM) and possibly fueling star formation in neighboring galaxies---or return to the host galaxy as part of a galactic fountain \citep{2015MNRAS.446..299M}. Characterizing such outflows is essential for understanding their role in regulating star formation and gas mixing, which are key processes in galaxy evolution \citep{2021MNRAS.508.2979P}. While outflows have been studied in many nearby and distant galaxies via their ultraviolet (UV) absorption signatures, such studies are often limited to only one or a few sightlines.

Although the importance of winds is widely recognized, the detailed
physics that govern their behavior and effects are not well understood. Indeed, one of the major findings of the last two decades concerns their complexity: winds can be driven by radiation from hot stars, supernova blast waves, cosmic rays, AGN, and all these processes in combination \citep{2005ApJ...618..569M, 2005ARAA..43..769V, 2017arXiv170109062H}. Galactic winds arise from local sources that cluster on 10~pc scale like stars and black holes, which can propagate to beyond 100~kpc from the galaxy. These multiple processes acting on small scales requires an observational approach that can resolve the flows at the right scale so that they can be correlated with the driving forces. We can resolve both the galactic winds and the internal stellar activity of the Large Magellanic Cloud (LMC) at these scales as it is located just $50\,\kpc$ away \citep{2014AJ....147..122D} and is viewed nearly face-on \citep{2022ApJ...927..153C}. Further, although \citet{2025ApJ...982..188H} finds evidence that the LMC hosts an AGN from hyper velocity stars traced back to this galaxy, the are no indications in the X-ray emission that it is active \citep{2012ApJ...746...27K}, suggesting stellar feedback can be isolated. The galactic outflows from LMC have been detected in emission (\hi: \citealt{2003MNRAS.339...87S}; H$\alpha$: \citealt{Ciampa2020}; [O\textsc{~iii}]:  \citealt{2003MNRAS.344..741R}) and UV absorption \citep{2002ApJ...569..214H, 2009ApJ...702..940L, 2016ApJ...817...91B, Ciampa2020, 2024arXiv240204313Z, 2025ApJ...984..161P}. 

Previous studies have shown that the neutral and ionized gas in the LMC outflows is highly structured, with large \hi\, voids likely created by stellar feedback \citep{2003MNRAS.339...87S} and that there is a strong concentration with the outflow signatures in \ha\ emission near the 30~Doradus starburst region \citep{Ciampa2020}. \citet{2024ApJ...974...22Z} find that Si\textsc{~iv} and C\textsc{~iv} traced outflows in the LMC have bulk velocities of ${\sim}20$--$60~\kms$---implying that most of the gas is gravitationally bound---and that the outflow column density correlates with the star formation rate surface density. However, some studies suggest that some of these winds are able to achieve speeds of  $|v_{\rm outflow}|\gtrsim150\,\kms$ \citep{2009ApJ...702..940L}, especially in the direction of the 30~Doradus starburst region \citep{2025ApJ...984..161P}. These fast moving clouds kinematically overlap with where the MW's high-velocity cloud (HVCs) population typically resides \citep{1981ApJ...243..460S, 1990A&A...233..523D, 1999Natur.402..386R, 2015A&A...584L...6R}. Spectroscopic studies of massive stars near R136 also reveal multiple kinematic components associated with both local winds and halo gas \citep{2013A&A...550A.108V}. The relative contribution of clouds in this velocity range that belong to the MW and to the LMC has yet to be disentangled.

Despite clear evidence for multiphase outflows from the LMC, the kinematic structure of these winds and their physical significance remain poorly constrained. Moreover, it is unclear which absorption components are associated with the wind, which arise from foreground or unrelated structures, and what role the fastest-moving gas plays in the feedback cycle. By kinematically resolving individual absorption components, we can characterize the physical nature of discrete gas clouds across the full velocity range, from low-velocity material to the fastest-moving components. This approach allows us to isolate high-velocity, low–column density gas, which---despite contributing little to the total mass---provides a sensitive probe of the (1) stellar feedback energetics, (2) ability of stellar activity to accelerate material to escape speeds, and (3) influence that a confining LMC halo or bow shock has on the wind. We use the Doppler line widths of absorption lines to constrain the physical processes shaping these clouds and ionic ratios, such as [Si\textsc{~ii}/O\textsc{~i}], to assess the ionization. Finally, we use photoionization modeling for individual absorbers to assess the metallicity and dust-depletion patterns that further inform the physical and chemical state and possible origin of foreground and wind-related clouds.

To investigate how well the LMC's winds spatially and kinematically correlate with the stellar activity occurring within this galaxy, we perform a component-by-component analysis of low ionization UV absorption features (O\textsc{~i}, S\textsc{~ii}, Si\textsc{~ii}) along \nsightlines\ stellar sightlines observed with \textit{Hubble Space Telescope's} (HST) Ultraviolet Legacy Library of Young Stars as Essential Standards (ULLYSES; \citealt{2025ApJ...985..109R}), a director’s discretionary initiative. These background OB stars lie across the LMC's disk, which enables us to characterize the gas distribution, ionization state, and kinematics of the LMC's galactic wind. We further assess the fast-moving component of the LMC’s wind using additional Fe\textsc{~ii} and Si\textsc{~iv} absorption and characterize its physical and chemical properties through photoionization modeling along 13 of our sightlines; this analysis has the added benefit of enabling us to explore the contribution of LMC winds, Magellanic HVCs, or MW HVCs that overlap in the $+90 \lesssim \vlsr \lesssim +175\,\kms$ velocity range\footnote{$\vlsr$ is velocity in local standard of rest. We adopt the kinematic LSR convention, in which the Sun is assumed to move at a velocity of $20~\mathrm{km~s^{-1}}$ toward $(\mathrm{R.A.},~\mathrm{Decl.})_{\mathrm{J2000}} = (18^{\mathrm{h}}03^{\mathrm{m}}50\fs29,~+30\degr00'16\farcs8)$.}. In addition, we incorporate H$\alpha$ emission line data to investigate the relation between the star-forming activities and the outflows and the H\textsc{~i}~21-cm emission-line data were used for constraining the kinematic width of the LMC. In Section~\ref{section:observations}, we describe the UV, radio, and optical observations and data reduction. Section~\ref{section:UV_Analysis} details the analysis of UV absorption-line. We present our efforts for characterizing the gas distributions, kinematics, ionization, and correalation with star-forming activities in Section~\ref{section:Prob_wind}. We explore the photoionization modeling for the ambiguous HVCs in Section~\ref{section:Cloudy} and discuss the high-ion contribution in Section~\ref{section:high-ions}. We provide discussion of our results in Section~\ref{section:Discussion} and summarize our main conclusions in Section~\ref{section:Summary}.

\section{Observations and Data Reduction} \label{section:observations}

\begin{figure*}
  \centering
  \includegraphics[width=0.85\textwidth]{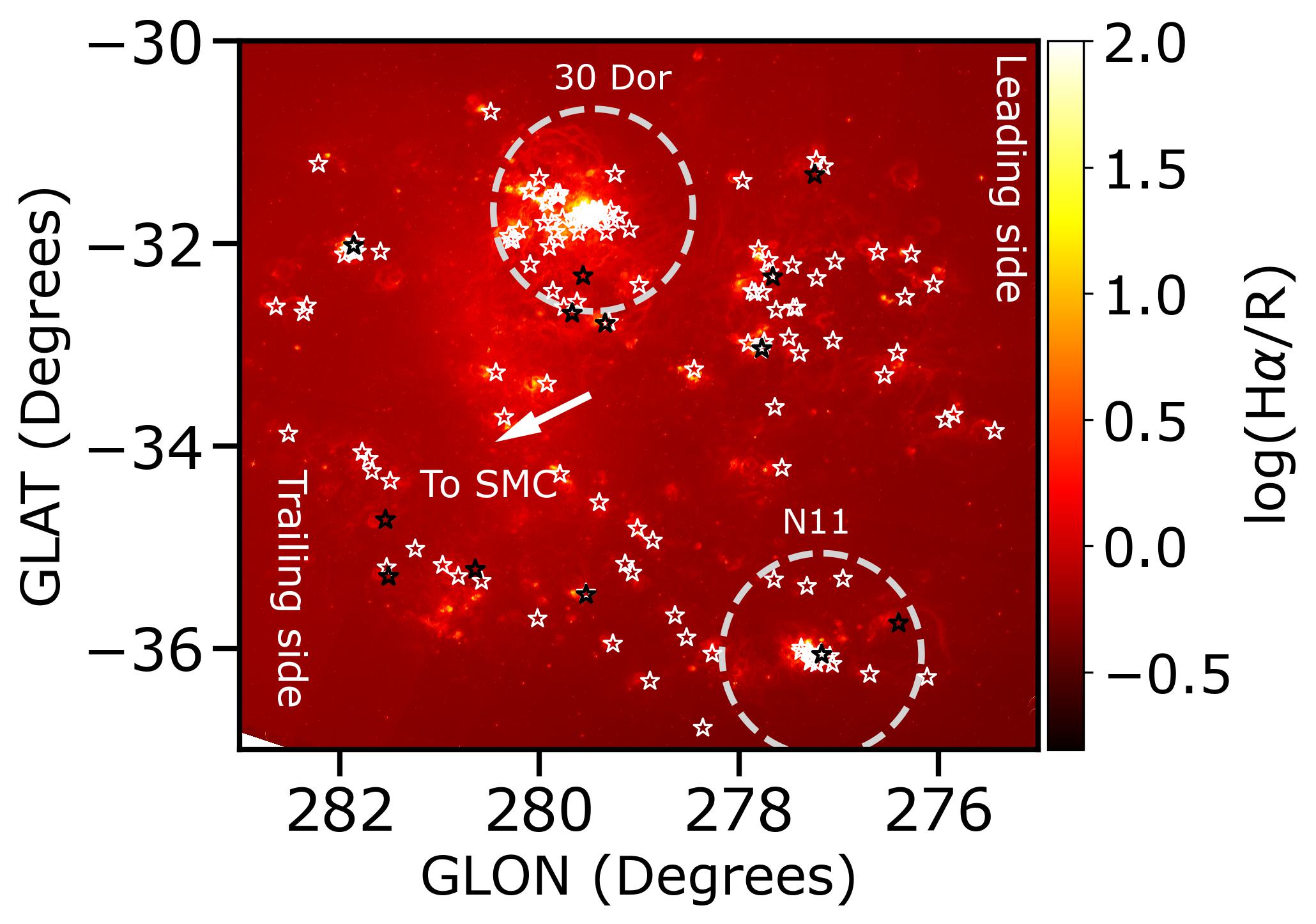}
  \caption{Locations of \nsightlines\, ULLYSES sightlines from the DR7 catalogue (star symbols) overlaid on the \ha\, emission map of the LMC's disk, using observations from the Magellanic Cloud Emission-line Survey (MCELS; \citealt{1999IAUS..190...28S}). The \ha\ flux is expressed in Rayleigh (R), where 1~R $= (10^6/4\pi)$ photons cm$^{-2}$ sr$^{-1}$ s$^{-1}$ and is equivalent to $5.7 \times 10^{-18}~\mathrm{erg\,s^{-1}\,cm^{-2}\,arcsec^{-2}}$
 at H$\alpha$. Two prominent star-forming regions, 30~Dor and N11, are located at the centers of dashed circles of radii $1^{\circ}$. The arrow points in the direction of SMC and we have labeled the approximate leading and trailing sides of the LMC. The black stars are the 13~sightlines for which we perform the photoionization modeling (see Section~\ref{section:Cloudy}); these targets are indicated in Table~\ref{table:targets}.}
  \label{fig:ha_map}
\end{figure*}

We explore the galactic wind in UV absorption, optical emission, and radio emission toward \nsightlines\, OB stellar background targets that are scattered across the LMC disk. We indicate the positions of these stars in Figure~\ref{fig:ha_map} overlaid on top of the LMC's \ha\ emission. For both the UV absorption and \hi\ emission observations toward these sightlines, we adopt the LSR and the LMC centered (LMCSR) velocity frames, with the latter defined such that the LMC disk lies at $\mathrm{v_{LMCSR}} = 0~\mathrm{km~s^{-1}}$. The transformation from the LSR to the LMCSR frame follows the method described in \citet{2025ApJ...984..161P} that uses the relationship provided by \citet{Ciampa2020}.

\subsection{UV Observations}

Our UV absorption-line dataset is drawn entirely from the 7th data release (DR7) of ULLYSES. The ULLYSES dataset includes a compilation of \textit{HST} observations that were taken with the Cosmic Origins Spectrograph (COS) and Space Telescope Imaging Spectrograph (STIS) instruments and archival Far Ultraviolet Space Explorer (FUSE) observations. This publicly available spectroscopic resource targets low and high mass stars in the Magellanic Clouds and in nearby dwarf galaxies and contains 182 sightlines across the disk of the LMC. These observations have a typical continuum signal-to-noise ratio per resolution element of $20 \lesssim S/N \lesssim 30$. Out of 182, we utilize 170 observations of OB stars embedded in the LMC, which we use as background sources to explore foreground gas clouds. Twelve of these sightlines (Sk$-$69$^{\circ}$220, Sk$-$70$^{\circ}$50, BAT99$-$20, LMC195$-$1, LH$-$41$-$1042, Sk$-$68$^{\circ}$145, LMC174$-$1, VFTS$-$731, LMCE078$-$3, Sk$-$69$^{\circ}$207, VFTS$-$102, and VFTS$-$285) were excluded as the absorption along these stars were affected by one or multiple situations which include complex stellar continuum, very low signal to noise and non coverage of the lines of interest. Out of 170~sightlines, 89~of them were observed with STIS/E140M grating which has a high spectroscopic resolution at a full width at half maximum of $\rm FWHM_{STIS}\approx 6.5\,\kms$. The rest of the sightlines were observed with COS with spectroscopic resolution of $\rm FWHM_{COS}\approx 17\,\kms$. Additionally, 44~of the sightlines were observed with STIS/E230M grating which has a spectroscopic resolution at a full width at half maximum of $\rm FWHM_{STIS}\approx 10\,\kms$ and covers the longer wavelength region. Although many of the targets observed with STIS were also observed with COS, we preferentially use the STIS datasets whenever possible due to their superior spectroscopic resolution as multiple absorbers tend to overlap in our kinematic region of interest. In addition, the ULLYSES archive includes science ready FUSE spectra. We incorporate the Far-UV dataset when available as it is particularly useful for accessing the O\,\textsc{i} $\lambda1039$ absorption line, especially in situations where the O\,\textsc{i} $\lambda1302$ line is saturated. Out of the total \nsightlines~sightlines we used in our sample, 86~were also observed by FUSE at a velocity resolution of $\rm FWHM_{FUSE} \approx 20\,\kms$ and a typical continuum $S/N \approx 10$. We list the names, spectral types, and positions of our stellar background targets in Table~\ref{table:targets} and additionally include the observing modes of the UV observations and their corresponding spectral resolutions.

The methodology used to process the ULLYSES data products is detailed in \citet{2020RNAAS...4..205R} and on the program’s official website.\footnote{\url{https://ullyses.stsci.edu}} We summarize the key data processing steps here. Spectra obtained with the STIS instrument were reduced using the CalSTIS pipeline and those obtained with COS were reduced using CalCOS pipeline. These pipelines perform wavelength and flux calibrations and correct for any spectral shifts to ensure that the spectral features for repeated observations align and then combine the exposures taken with the same instrument and grating configuration. In cases where data were collected using different gratings or instruments, spectra were either spliced at non-overlapping wavelengths, leading to discontinuities in dispersion, or were trimmed and joined at a selected overlap region to ensure a smooth transition; therefore, we only used the spectra products of combined diffraction gratings during the exploratory stages and the spectra products that only combine observations taken with one grating for analysis.

\startlongtable 
\begin{deluxetable*}{ccccccc}
\tabletypesize{\scriptsize}
\tablewidth{0pt}
\tablecaption{Properties of $170$ ULLYSES DR7 Stellar Sightlines in the LMC Disk\label{table:targets}} 
\tablehead{
\colhead{Background}  & \colhead{$l$} & \colhead{$b$} & \colhead{Stellar} & \colhead{UV Observing}  \\ [ -0.3cm] 
\colhead{Target}  & \colhead{$(^\circ)$} & \colhead{$(^\circ)$} & \colhead{Type} & \colhead{Modes$^{m}$}}
\startdata
SK$-$65$^{\circ}$47 & 275.44 & $-$33.85 & O4 I(n)f+p & E140M;FUSE \\
SK$-$65$^{\circ}$55 & 275.85 & $-$33.69 & WN6 h & E140M;FUSE \\
LMCE169$-$1 & 275.93 & $-$33.74 & WN3/O3 & G160M \\
SK$-$66$^{\circ}$152 & 276.05 & $-$32.4 & O7 Ib(f) & E140M;G130M \\
SK$-$65$^{\circ}$2 & 276.11 & $-$36.28 & B1 V & E140M;FUSE \\
SK$-$66$^{\circ}$172 & 276.27 & $-$32.1 & O2 III(f*) + OB & E140M;E230M;FUSE \\
LMCX$-$4 & 276.33 & $-$32.53 & O8 III & E140M;FUSE \\
SK$-$65$^{\circ}$22$^{\textcolor{blue}{C1}}$ & 276.4 & $-$35.75 & O6 Iaf+ & E140M;E230M;FUSE \\
SK$-$66$^{\circ}$97 & 276.41 & $-$33.08 & sgB[e] & E140M;FUSE \\
LMCE159$-$1 & 276.54 & $-$33.3 & WN3/O3 & G160M \\
SK$-$66$^{\circ}$171 & 276.61 & $-$32.08 & O9 Ia & E140M;E230M;FUSE \\
SK$-$66$^{\circ}$18 & 276.69 & $-$36.25 & O6 V((f)) & E140M;FUSE \\
NGC1818$-$ROB$-$D1 & 276.96 & $-$35.31 & B1 V & G130M;G160M \\
SK$-$67$^{\circ}$216 & 277.04 & $-$32.18 & B0.5 V & E140M;FUSE \\
SK$-$66$^{\circ}$21 & 277.06 & $-$36.15 & WC4 & G130M;G160M \\
SK$-$66$^{\circ}$100 & 277.06 & $-$32.96 & O6 II(f) & E140M;FUSE \\
N11$-$ELS$-$026 & 277.09 & $-$36.08 & O2.5 III(f*) & E140M;G130M \\
N11$-$ELS$-$051 & 277.13 & $-$36.11 & O5 Vn((f)) & G130M;G160M \\
SK$-$67$^{\circ}$261 & 277.15 & $-$31.24 & O8.5 III & E140M;G130M \\
N11$-$ELS$-$032$^{\textcolor{blue}{C2}}$ & 277.17 & $-$36.06 & O7 II(f) & FUSE;G130M;G160M \\
N11$-$ELS$-$013 & 277.17 & $-$36.04 & O8 V & E140M;E230M;FUSE \\
N11$-$ELS$-$048 & 277.18 & $-$36.05 & O6.5 V((f)) & E140M \\
N11$-$ELS$-$018 & 277.19 & $-$36.08 & O6 II(f+) & FUSE;G130M;G160M \\
N11$-$ELS$-$060 & 277.19 & $-$36.07 & O3 V((f*)) & G130M;G160M \\
PGMW$-$3120 & 277.19 & $-$36.07 & O5.5 V((f*)) & E140M;E230M \\
PGMW$-$3070 & 277.2 & $-$36.07 & O6 V & E140M;FUSE \\
N11$-$ELS$-$038 & 277.2 & $-$36.07 & O5 III(f+) & G130M;G160M \\
N11$-$ELS$-$031 & 277.2 & $-$36.07 & ON2 III(f*) & G130M;G160M \\
SK$-$67$^{\circ}$195 & 277.22 & $-$32.34 & B6 I & E230M;G130M;G160M \\
SK$-$67$^{\circ}$266 & 277.22 & $-$31.17 & WN11h & E140M;FUSE \\
SK$-$66$^{\circ}$19 & 277.22 & $-$36.15 & O7 V & E230M;G130M;G160M \\
BI$-$272$^{\textcolor{blue}{C3}}$ & 277.24 & $-$31.32 & O7 II & E140M;FUSE \\
BAT99$-$10 & 277.27 & $-$36.07 & WC4 & E140M \\
N11$-$ELS$-$049 & 277.27 & $-$36.08 & O7.5 V & E140M;G130M \\
N11$-$ELS$-$033 & 277.28 & $-$36.11 & B0 IIIn & G130M;G160M \\
SK$-$66$^{\circ}$17 & 277.29 & $-$36.13 & OC9.5 II & G130M;G160M \\
PGMW$-$1363 & 277.29 & $-$36.05 & O8.5 Iaf & E140M \\
N11$-$ELS$-$020 & 277.31 & $-$36.03 & O5 Inf+p & E140M;G130M \\
SK$-$66$^{\circ}$51 & 277.32 & $-$35.39 & WN 7h & E140M;FUSE \\
N11$-$ELS$-$046 & 277.38 & $-$36.03 & O9.5 V & E140M;G130M \\
SK$-$66$^{\circ}$35 & 277.38 & $-$35.99 & BC1 Ia & E140M;E230M;FUSE \\
SK$-$67$^{\circ}$105 & 277.39 & $-$33.08 & O4 f + O6 V & E140M;E230M;FUSE \\
NGC2004$-$ELS$-$003 & 277.42 & $-$32.63 & B5 Ia & E230M;G130M;G160M \\
SK$-$67$^{\circ}$207 & 277.46 & $-$32.22 & B9 Ia & E230M;G130M;G160M \\
NGC2004$-$ELS$-$026 & 277.46 & $-$32.64 & B2 II & FUSE;G130M;G160M \\
SK$-$67$^{\circ}$118 & 277.5 & $-$32.93 & O7 V & E140M;FUSE \\
SK$-$67$^{\circ}$69 & 277.57 & $-$34.22 & O4 III(f) & E140M;FUSE \\
SK$-$67$^{\circ}$144 & 277.63 & $-$32.66 & WC4 & FUSE;G130M;G160M \\
SK$-$67$^{\circ}$78 & 277.64 & $-$33.62 & B3 Ia & E230M;FUSE;G130M;G160M \\
SK$-$66$^{\circ}$50 & 277.65 & $-$35.32 & B8 Ia+ & E230M;G130M;G160M \\
SK$-$67$^{\circ}$191$^{\textcolor{blue}{C4}}$ & 277.66 & $-$32.33 & O8 V & E140M;E230M;FUSE \\
SK$-$67$^{\circ}$197 & 277.69 & $-$32.29 & B7 I & E230M;G130M;G160M \\
SK$-$67$^{\circ}$211 & 277.7 & $-$32.16 & O2 III(f*) & E140M;E230M;FUSE \\
SK$-$67$^{\circ}$111 & 277.75 & $-$32.97 & O6 Ia(n)fpv & E140M;FUSE \\
SK$-$67$^{\circ}$107 & 277.76 & $-$33.01 & O9 Ib(f) & E140M;FUSE \\
SK$-$67$^{\circ}$168 & 277.76 & $-$32.48 & O8 I(f)p & E140M;FUSE \\
SK$-$67$^{\circ}$106 & 277.77 & $-$33.02 & O8 III((f)) & E140M;FUSE \\
SK$-$67$^{\circ}$104$^{\textcolor{blue}{C5}}$ & 277.77 & $-$33.04 & WC4 + abs & E140M;FUSE \\
SK$-$67$^{\circ}$101 & 277.78 & $-$33.05 & O8 II((f)) & E140M;E230M;FUSE \\
BI$-$237 & 277.8 & $-$32.06 & O2 V((f*)) & E230M;FUSE;G130M;G160M \\
SK$-$67$^{\circ}$166 & 277.84 & $-$32.49 & O4 If+ & E140M;FUSE \\
SK$-$67$^{\circ}$167 & 277.87 & $-$32.47 & O4 Inf+ & E140M;FUSE \\
SK$-$67$^{\circ}$108 & 277.91 & $-$32.99 & O4-5 III & E140M;FUSE \\
LH$-$114$-$7 & 277.96 & $-$31.38 & O2 III(f*) + OB? & E140M;FUSE \\
SK$-$67$^{\circ}$14 & 278.27 & $-$36.05 & B1.5 Ia & E140M;E230M;FUSE \\
SK$-$67$^{\circ}$2 & 278.36 & $-$36.79 & B1 Ia+ (N wk) & E140M;E230M;FUSE \\
SK$-$68$^{\circ}$73 & 278.45 & $-$33.24 & WN9pec & E230M;FUSE;G130M;G160M \\
SK$-$67$^{\circ}$20 & 278.52 & $-$35.89 & WN4 b & E140M;FUSE \\
SK$-$67$^{\circ}$22 & 278.64 & $-$35.67 & O2 If*/WN5 & E140M;FUSE \\
LMC277$-$2 & 278.86 & $-$34.93 & WN3/O3 & G160M \\
SK$-$67$^{\circ}$5 & 278.89 & $-$36.32 & O9.7 Ib & E140M;E230M;FUSE \\
SK$-$68$^{\circ}$112 & 279.0 & $-$32.41 & O7.5(n)(f)p & E140M \\
SK$-$68$^{\circ}$41 & 279.02 & $-$34.82 & B0.5 Ia & E140M;E230M;FUSE \\
SK$-$68$^{\circ}$23A & 279.07 & $-$35.25 & B1 III & G130M;G160M \\
SK$-$68$^{\circ}$133 & 279.1 & $-$31.86 & OC3.5 III(f*) & G130M;G160M \\
SK$-$68$^{\circ}$26 & 279.14 & $-$35.17 & BC2 Ia & E230M;FUSE;G130M;G160M \\
SK$-$68$^{\circ}$137 & 279.21 & $-$31.72 & O2 III(f*) & FUSE;G130M;G160M \\
SK$-$68$^{\circ}$155 & 279.24 & $-$31.31 & B0.5 I & E230M;FUSE;G130M;G160M \\
BI$-$13 & 279.26 & $-$35.95 & O6.5 V & E140M;FUSE \\
SK$-$68$^{\circ}$135 & 279.26 & $-$31.77 & ON9.7 Ia+ & E140M;E230M;FUSE \\
SK$-$68$^{\circ}$140 & 279.28 & $-$31.67 & B0.7 Ib-Iab Nwk & E230M;FUSE;G130M;G160M \\
VFTS$-$355 & 279.3 & $-$31.71 & O4 V((n))((fc))z & G130M;G160M \\
LH$-$58$-$496 & 279.3 & $-$32.78 & O5 V((f)) & E140M;G130M \\
SK$-$68$^{\circ}$129 & 279.33 & $-$31.89 & B1 I & E230M;FUSE;G130M;G160M \\
SK$-$68$^{\circ}$80$^{\textcolor{blue}{C6}}$ & 279.34 & $-$32.79 & WC4 + O6 III/V & E140M;FUSE \\
VFTS$-$72 & 279.38 & $-$31.78 & O2 V-III(n)((f*)) & E230M;FUSE;G130M;G160M \\
SK$-$69$^{\circ}$246 & 279.38 & $-$31.66 & WN5/6h + WN6/7h & E140M;E230M;FUSE \\
SK$-$68$^{\circ}$52 & 279.4 & $-$34.56 & B0 Ia & E140M;E230M;FUSE \\
VFTS$-$586 & 279.4 & $-$31.67 & O4 V((n))((fc))z & G130M;G160M \\
VFTS$-$406 & 279.43 & $-$31.69 & O6 Vnn & G130M;G160M \\
VFTS$-$507 & 279.45 & $-$31.67 & WN6 & E140M \\
VFTS$-$506 & 279.45 & $-$31.67 & ON2 V((n))((f*)) & G130M;G160M \\
VFTS$-$66 & 279.45 & $-$31.78 & O9 V + B0.2 V & G130M;G160M \\
VFTS$-$440 & 279.45 & $-$31.68 & O6-6.5 II(f) & E140M \\
SK$-$69$^{\circ}$235 & 279.46 & $-$31.75 & WC4 + B1 Ia & G130M;G160M \\
VFTS$-$542 & 279.46 & $-$31.67 & O2 If*/WN5 + B0 V & G130M;G160M \\
BAT99$-$105 & 279.46 & $-$31.67 & O2 If* & E140M;FUSE \\
VFTS$-$482 & 279.46 & $-$31.68 & O2.5 If*/WN6 & G130M;G160M \\
VFTS$-$545 & 279.47 & $-$31.67 & O2 If*/WN5 & G130M;G160M \\
VFTS$-$244 & 279.47 & $-$31.72 & O5 III(n)(fc) & E140M;G130M;G160M \\
VFTS$-$267 & 279.5 & $-$31.71 & O3 III-I(n)f* & G130M;G160M \\
VFTS$-$87 & 279.51 & $-$31.77 & O9.7 Ib-II & E140M \\
SK$-$68$^{\circ}$16 & 279.53 & $-$35.45 & O7 III & E140M;FUSE \\
SK$-$68$^{\circ}$15$^{\textcolor{blue}{C7}}$ & 279.53 & $-$35.47 & WC4 & E140M;FUSE \\
VFTS$-$404 & 279.54 & $-$31.68 & O3.5 V(n)((fc)) & G130M;G160M \\
VFTS$-$180 & 279.55 & $-$31.74 & O3 If* & E140M;G130M;G160M \\
SK$-$69$^{\circ}$175$^{\textcolor{blue}{C8}}$ & 279.56 & $-$32.32 & WN 11h & E140M;E230M;FUSE \\
VFTS$-$169 & 279.56 & $-$31.74 & O2.5 V(n)((f*)) & G130M;G160M \\
VFTS$-$352 & 279.57 & $-$31.68 & O4.5 V(n) & G130M;G160M \\
SK$-$69$^{\circ}$224 & 279.6 & $-$31.82 & B1 Ia+ & FUSE;G130M;G160M \\
SK$-$69$^{\circ}$212 & 279.61 & $-$31.89 & O5n(f)p & E140M;G130M \\
VFTS$-$190 & 279.61 & $-$31.73 & O7 Vnn((f))p & G130M;G160M \\
LMC199$-$1 & 279.62 & $-$32.58 & WN3/O3 V & G160M \\
SK$-$69$^{\circ}$234 & 279.64 & $-$31.74 & WC4 & G130M;G160M \\
LMCE078$-$1 & 279.65 & $-$31.76 & O6 Ifc & G130M;G160M \\
BI$-$173$^{\textcolor{blue}{C9}}$ & 279.67 & $-$32.69 & O8 II: & E140M;E230M;FUSE \\
SK$-$69$^{\circ}$140 & 279.75 & $-$32.63 & B4 I & E230M;G130M;G160M \\
SK$-$69$^{\circ}$231 & 279.77 & $-$31.75 & WC4 & G130M;G160M \\
HV$-$5622 & 279.79 & $-$34.28 & B0 V & FUSE;G130M;G160M \\
BAT99$-$127 & 279.8 & $-$31.51 & WC4 + O & E140M \\
SK$-$69$^{\circ}$255 & 279.81 & $-$31.53 & WC4 (+OB) & G130M;G160M \\
LMC172$-$1 & 279.81 & $-$31.97 & WN3/O3 V & G160M \\
ST92$-$4$-$18 & 279.81 & $-$31.54 & O5 If & G130M;G160M \\
UCAC3$-$42$-$33014 & 279.83 & $-$31.5 & O7.5n(f)p & G130M;G160M \\
UCAC3$-$42$-$30814 & 279.84 & $-$31.89 & O6(f)np & G130M;G160M \\
LMC170$-$2 & 279.86 & $-$32.47 & WN3/O3 V & G160M \\
SK$-$69$^{\circ}$222 & 279.88 & $-$31.79 & WC4 (+OB) & E140M \\
BI$-$214 & 279.9 & $-$32.04 & O6.5(n)(f)p & E140M \\
H$^{\circ}$$-$269927C & 279.91 & $-$31.6 & WN9h & E140M;FUSE \\
SK$-$69$^{\circ}$104 & 279.92 & $-$33.39 & O6 Ib(f) & E140M;E230M;FUSE \\
ST92$-$5$-$27 & 279.93 & $-$31.58 & O3 V((f)) & G130M;G160M \\
ST92$-$5$-$31 & 279.94 & $-$31.58 & O2-3(n)f*p & G130M;G160M \\
SK$-$69$^{\circ}$279 & 280.0 & $-$31.35 & O9.2 Iaf & E230M;FUSE;G130M;G160M \\
SK$-$68$^{\circ}$8 & 280.02 & $-$35.71 & B5 Ia+ & E230M;G130M;G160M \\
SK$-$69$^{\circ}$178 & 280.09 & $-$32.21 & O9.2 II & G130M;G160M \\
FARINA$-$88 & 280.09 & $-$31.48 & O4 III(f) & G130M;G160M \\
BI$-$265 & 280.11 & $-$31.49 & O5 III(fc) & E140M;G130M \\
GAIA$-$DR3$-$4657 & 280.2 & $-$31.87 & O3 If* & G130M;G160M \\
W61$-$28$-$5 & 280.26 & $-$31.96 & O4 V((f+)) & G130M;G160M \\
SK$-$69$^{\circ}$191 & 280.28 & $-$31.97 & WC4 & E140M;FUSE \\
W61$-$28$-$23 & 280.3 & $-$31.92 & O3.5 V((f+)) & G130M;G160M \\
SK$-$69$^{\circ}$83 & 280.35 & $-$33.72 & O7.5 Iaf & E140M;G130M \\
SK$-$69$^{\circ}$106 & 280.43 & $-$33.27 & WC4 + O5-6III-V + O & E140M;FUSE \\
SK$-$70$^{\circ}$115 & 280.49 & $-$30.7 & O6 If & E140M;E230M;FUSE \\
SK$-$69$^{\circ}$43 & 280.58 & $-$35.33 & B0.5 Ia & E140M;E230M;FUSE \\
SK$-$69$^{\circ}$50$^{\textcolor{blue}{C10}}$ & 280.64 & $-$35.22 & O7(n)(f)p & E140M;FUSE \\
SK$-$69$^{\circ}$42 & 280.81 & $-$35.29 & WC4 & G130M;G160M \\
LMCE055$-$1 & 280.97 & $-$35.18 & WN4 /O4 & G130M;G160M \\
SK$-$69$^{\circ}$52 & 281.24 & $-$35.02 & B2 Ia & E230M;FUSE;G130M;G160M \\
SK$-$70$^{\circ}$60 & 281.49 & $-$34.35 & O4-5 V((f))pec & E140M;FUSE \\
SK$-$70$^{\circ}$13$^{\textcolor{blue}{C11}}$ & 281.51 & $-$35.29 & O9 V & E140M;FUSE \\
SK$-$70$^{\circ}$16 & 281.53 & $-$35.2 & B4 I & G130M;G160M \\
SK$-$70$^{\circ}$32$^{\textcolor{blue}{C12}}$ & 281.54 & $-$34.73 & O9.5 II: & E140M;FUSE \\
N206$-$FS$-$170 & 281.59 & $-$32.08 & B1 IV & G130M;G160M \\
SK$-$70$^{\circ}$69 & 281.68 & $-$34.25 & O5.5 V((f)) & E140M;FUSE \\
SK$-$70$^{\circ}$79 & 281.71 & $-$34.13 & B0 III & E140M;E230M;FUSE \\
LMC079$-$1 & 281.77 & $-$34.06 & WN3/O3 V & G160M \\
SK$-$71$^{\circ}$38 & 281.83 & $-$32.08 & WC4 + OB & FUSE;G130M;G160M \\
BI$-$184 & 281.84 & $-$32.09 & O8 (V)e & E230M;FUSE;G130M;G160M \\
SK$-$71$^{\circ}$46 & 281.84 & $-$31.98 & O4 If & FUSE;G130M;G160M \\
BI$-$189 & 281.86 & $-$32.05 & O8 IV((f))e & G130M;G160M \\
SK$-$71$^{\circ}$45$^{\textcolor{blue}{C13}}$ & 281.86 & $-$32.02 & O4-5 III(f) & E140M;E230M;FUSE;G130M \\
N206$-$FS$-$194 & 281.86 & $-$32.01 & O8.5 V & G130M;G160M \\
SK$-$71$^{\circ}$41 & 281.9 & $-$32.07 & O9.7 Iab & E140M;FUSE \\
SK$-$71$^{\circ}$35 & 281.96 & $-$32.11 & B1 II & G130M;G160M \\
SK$-$71$^{\circ}$50 & 282.21 & $-$31.21 & O6.5 III & E140M;E230M;FUSE \\
SK$-$71$^{\circ}$26 & 282.33 & $-$32.61 & WC4 + abs & FUSE;G130M;G160M \\
SK$-$71$^{\circ}$19 & 282.36 & $-$32.68 & O6 III & FUSE;G130M;G160M \\
SK$-$71$^{\circ}$8 & 282.51 & $-$33.88 & O9 II & E140M;FUSE \\
SK$-$71$^{\circ}$21 & 282.64 & $-$32.62 & WN6(h) & E140M;FUSE \\
\enddata
\tablenotetext{\textcolor{blue}{Ci}}{\,Where i=1,2,...13 represents indices of the sightlines for which we conducted photoionization modeling (see Section~\ref{section:Cloudy} and  Table~\ref{tab:cloudy_results}).}
\tablenotetext{m}{E140M and E230M are gratings associated with the HST/STIS with instrumental resolutions of $\rm FWHM_{\rm res}\approx6.5\,\kms$ ($R\approx45{,}800$) and $\rm FWHM_{\rm res}\approx10\,\kms$ ($R\approx30{,}000$), respectively. G130M and G160M are gratings associated with the HST/COS with instrumental resolution of $\rm FWHM_{\rm res}\approx16.5\,\kms$ ($R\approx18{,}000$). FUSE has a resolution of $\rm FWHM_{\rm res}\approx20\,\kms$ ($R\approx15{,}000$.)}
\tablecomments{The sightlines are in the order of increasing Galactic longitudes ($l$).}
\end{deluxetable*}

\subsection{Radio Observations}

To investigate the neutral hydrogen gas phase in the LMC and its wind, we used the H\textsc{~i} 21-cm emission-line data from two different surveys: (1) Galactic Australian Square Kilometre Array Pathfinder (GASKAP) survey \citep{2013PASA...30....3D,2022PASA...39....5P}, which has a sensitivity of $\log{\left(N_{\rm H\textsc{~i},\,3\sigma}/\cm^{-2}\right)}=20.0$ for typical clouds with $\rm FWHM=30\,\kms$ and an angular resolution of $\Delta\theta_{\rm res}=30\arcsec$. And, (2) Parkes Galactic All-Sky Survey (GASS; \citealt{2009ApJS..181..398M, 2010AA...521A..17K}), which is 63 times more sensitive than GASKAP at $\log{\left(N_{\rm H\textsc{~i},\,3\sigma}/\cm^{-2}\right)}=18.2$ and has a much larger spatial resolution of $\Delta\theta_{\rm res}=16\arcmin$. The datacubes provided by both of these surveys are already reduced and science ready.

\subsection{Optical Observations}
To characterize the spatial distribution of star-forming activity across the LMC disk, we use optical H$\alpha$ emission-line data from both the Magellanic Cloud Emission-line Survey (MCELS; \citealt{1999IAUS..190...28S}) and the Southern H$\alpha$ Sky Survey Atlas (SHASSA; \citealt{2001PASP..113.1326G}). These two surveys provide complementary strengths in angular resolution and absolute flux calibration. MCELS provides wide-field narrowband imaging of the Magellanic Clouds in several key optical emission lines, including H$\alpha$, [S\,\textsc{ii}], and [O\,\textsc{iii}], with a moderate angular resolution of $\theta \lesssim 5''$. The survey reaches a $1\sigma$ H$\alpha$ surface brightness sensitivity of $I_{\rm H\alpha,\,1\sigma} \approx 1.3~{\rm R}$ \citep{2012ApJ...755...40P}, enabling the detection of warm ($T_e \approx 10^4$~K) ionized gas with emission measures $EM\gtrsim 4~{\rm cm^{-6}\,pc}$. Owing to its relatively high spatial resolution, we primarily use the MCELS H$\alpha$ data to construct detailed emission maps and to examine the morphology of ionized gas structures across the LMC disk (see Figure~\ref{fig:ha_map}). To estimate the star-formation rate (SFR) surface density, $\Sigma_{\rm SFR}$, we rely on the SHASSA \ha\ data, which provide flux-calibrated, continuum-subtracted \ha\ emission maps over the full extent of the LMC. SHASSA has an angular resolution of $\theta\approx0\farcm8$ and a sensitivity of $I_{\rm H\alpha}\approx2$~R, corresponding to a surface brightness limit of $I_{\rm H\alpha}\approx1.2\times10^{-17}~{\rm erg\,cm^{-2}\,s^{-1}\,arcsec^{-2}}$, and have been corrected for flux calibration using the [N\,\textsc{ii}] $\lambda\lambda6549,6585$ doublet \citep{2001PASP..113.1326G}.


\section{UV Absorption-lines Analysis}\label{section:UV_Analysis}

We explore LMC outflow features associated with the following spectral transitions that are covered by the \textit{HST}/STIS and \textit{HST}/COS observations: O\textsc{~i}\,$\lambda1302$, S\textsc{~ii}\,$\lambda$$\lambda1250$, 1253, and Si\textsc{~ii}\,$\lambda$$\lambda1304$, 1526, 1808. 
In some cases, we also explore the weaker O\textsc{~i}\,$\lambda1039$ absorption features covered by FUSE because the O\textsc{~i}\,$\lambda1302$ absorption lines are often saturated. Overall, we analyzed Si\textsc{~ii} absorption along 165 sightlines, O\textsc{~i} along 156 sightlines, and S\textsc{~ii} along 150 sightlines. Additionally, for the 13~sightlines for which we perform photoionization modeling (see Table~\ref{table:targets}), we also analyze the Fe\textsc{~ii} $\lambda1608$ and Si\textsc{~iv} $\lambda\lambda1393,1402$ transitions.

\subsection{Fitting of the Spectral continuum}
\label{subsec:cont}

For each explored low-ion transition, we created smaller spectral snippets that spanned $-1{,}500\le v_{\rm LSR}\le+1{,}500\,\kms$. We shifted, extended, or truncated these spectral snippets in cases where spectra were contaminated by neighboring line transitions or stellar spectral features. We then mask all remaining spectral absorption features and fit the stellar continuum using polynomials with a typical order of $3\le n\le 8$. We established best-fit polynomial orders using goodness-of-fit tests by minimizing the reduced Chi-square ($\tilde{\chi}^2$) and used the fit with smallest order that resulted in a marginal change in the $\tilde{\chi}^2$. Further details of the continuum-fitting process, including that for high-ions, are described in \citet{2025ApJ...984..161P}. We additionally provide plotstacks of the normalized and non-normalized spectral snippets for each sightline and fitted line transition in a publicly available GitHub repository at {\url{https://github.com/sjpoudel/LMC_Winds}}.

Continuum placement is an important source of systematic uncertainty in UV absorption-line studies of massive stars, especially for OB and WR spectra that contain broad photospheric and wind features. The  stellar winds of massive stars can imprint strong P-Cygni line features on the stellar continuum, though this primarily affects the high ionization species (i.e., Si\textsc{~iv}, C\textsc{~vi}, N\textsc{~v}, O\textsc{~vi}; see \citealt{1982MNRAS.198..897W}). In contrast, low-ionization species primarily exist close to the stellar surface, where high densities permit rapid recombination. Consequently, they remain confined to the low-velocity base and do not trace the extensive velocity gradients associated with the accelerating stellar wind. In this study, we primarily explore the LMC's galactic wind via low ionization species and only include Si\textsc{~iv} along the 13~sightlines in which we conducted Cloudy radiative transfer simulations (see Table~\ref{table:targets}); further, we do not provide estimates for Si\textsc{~iv} column densities or limits when P-Cygni line features overwhelm the stellar continuum (see Table~\ref{tab:cloudy_results}).

Polynomial continuum fitting cannot perfectly reproduce all stellar features, and uncertainties in the adopted continuum can affect the inferred depth, width, and column density of individual absorption components. However, local empirical continuum fitting with low- to moderate-order polynomials is a standard and widely used approach in ISM and CGM UV absorption-line studies based on \textit{HST}/COS, \textit{HST}/STIS, and \textit{FUSE} spectra \citep[e.g.,][]{2002ApJ...569..214H, 2009ApJ...702..940L, 2011MNRAS.412.1105P, 2016ApJ...817...91B, 2017MNRAS.464.4927S, 2025ApJ...984..161P, 2025ApJ...993..195H}. Our goal is not to model the stellar atmosphere globally, but to estimate the local continuum around each interstellar absorption feature. To reduce continuum-related systematics, we masked obvious absorption and unrelated spectral features, avoided unnecessarily high-order polynomials that could artificially follow absorption 
structure, and excluded transitions or sightlines where the continuum was too uncertain, the signal-to-noise was too low, or the absorption was severely blended with stellar or unrelated features. For ions with multiple available transitions, we used consistency among independent lines as an additional check on the continuum placement and column-density measurements.

\subsection{Metal-line Fitting}
\label{subsec:Voigt}

We used the Voigt profile fitting software \textsc{VoigtFit} (version 3.14.1.1, \citealt{2018arXiv180301187K}) to analyze the multi-component absorption features in the stellar spectra. This Python program fits one or more absorption lines with Voigt-line profiles using a best-fit non-linear least-squares minimization. We use this program to simultaneously fit all components associated with multiple transitions of the same ionic species to estimate the Doppler parameters ($b$), line centers ($v_{\rm cen}$), and column densities ($N$). For example, we often simultaneously fit the weaker Si\textsc{~ii} $\lambda1808$ transition along with relatively stronger Si\textsc{~ii} $\lambda1526$ and Si\textsc{~ii} $\lambda1304$ lines to estimate the Si\textsc{~ii} column density. Similarly, we fit S\textsc{~ii}\,$\lambda1250$ and S\textsc{~ii} $\lambda1253$ transitions simultaneously to characterize S\textsc{~ii}. \textsc{VoigtFit} uses atomic data from both the Vienna Atomic Line Data Base (VALD: \citealt{1995A&AS..112..525P}) and \citet{2017ApJS..230....8C} for the oscillator strengths ($f$) when calculating column densities using the line areas. 

For the STIS observations, we accounted for the instrumental broadening using a Gaussian shape line spread function (LSF). We have discussed the consequences of using a Gaussian LSF instead of the tabulated LSFs for STIS in \citep{2025ApJ...984..161P}, where we have found that the differences in the $\log\, N$ values were less than $0.1\, \rm dex$. Similarly, Doppler $b$-values were also consistent within $<0.5\, \sigma$. However, we apply the tabulated LSFs for HST/COS to account for the broadening associated with the detector (see the Table~\ref{table:targets} notes for resolution information for all the instruments and gratings used in the analysis). We list the summary of Voigt-profile fitting results used for our analysis in Table~\ref{tab:Voigt_results}.

For saturated components, we report only the lower limits on the column density. The column density determined using the Voigt profile fitting depends on the modeled $b$-value, which is not well constrained in cases of strong saturation or blending. Therefore, we often use the apparent optical depth (AOD) method \citep{1991ApJ...379..245S} to numerically integrate the absorption to estimate lower limits in the column densities. Additionally, we also use AOD method to explore global column density trends across different kinematic ranges. We assess unresolved saturation in the absorption by comparing the resultant column densities of ions with multiple transitions. If we found that their column densities were inconsistent, then we would adopt the larger of the two column densities to minimize the saturation effect. We have discussed the AOD method in detail in \citet{2025ApJ...984..161P}.  

\begin{figure*}
    \centering
    \includegraphics[scale=0.70, trim=0 0 0 0, clip]{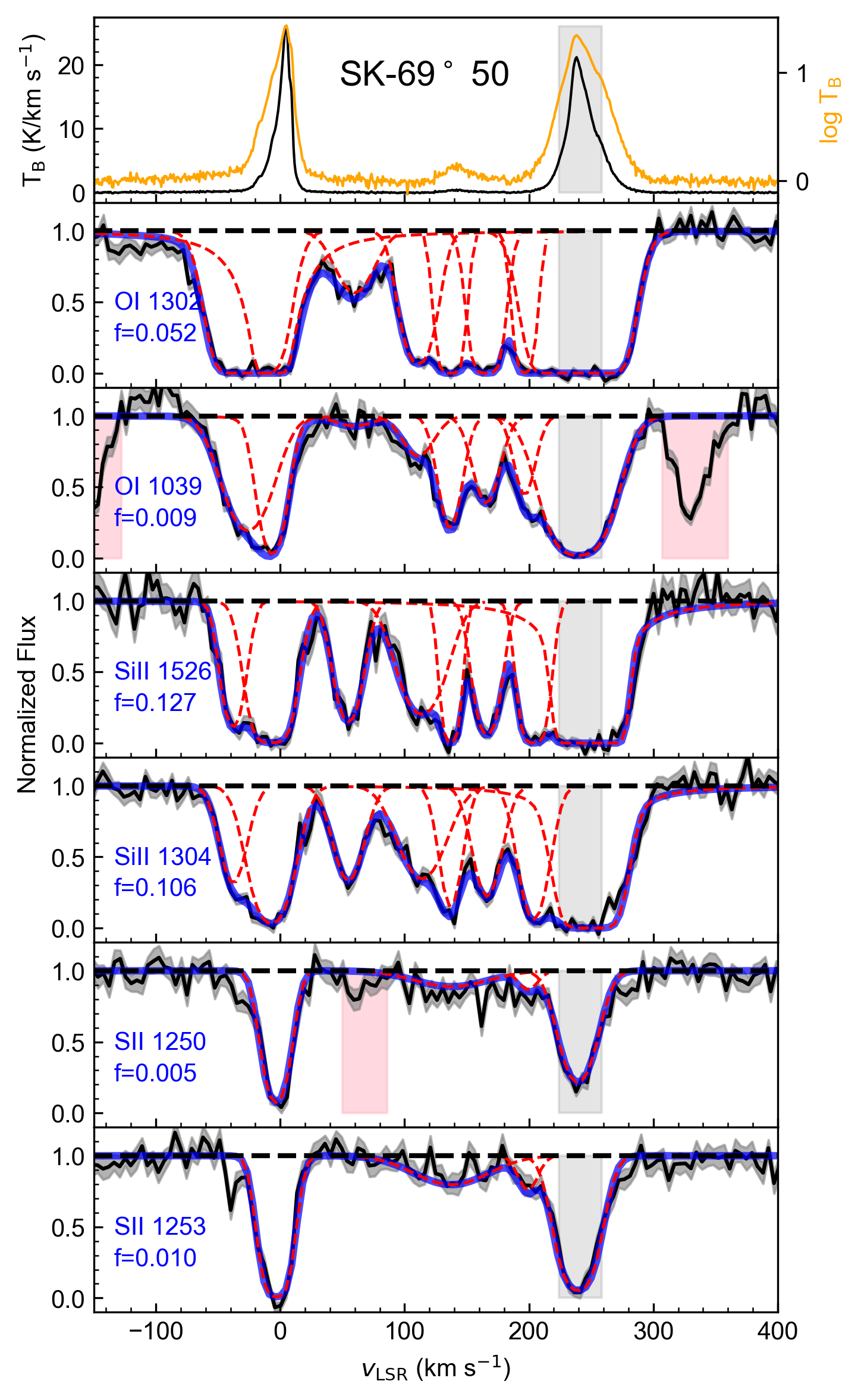}
    \includegraphics[scale=0.70, trim=0 0 0 0, clip]{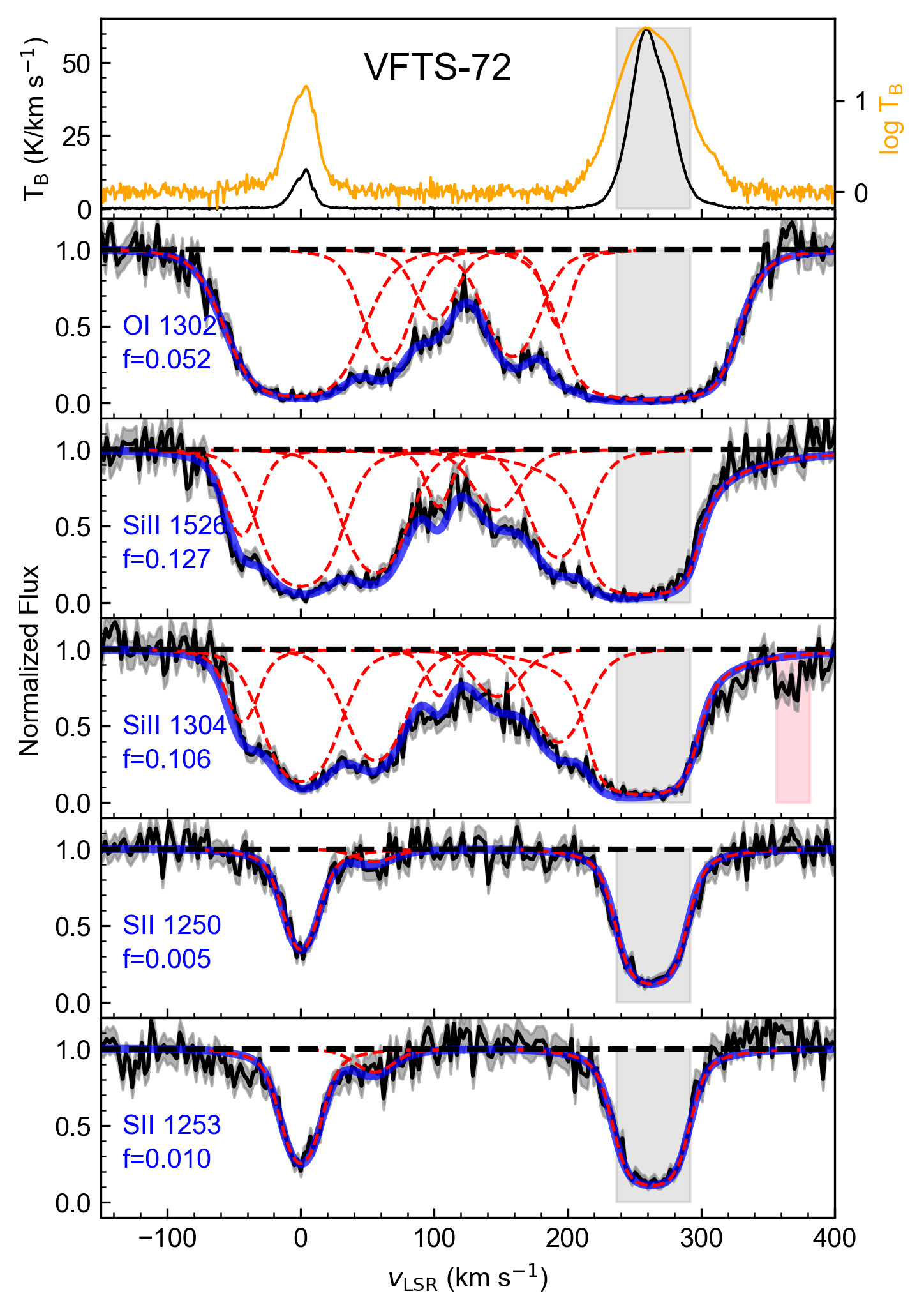}
    \caption{Velocity plots of H\textsc{~i} 21-cm emission (top panels) from the Parkes Galactic All-Sky Survey (GASS; \citealt{2010AA...521A..17K}) and UV absorption lines (bottom panels) for two representative sightlines. 
The \textit{left} panel (SK$-$69$^\circ$50) shows \textit{HST}/STIS spectra, except for O\textsc{~i} $\lambda1039$, which is taken from \textit{FUSE}, while the \textit{right} panel (VFTS$-$72) shows example of \textit{HST}/COS spectra. 
For H\textsc{~i}, an additional logarithmic-scale spectrum (orange) highlights faint features. 
The shaded gray region indicates the velocity range of the LMC H\textsc{~i} disk. 
Voigt-profile fits are shown for the metal absorption lines, with oscillator strengths listed in each panel. 
The normalized flux is plotted in black, with the 1$\sigma$ uncertainty shown as a gray envelope. 
Dashed red curves represent individual Voigt components, and the blue curves show their combined fit. 
Unrelated absorption (not included in the Voigt fits) is shaded in pink and the horizontal dashed line in black represents the continuum level. Similar plot stacks for all sightlines analyzed in this work are provided in a publicly available GitHub repository ({\url{https://github.com/sjpoudel/LMC_Winds}}).}
\label{fig:voigt_figs}
\end{figure*}

\begin{table}
\centering
\caption{Summary of Voigt profile-fitting results} 
\label{tab:Voigt_results}
\begin{tabular}{cccccc}
\hline
\hline
Ion & $v_{\rm LSR}$ & $v_{\rm LMCSR}$  &  $\log{\left(N_x/\cm^{-2}\right)}$ & $b$  \\
    & (\kms) & (\kms) &   & (\kms)  \\
\hline
\multicolumn{5}{c}{\textbf{Sk$-$68\ensuremath{^\circ}80}}\\
\multicolumn{5}{c}{LMC \hi\, disk: $+228 < v_{\rm LSR} < +272$;  $-37 < v_{\rm LMCSR} < +7$} \\
\tableline
O~\textsc{i} & $-23.8 \pm 10.2$ & $-289.3$ & $>15.20$ & $30.1 \pm 4.4$ \\
 & $+6.0 \pm 12.0$ & $-259.5$ & $14.83 \pm 0.73$ & $12.9 \pm 8.2$ \\
 & $+56.8 \pm 1.5$ & $-208.7$ & $14.69 \pm 0.03$ & $24.4 \pm 2.1$ \\
 & $+121.3 \pm 0.6$ & $-144.2$ & $14.85 \pm 0.03$ & $18.9 \pm 0.8$ \\
 & $+173.8 \pm 1.5$ & $-91.7$ & $14.40 \pm 0.12$ & $9.0 \pm 1.5$ \\
 & $+227.9 \pm 4.8$ & $-37.6$ & $>15.69$ & $26.8 \pm 5.2$ \\
 & $+268.7 \pm 8.1$ & $+3.2$ & $14.04 \pm 0.94$ & $5.9 \pm 7.2$ \\
\multicolumn{5}{c}{}\\[-1.2ex]
Si~\textsc{ii} & $-37.1 \pm 1.5$ & $-302.6$ & $14.13 \pm 0.05$ & $16.1 \pm 1.1$ \\
 & $-1.2 \pm 0.9$ & $-266.7$ & $>14.61$ & $17.9 \pm 1.3$ \\
 & $+55.3 \pm 0.6$ & $-210.2$ & $14.33 \pm 0.01$ & $21.3 \pm 0.9$ \\
 & $+118.2 \pm 0.3$ & $-147.3$ & $14.34 \pm 0.01$ & $20.6 \pm 0.6$ \\
 & $+172.5 \pm 0.6$ & $-93.0$ & $13.89 \pm 0.04$ & $8.5 \pm 0.8$ \\
 & $+219.3 \pm 1.8$ & $-46.2$ & $>14.84$ & $24.7 \pm 2.6$ \\
 & $+261.1 \pm 4.8$ & $-4.4$ & $13.92 \pm 0.18$ & $18.4 \pm 3.0$ \\
\multicolumn{5}{c}{}\\[-1.2ex]
S~\textsc{ii} & $-17.7 \pm 6.6$ & $-283.2$ & $14.87 \pm 0.07$ & $48.1 \pm 5.9$ \\
 & $+1.0 \pm 0.3$ & $-264.5$ & $15.48 \pm 0.03$ & $9.9 \pm 0.4$ \\
 & $+60.7 \pm 4.2$ & $-204.8$ & $14.31 \pm 0.15$ & $21.8 \pm 6.7$ \\
 & $+122.9 \pm 2.7$ & $-142.6$ & $14.44 \pm 0.06$ & $24.2 \pm 4.3$ \\
 & $+176.4 \pm 2.1$ & $-89.1$ & $14.32 \pm 0.09$ & $12.8 \pm 3.3$ \\
 & $+198.1 \pm 1.2$ & $-67.4$ & $14.21 \pm 0.13$ & $6.0 \pm 2.2$ \\
 & $+225.7 \pm 0.6$ & $-39.8$ & $15.18 \pm 0.01$ & $16.9 \pm 0.7$ \\  
\multicolumn{5}{c}{}\\[-1.2ex]
Si~\textsc{iv} & $-2.2 \pm 0.7$ &   $-267.7$   &  $13.13 \pm 0.02$ &   $14.9 \pm 1.0$ \\
& $+41.5 \pm 1.1$ &  $-224.0$   &  $13.07 \pm 0.03$ &   $19.3 \pm 1.8$ \\
& $+110.3 \pm 1.3$ & $-155.2$   &  $12.99 \pm 0.03$ &   $23.9 \pm 2.3$ \\
& $+160.2 \pm 0.8$ & $-105.3$   &  $13.05 \pm 0.04$ &   $7.7 \pm 1.3$ \\
& $+185.3 \pm 0.5$ & $-80.2$    &  $13.74 \pm 0.04$ &   $9.0 \pm 0.9$ \\
& $+236.3 \pm 1.9$ & $-29.2$    &  $14.20 \pm 0.04$ &   $26.6 \pm 1.8$ \\
& $+267.1 \pm 1.3$ & $+1.6$     &  $15.99 \pm 0.25$ &   $5.2 \pm 0.5$ \\
\hline     
\end{tabular}
\tablecomments{An example of the Voigt profile fitting outcomes for all the components spanning from MW to LMC across various ion species along the sightline Sk$-$68$^\circ$80. The errors in $v_{\rm LSR}$ also represent the errors in $v_{\rm LMCSR}$. Additionally, we also list the kinematic range of the LMC \hi\, disk in both the LSR and LMCSR velocity frames. The results for all of the sightlines are available in machine-readable form in the online article.}
\end{table}

\section{Identifying and probing the LMC outflows}
\label{section:Prob_wind}
In order to characterize the LMC's galactic winds, we must first separate them from the LMC's disk. For this, we start by comparing the \hi\, 21-cm emission-line data from GASKAP with the UV absorption lines from ULLYSES. We generally define the kinematic extent of the \hi\, gas in its disk to be where the signal is at least $3\sigma$ above the RMS noise in the continuum. However, we also compare this \hi\, derived disk width with low-ion UV absorption for weak transitions and make small adjustments when necessary. These criteria are consistent with the \hi\, column density per pixel at the disk boundary being $\log{\left(N_{\mathrm{H\textsc{~i}}}/\mathrm{cm}^{-2}\right)} \approx 18.9$ for most of the sightlines. We list the adopted kinematic range of the LMC \hi\, disk in both the LSR and LMCSR velocity frames for each sightline in Table~\ref{tab:Voigt_results}. We then identify absorbers associated with the nearside outflows as those that are blueshifted with respect to the LMC disk. 

\subsection{Kinematic distribution of column densities}
\label{subsection:kinematic_dist}

\begin{figure*}[ht!]
  \centering
  \includegraphics[width=0.75\textwidth]{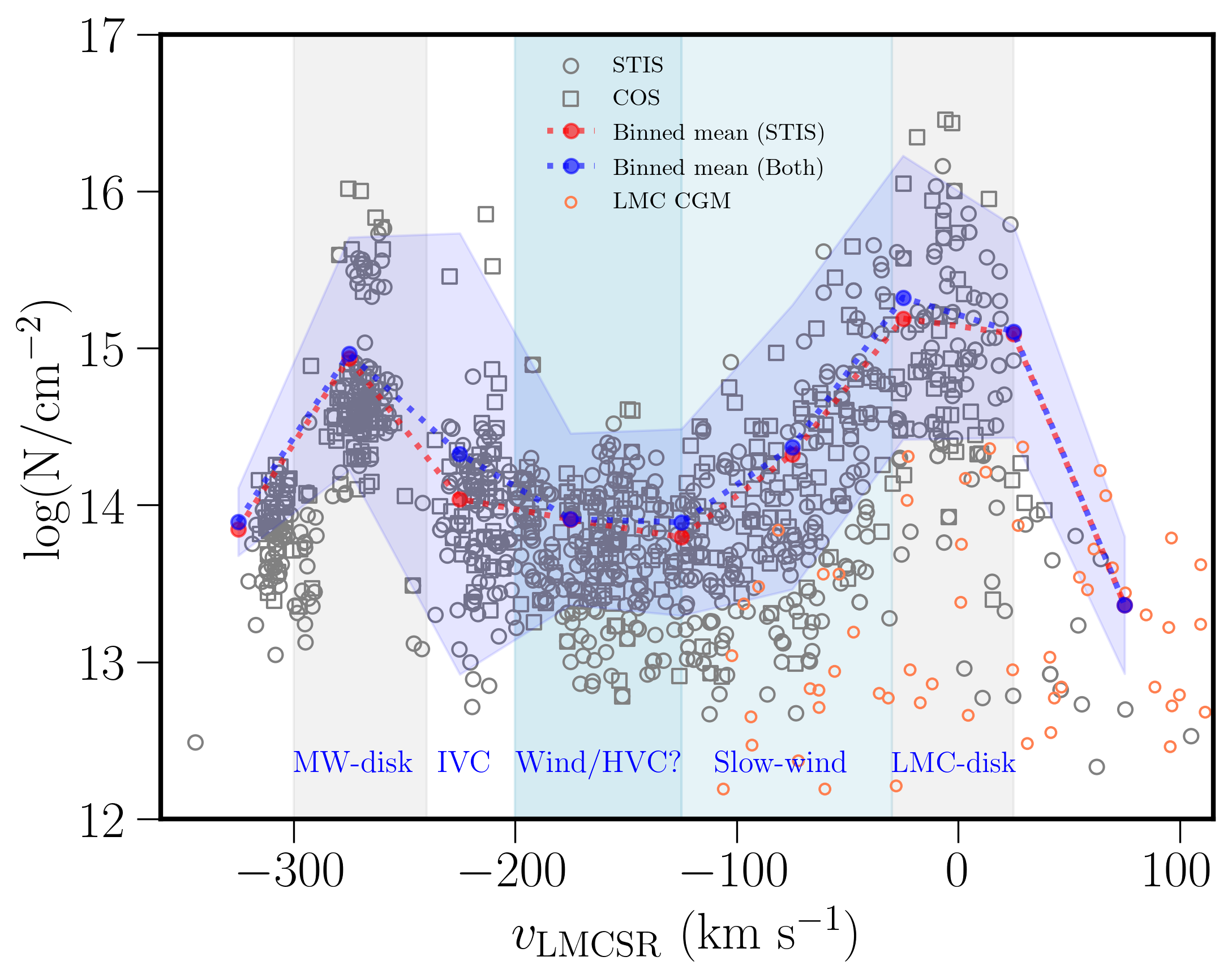}
  \caption{Column density distributions of individual Voigt-fitted components for Si\textsc{~ii} as a function of velocity with respect to the LMC. The gray circles and squares represent individual $\log\rm N$ values observed with STIS and COS, respectively. The binned averages, computed in $50\, \kms$ intervals, are shown as red dots for STIS and as blue dots for STIS+COS. The light shade in blue around the binned averages represents the $1\, \sigma$ scatter from the binned means for STIS+COS. The gray vertically shaded region in the right that is centered at $v_{\rm LMCSR}\approx0\,\kms$ represents the mean disk width of LMC estimated from \hi\, emission from GASKAP for the entire sightlines. The gray vertically shaded region in the left at $v_{\rm LMCSR}\approx-270\,\kms$ indicates the approximate width of the MW disk. Approximate kinematic widths for various regions of interest are also vertically shaded and labeled above. For comparison, $\log{N_{\rm Si\textsc{~ii}}}$ values for LMC's CGM from \citet{2024ApJ...976L..28M} are plotted as orange circles and are generally located in the bottom right.}
  \label{fig:SiII_logN}
\end{figure*}

\begin{figure}[ht!]
  \centering
  \includegraphics[width=0.48\textwidth]{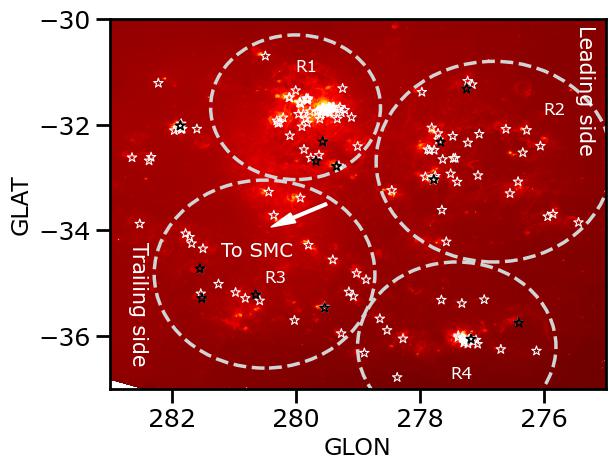}
  \caption{Similar to Figure~\ref{fig:ha_map}, but showing the division of the LMC into four regions (R1–R4) used to study the spatial variation of gas properties. Regions R1 and R4 are centered on the well-known star-forming regions 30~Doradus and N11, respectively, while R2 and R3 correspond to the leading and trailing sides of the LMC. The \ha\, emission map is from MCELS  \citep{1999IAUS..190...28S}.}
  \label{fig:hi_region_map}
\end{figure}

\begin{figure*}[ht!]
  \centering
  \includegraphics[scale=0.55, trim=0 55 0 0, clip]{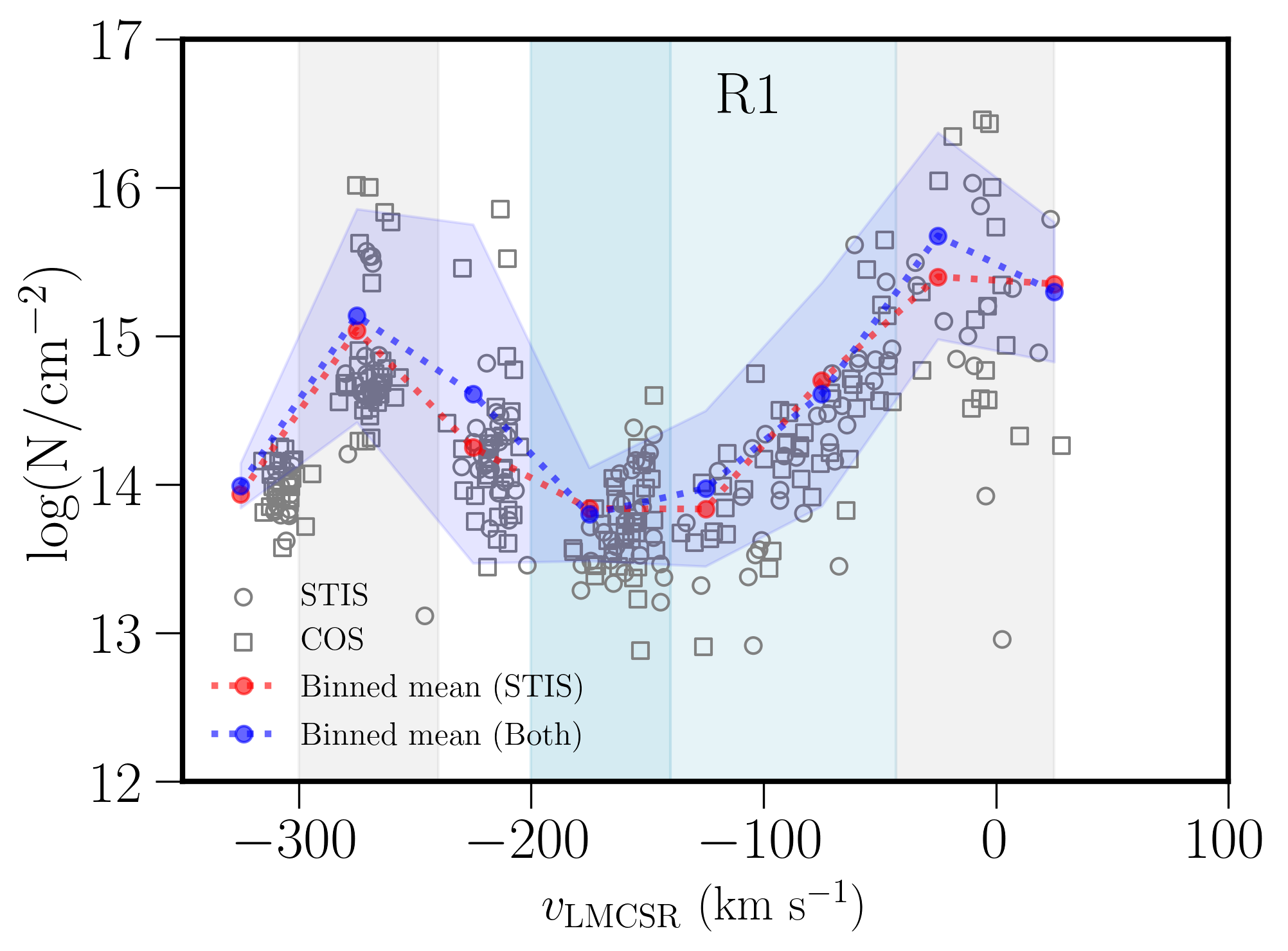}
  \hspace{-10pt}
  \includegraphics[scale=0.55, trim=32 30 0 0, clip]
  {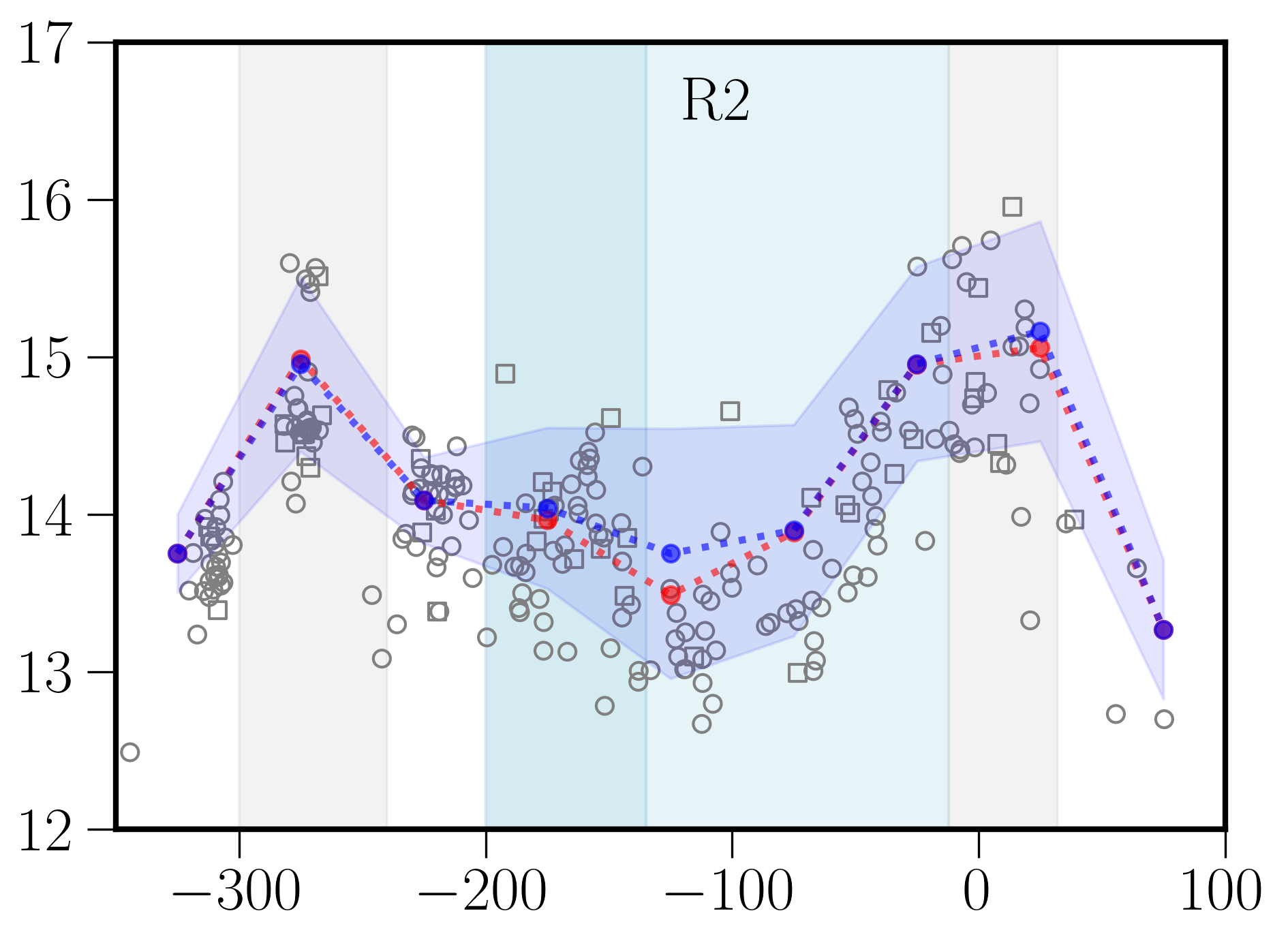}
  \includegraphics[scale=0.55, trim=0 0 0 0, clip]{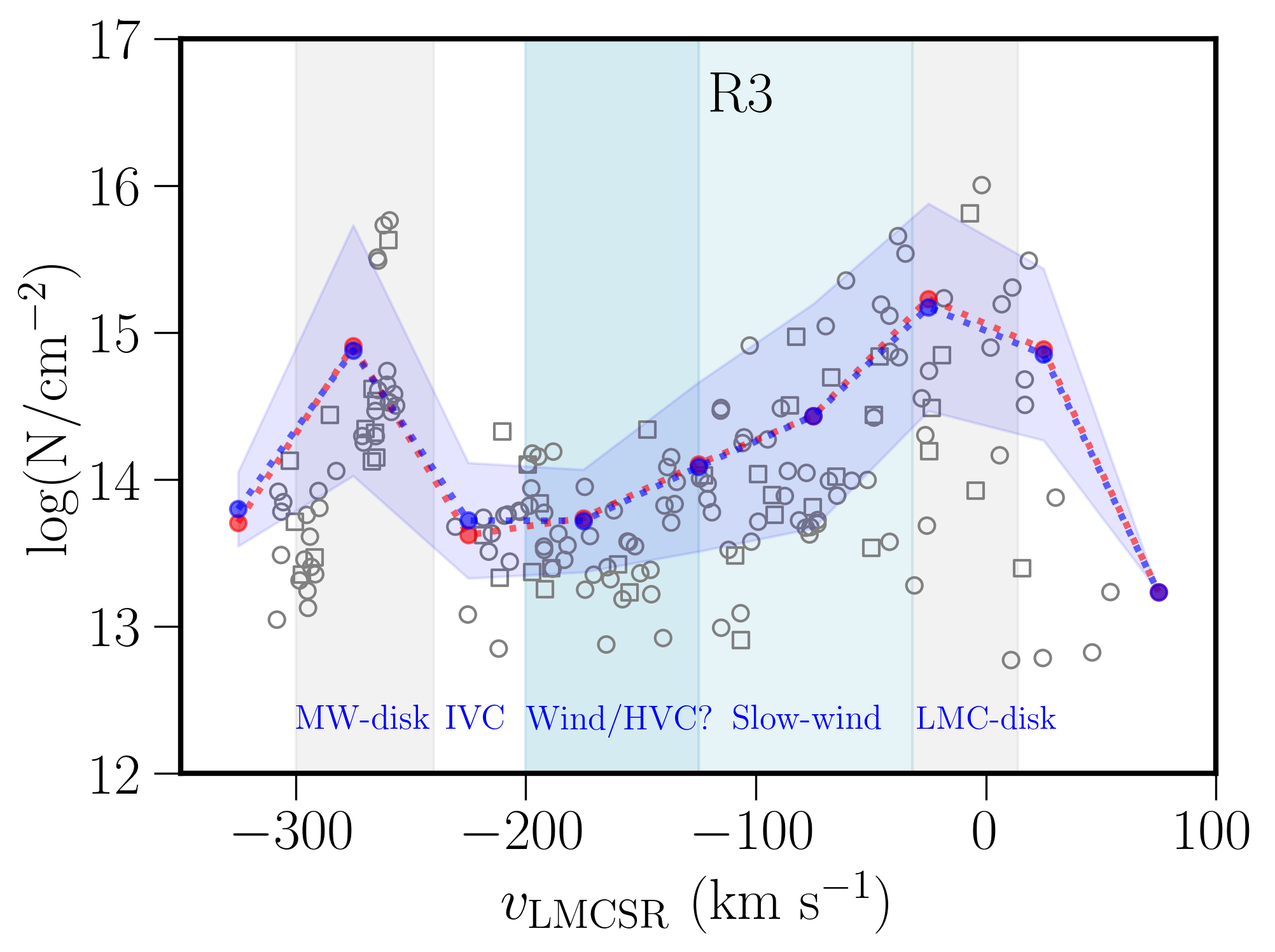}
  \hspace{-10pt}
  \includegraphics[scale=0.55, trim=32 0 0 0, clip]
  {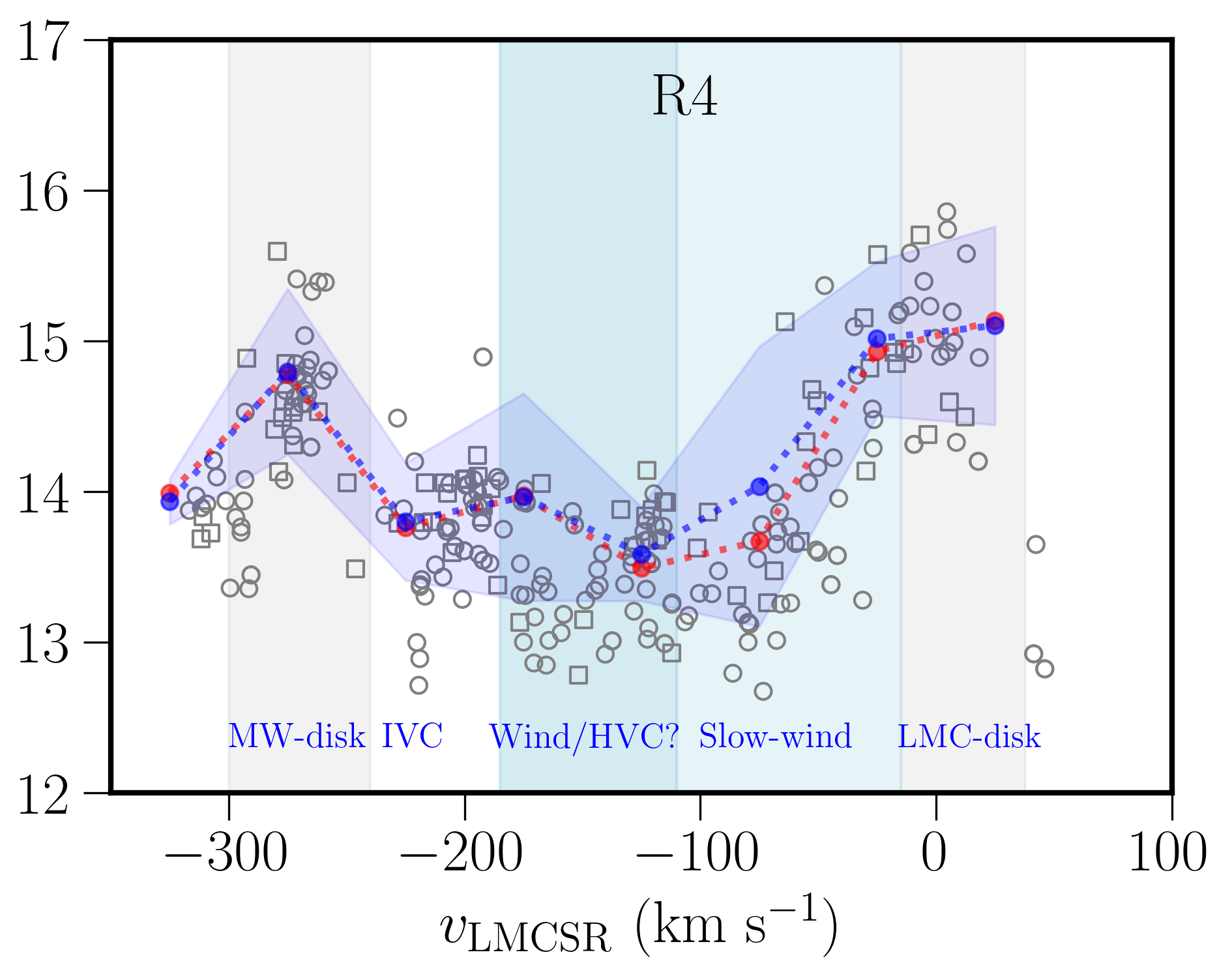}
  \caption{Column density distributions of individual Voigt-fitted components for Si\textsc{~ii} as a function of velocity with respect to the LMC for 4 different regions defined in Figure~\ref{fig:hi_region_map}. The gray circles and squares represent individual $\log\rm N$ values observed with STIS and COS, respectively. The binned averages, computed in $50\, \kms$ intervals, are shown as red dots for STIS and as blue dots for STIS+COS. The light shade in blue around the binned points represents the $1\, \sigma$ scatter from the binned means for STIS+COS. The gray shaded region in the right indicates the mean disk width of LMC estimated from \hi\, emission from GASKAP for the sightlines inside each regions defined.}
  \label{fig:SiII_logN_regions}
\end{figure*}

We investigate the distribution of column densities derived from Voigt-profile fitting of individual components spanning the full kinematic range from the MW to the LMC. We focus on the Si\textsc{~ii} ion for this analysis because we have access to multiple line transitions of different strengths that aid us measuring the column density even if one of the transitions is too weak or saturated. We include the following lines in this analysis: Si\textsc{~ii}\,$\lambda$1526, Si\textsc{~ii}\,$\lambda$1304, and, in several cases, Si\textsc{~ii}\,$\lambda$1808. The inclusion of the weaker Si\textsc{~ii}\,$\lambda$1808 line allows for more robust column density measurements and improved deblending of individual absorption components.

We find that there is a global trend in which the absorbers moving at speeds similar to the MW and LMC have the highest Si\textsc{~ii} column densities, which drops with increasing kinematic offset for either galaxy  (see Figure~\ref{fig:SiII_logN}). The average Si\textsc{~ii} column density reaches a minimum between $v_{\mathrm{LMCSR}}\approx-200$ and $-125$\,km\,s$^{-1}$, where the MW resides at $v_{\mathrm{LMCSR}}\approx-270\,\kms$ and the LMC at $v_{\mathrm{LMCSR}}=0\,\kms$. Therefore, the column density for nearside winds of the LMC steeply decline with increasing outflow speed. Assuming that these faster lower column density absorbers are diffused and have less mass, they may be more easily accelerated to higher velocities in galactic outflows. This scenario is consistent with the \citet{2002ASPC..254..292H} model for accelerating clouds in a wind in which the cloud's speed varies with its column density as $v_{\rm cloud} \propto \frac{1}{\sqrt{N}}$. We found similar behavior for the blueshifted absorbers that lie in the foreground of the 30~Doradus region \citep{2025ApJ...984..161P}. This trend persists whether we include only the STIS or both the COS and STIS dataset. Including the kinematically less resolved COS measurements tend to increase the total binned mean values only by a very small amount. This difference is likely due to more component blending that result in higher inferred column densities for unresolved components. 

The monotonic decline of column density with increasing blueshift velocity is qualitatively consistent with models of multiphase galactic winds in which dense clouds are entrained and accelerated by a lower-density hot outflow. In such scenarios, clouds with lower column densities experience less inertia and can be accelerated more efficiently by ram pressure, producing a velocity–column density relation similar to the trend observed here \citep{2002ASPC..254..292H}. Numerical simulations of winds also show that clouds interacting with a hot outflow can undergo fragmentation, mixing, and mass loss as they are accelerated away from the disk \citep{2016ApJ...819...29B, 2020ApJ...893...29B}. These processes can produce progressively lower-column-density clouds at larger velocities and distances from the disk. The observed trend may therefore reflect a combination of dynamical acceleration and cloud evolution within the LMC wind. While the inverse relation between cloud speed and column density provides a useful empirical framework for ram-pressure acceleration \citep{2002ASPC..254..292H}, more recent models of mixing-driven cloud acceleration
show that clouds of different initial masses can evolve toward similar
velocities once they are entrained in a turbulent wind \citep[e.g.,][]{2022ApJ...924...82F}. In addition, low-column-density clouds may be disrupted if their mixing time is shorter than their cooling time.

To compare the spatial variation in the Si\textsc{~ii} column densities across the LMC disk, we defined four distinct regions that we outline in Figure~\ref{fig:hi_region_map}. The regions include: (1) R1 with radius $1.4^{\circ}$, centered on the most active star-forming region, 30~Doradus; (2) R2 with radius $1.9^{\circ}$, which contains the star-forming regions along the leading edge of the LMC; (3) R3 with radius $1.8^{\circ}$, with star-forming regions along the trailing edge of the LMC; and (4) R4 with radius $1.6^{\circ}$, centered on the second most prominent star-forming region, N11. Across all four regions, the general relation between column density for the low ionization species and velocity observed for the global sightlines is preserved, despite varying degrees of scatter among these regions (see Figure~\ref{fig:SiII_logN_regions}). Region~R1, which hosts 30~Doradus at its center, exhibits the least scatter, possibly indicating a tighter velocity and column density relation than other regions.

The continuity of the lower-column-density absorbers with the slow-wind sequence, together with the presence of higher-column-density absorbers offset from this trend, suggests that the ambiguous-velocity regime may include gas with multiple origins or evolutionary states; we therefore test this possibility quantitatively by measuring the offsets of these components from the extrapolated slow-wind relation. We fitted the Si\textsc{~ii} component column densities in the slow-wind interval and extrapolated this relation into the ambiguous-velocity range (see Figure~\ref{fig:population_test} in the Appendix). The best-fit
relation is
\begin{equation}\label{eq:bestfitsii}
\log\left(N_{\rm Si\,II}/\cm^{-2}\right) = 0.013\,\left(v_{\rm LMCSR}/\kms\right) + 15.16,
\end{equation}
with a residual scatter of $0.58\,\rm dex$ about the slow-wind trend. We then
computed the offset of each ambiguous-velocity component from this relation,
\[
\Delta \log N_{\rm Si\,II} =
\log N_{\rm Si\,II,obs} -
\log N_{\rm Si\,II,trend}(v_{\rm LMCSR}).
\]

Out of 349~components in this interval, 190~lie within the adopted near-trend region, whereas 159~have positive offsets larger than the residual scatter about the slow-wind trend. The median
offset is $\Delta \log N_{\rm Si\,II}=0.5$ dex, and the maximum offset reaches $2.2$ dex. Because the median offset is smaller than the intrinsic scatter of the slow-wind relation, these data do not provide strong evidence for two statistically distinct populations. Instead, they suggest that the ambiguous
velocity interval contains a broad distribution of Si\textsc{~ii} column
densities, with a high-column-density tail. Thus, we do not treat the offset distribution as a formal population division; instead, we use it as an empirical diagnostic that may point to mixed origins or different evolutionary states within the ambiguous-velocity gas.

We compare our observed Si\textsc{~ii} column densities trends with that observed by \citet{2024ApJ...976L..28M} for the LMC's CGM (see orange circles in Figure~\ref{fig:SiII_logN}). Using UV absorption-line observations toward background quasars that probe the LMC halo, \citet{2024ApJ...976L..28M} found that the LMC hosts a compact and cool CGM that exhibits a clear truncation at 17~kpc; their findings are consistent with ram-pressure stripping of the LMC halo as this galaxy travels through the Milky Way's halo. At comparable velocities, the CGM absorbers preferentially populate lower Si\textsc{~ii} column densities, often lying below the binned means defined by the combined STIS+COS sample. In contrast, our absorption components---particularly in the slow-wind and LMC-disk regimes---exhibit systematically higher column densities and a well-defined velocity-dependent structure. This indicates that the CGM gas traced by \citet{2024ApJ...976L..28M} is more diffuse and less centrally concentrated than the gas probed by stellar sightlines through the LMC disk.

\subsection{Comparisons with Stellar Activity}
\label{subsection:Gas_distribution}

We examine whether the foreground gas column density associated with low-ions correlates with local star-formation activity in the LMC, as traced by the star-formation rate surface density ($\Sigma_{\rm SFR}$). Such a correlation would provide insight into the connection between stellar feedback and the kinematics and ionization structure of foreground gas. To isolate physically distinct outflow components, we perform this analysis separately over two velocity intervals that are motivated by previous studies of LMC outflows and foreground high-velocity gas: (1) from the LMC disk boundary to $v_{\rm LMCSR} = -125~\kms$, where absorption is dominated by the slower component of the LMC wind (see also Section~\ref{subsection:kinematic_dist}); and (2) the more extreme range $-200 \le v_{\rm LMCSR} \le -125~\kms$, which encompasses the faster LMC outflow component and overlaps with the velocity regime occupied by the Milky Way HVC population \citep{2025ApJ...984..161P,2015A&A...584L...6R,Ciampa2020}. We focus on Si\textsc{~ii} for this comparison because we are able to more consistently measure its column density in comparison to other common low-ion tracers. O\textsc{~i} is frequently saturated in our data, while S\textsc{~ii} is often too weak to be reliably detected at the velocities where the LMC wind and Milky Way HVCs are expected to overlap. Although the strongest Si\textsc{~ii} transitions can also be saturated, the availability of multiple weaker lines (Si\textsc{~ii} $\lambda\lambda1808,1304$) allows us to robustly characterize the absorption across the relevant velocity ranges. To measure total Si\textsc{~ii} column densities, we adopt the AOD method rather than summing individual Voigt components. This approach enables integration over fixed velocity intervals and provides more reliable lower limits when portions of the absorption profile are saturated.

To estimate $\Sigma_{\rm SFR}$, we measure the average H$\alpha$ flux of the LMC interstellar medium within a circular aperture of radius $75\arcsec$ centered on each UV background target. Given the angular resolution of the SHASSA data ($\Delta\theta_{\rm SHASSA} = 48\arcsec$), this aperture provides a representative sampling of the local H$\alpha$ emission. We correct the observed H$\alpha$ flux for extinction using the prescription from \citet{2023ApJ...948..118S},
\begin{equation}
A(\mathrm{H}\alpha) = 2.31 \times 10^{-22} \langle N_{\mathrm{H\,I}} \rangle~\mathrm{cm^{2}~atoms^{-1}~mag},
\end{equation}
where $\langle N_{\mathrm{H\textsc{~i}}} \rangle$ is obtained by integrating the GASKAP H\textsc{~i} spectrum over the LMC disk velocity range defined in Section~\ref{section:Prob_wind}. The intrinsic H$\alpha$ luminosity is then calculated as
\begin{equation}
L_{\mathrm{int}}(\mathrm{H}\alpha) = \frac{L_{\mathrm{obs}}(\mathrm{H}\alpha)}{10^{-0.4A_{\mathrm{H\alpha}}}},
\end{equation}
and converted to a star-formation rate using the following \citet{2012ARA&A..50..531K} relationship:
\begin{equation}
\log \left(\frac{\mathrm{SFR}}{M_\odot\,\yr^{-1}}\right) = \log \left(\frac{L_{\mathrm{int}}(\mathrm{H}\alpha)}{\rm ergs\,s^{-1} }\right) - 41.27.
\end{equation}

\begin{figure*}[ht!]
  \centering
  \includegraphics[scale=0.45, trim=0 0 0 0, clip]{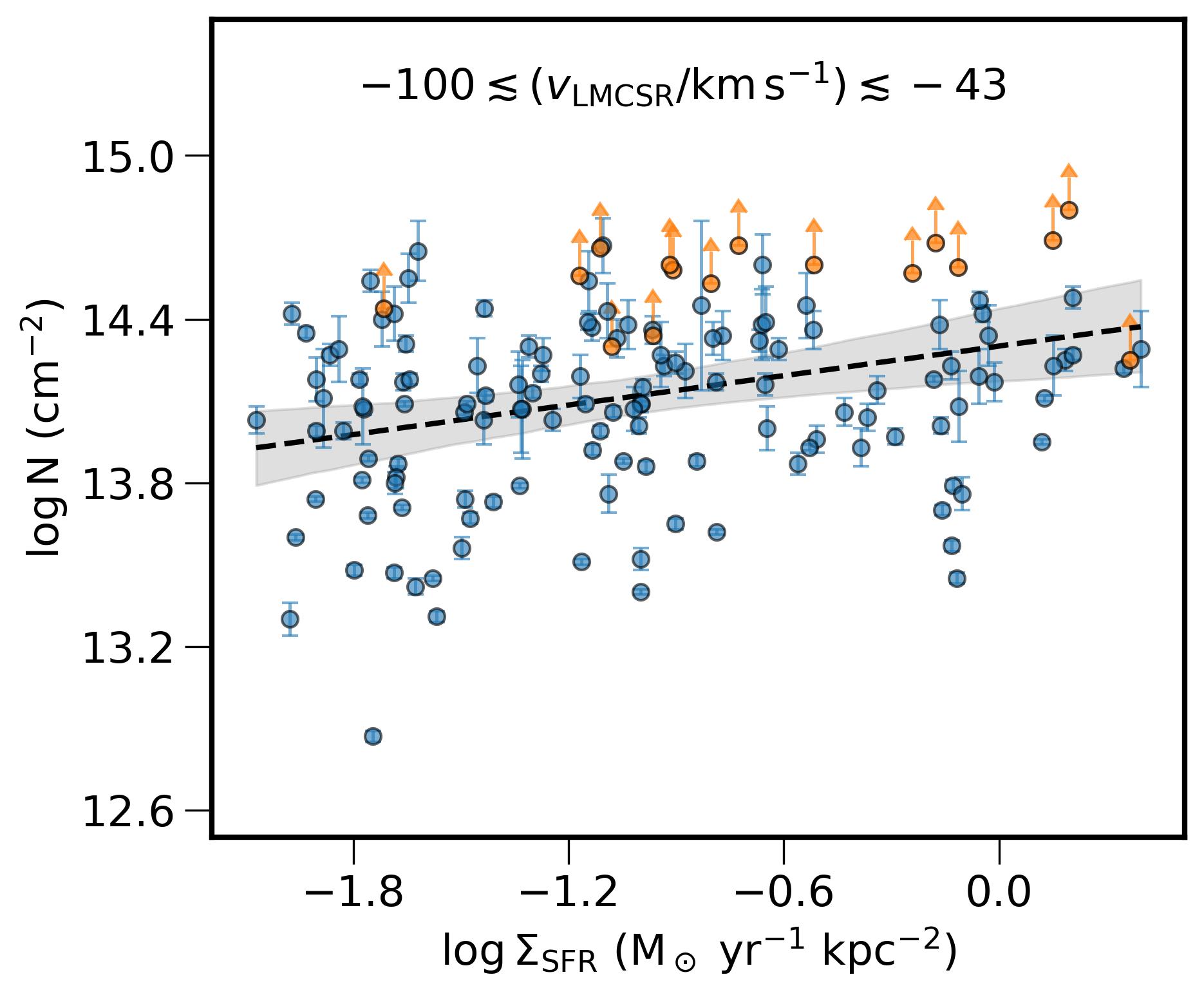}
  \includegraphics[scale=0.45, trim=70 0 0 0, clip]
  {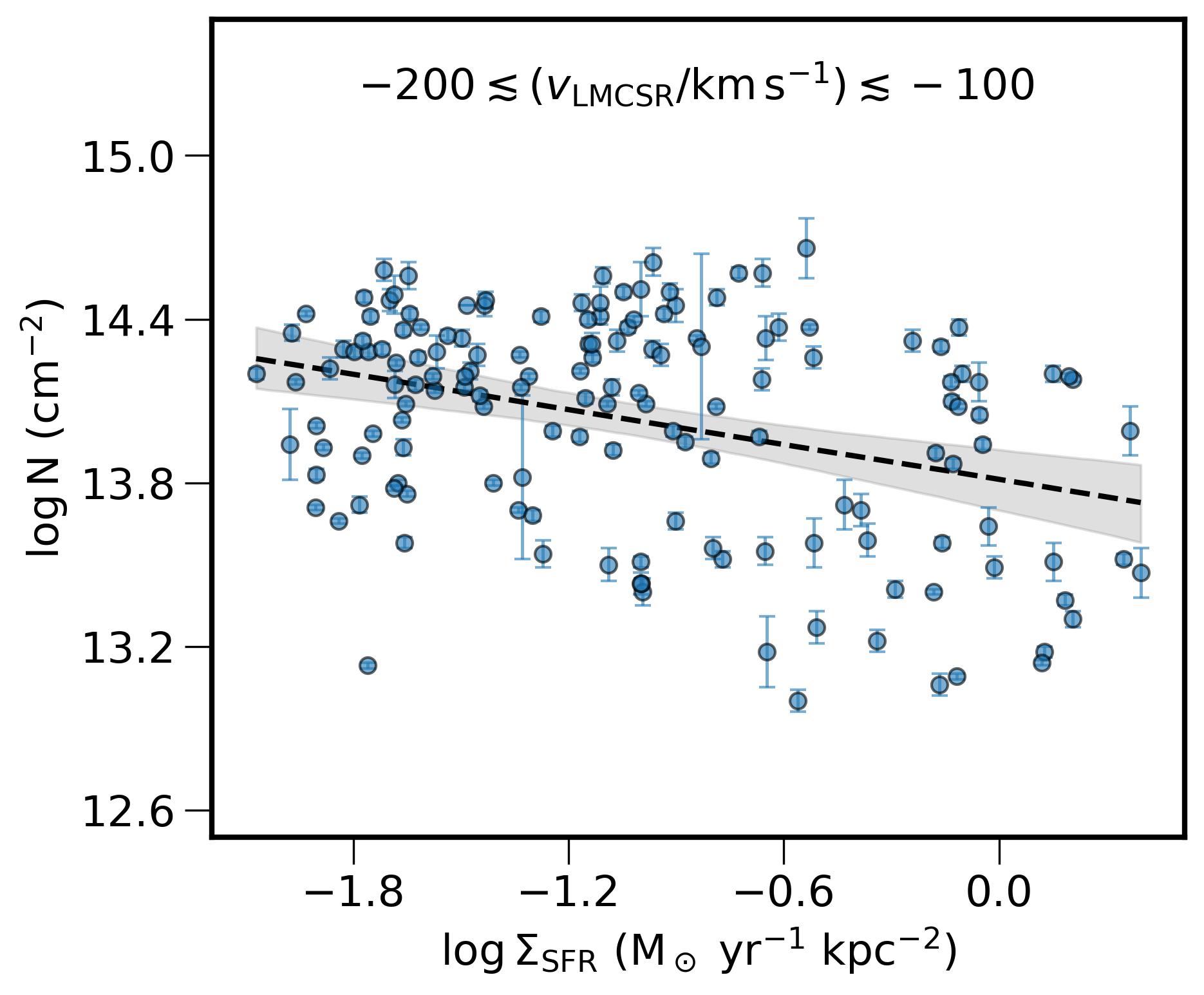}
  \caption{Si\textsc{~ii} column densities for foreground LMC absorbers plotted against star formation rate surface density for the LMC's ISM. Each point represents an individual stellar sightline, with Si\textsc{~ii} detections shown as circles and lower limits as upward-pointing arrows,  
  determined from the AOD method. The left panel shows the low-velocity range $-100 < v_{\rm LMCSR} < v_{\rm DISK}~\kms$, primarily tracing the slower LMC wind near the disk boundary. The right panel shows the $-200 < v_{\rm LMCSR} < -100~\kms$, tracing both Milky Way high-velocity clouds and LMC outflows. We perform Bayesian linear regression between log~$\Sigma_{\mathrm{SFR}}$ and log~$N_{\rm Si\textsc{~ii}}$ taking into account the lower limits as well. The fits account for measurement uncertainties and censored (saturated) data using a survival likelihood. Each panel shows the median posterior regression line (black) and the 95\% credible interval (gray band) derived from the \texttt{PyMC} model.}
  \label{fig:ha_logN_cor}
\end{figure*}

Our derived $\Sigma_{\rm SFR}$ values are broadly consistent with those inferred from resolved stellar population studies \citep{2021MNRAS.508..245M}, which trace recent star formation over similar spatial scales. Considering all sightlines together, we find that for the slower-moving gas ($-100\,\kms < v_{\rm LMCSR} < v_{\rm Disk}$) there is a weak but suggestive positive correlation between $\Sigma_{\rm SFR}$ and log~$N_{\rm Si\textsc{~ii}}$ (see left panel in Figure~\ref{fig:ha_logN_cor}), consistent with an origin in stellar-driven outflows from the LMC. Although the data exhibit substantial scatter, our Bayesian regression that incorporates survival analysis to account for lower limits indicates that a positive trend is statistically favored over no correlation. In contrast, the higher-velocity bin ($-200 < v_{\rm LMCSR} < -100~\kms$) has an apparent negative correlation between $\Sigma_{\rm SFR}$ and log~$N_{\rm Si\textsc{~ii}}$, accompanied by large scatter (see right panel in Figure~\ref{fig:ha_logN_cor}). At first glance, such a trend would be unexpected if the gas in this velocity range were produced solely by recent local star formation. However, as discussed in Section~\ref{subsection:kinematic_dist}, this velocity interval contains a broad distribution of Si\textsc{~ii} column-density offsets relative to the extrapolated slow-wind relation, including a high-column-density tail. Thus, the apparent negative trend may reflect a mixture of components with different origins and physical conditions, rather than a true anti-correlation between star-formation activity and outflow strength.

\begin{figure*}[ht!]
  \centering
  \includegraphics[scale=0.40, trim=0 30 0 0, clip]{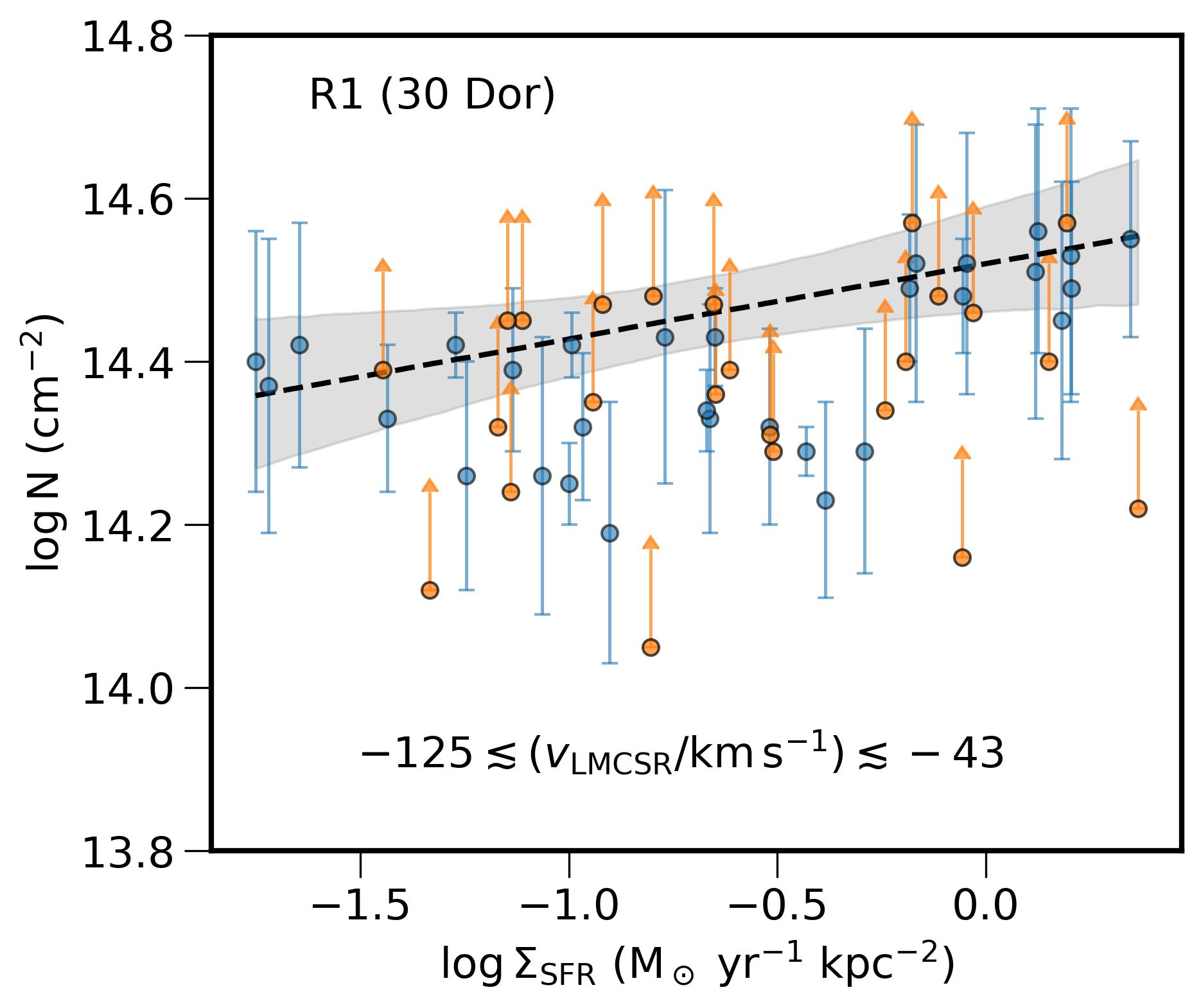}
  \includegraphics[scale=0.40, trim=30 30 0 0, clip]
  {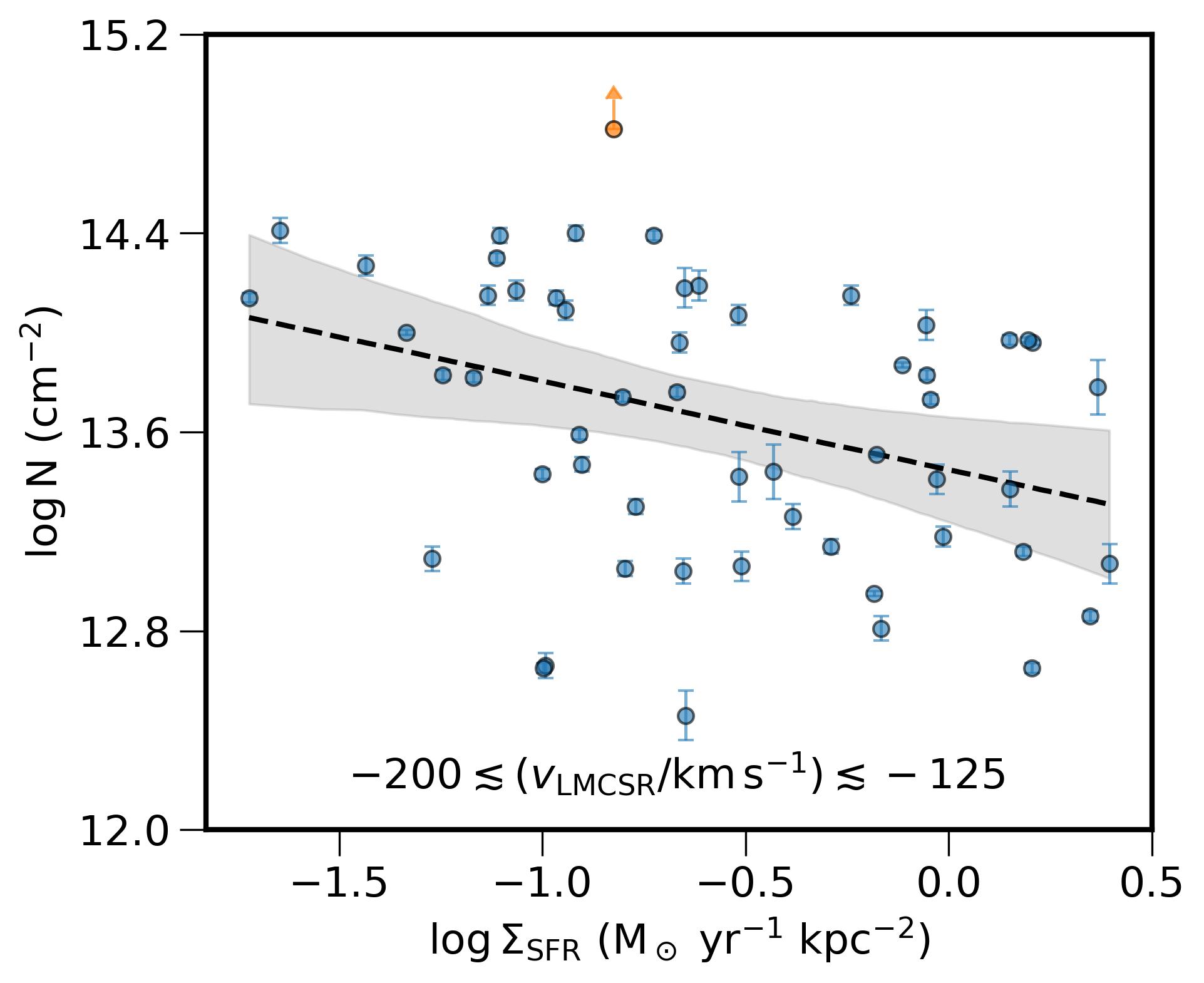}
  \includegraphics[scale=0.40, trim=0 30 0 0, clip]{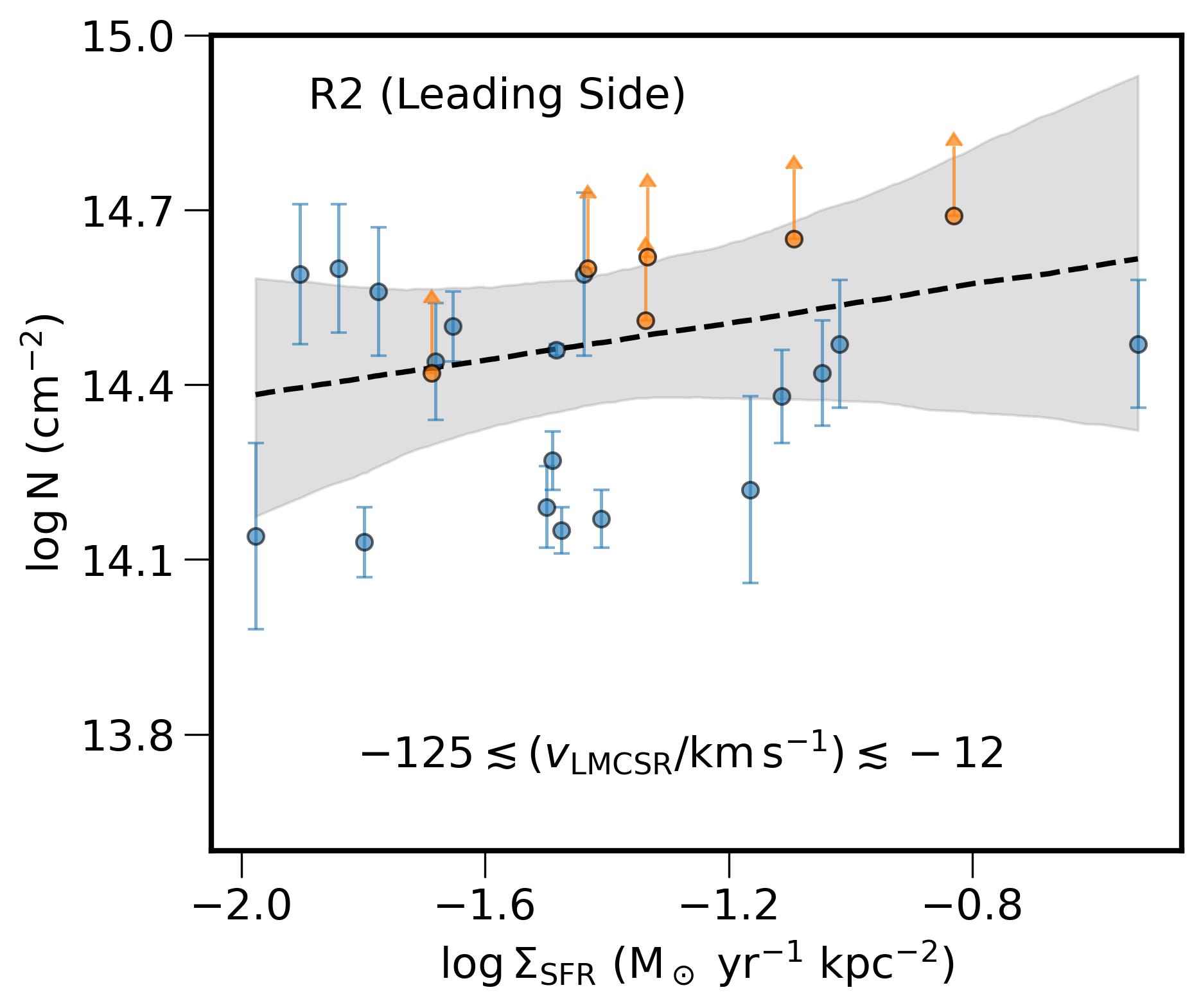}
  \includegraphics[scale=0.40, trim=30 30 0 0, clip]
  {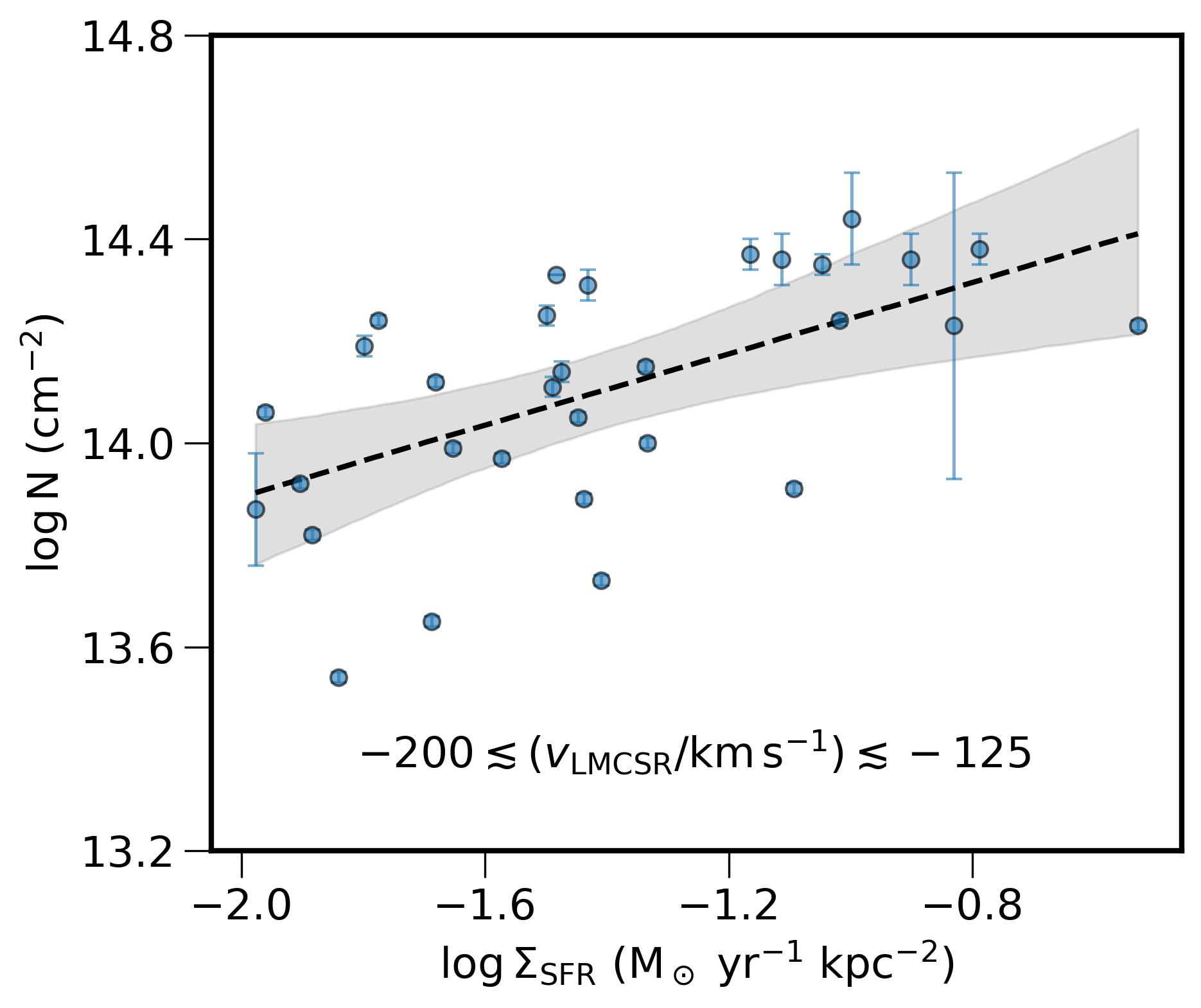}
  \includegraphics[scale=0.40, trim=0 30 0 0, clip]{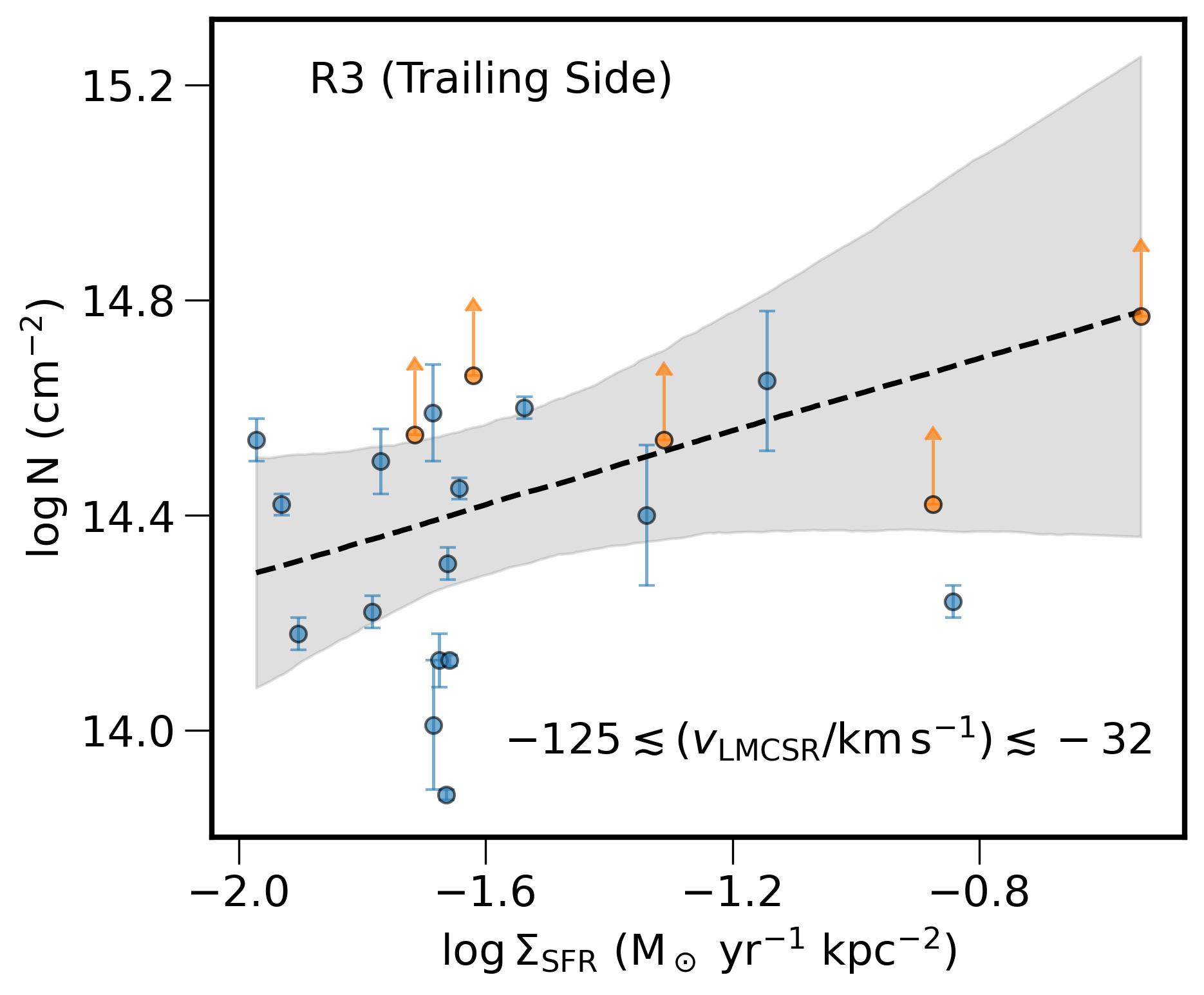}
  \includegraphics[scale=0.40, trim=30 30 0 0, clip]
  {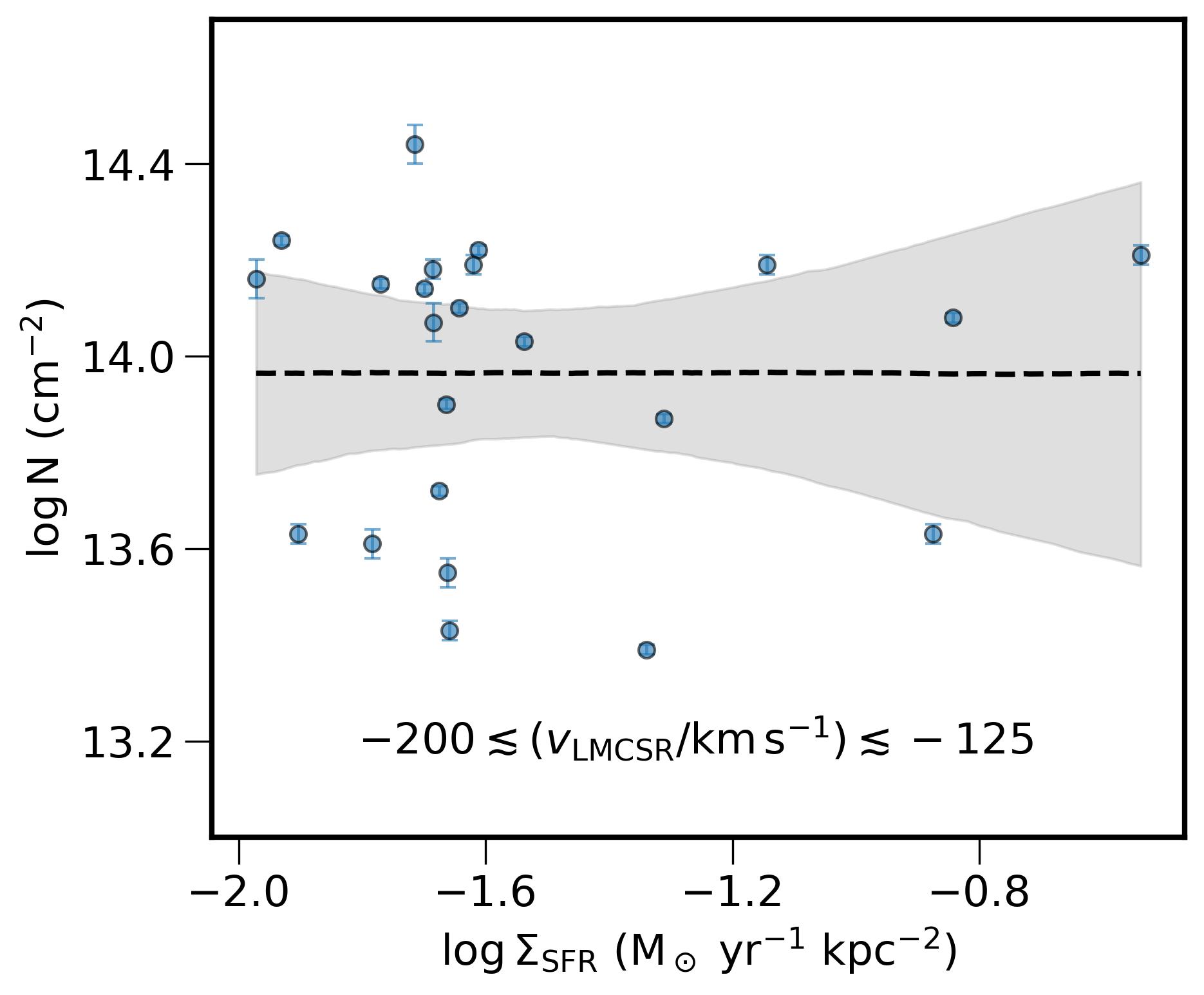}
  \includegraphics[scale=0.40, trim=0 0 0 0, clip]{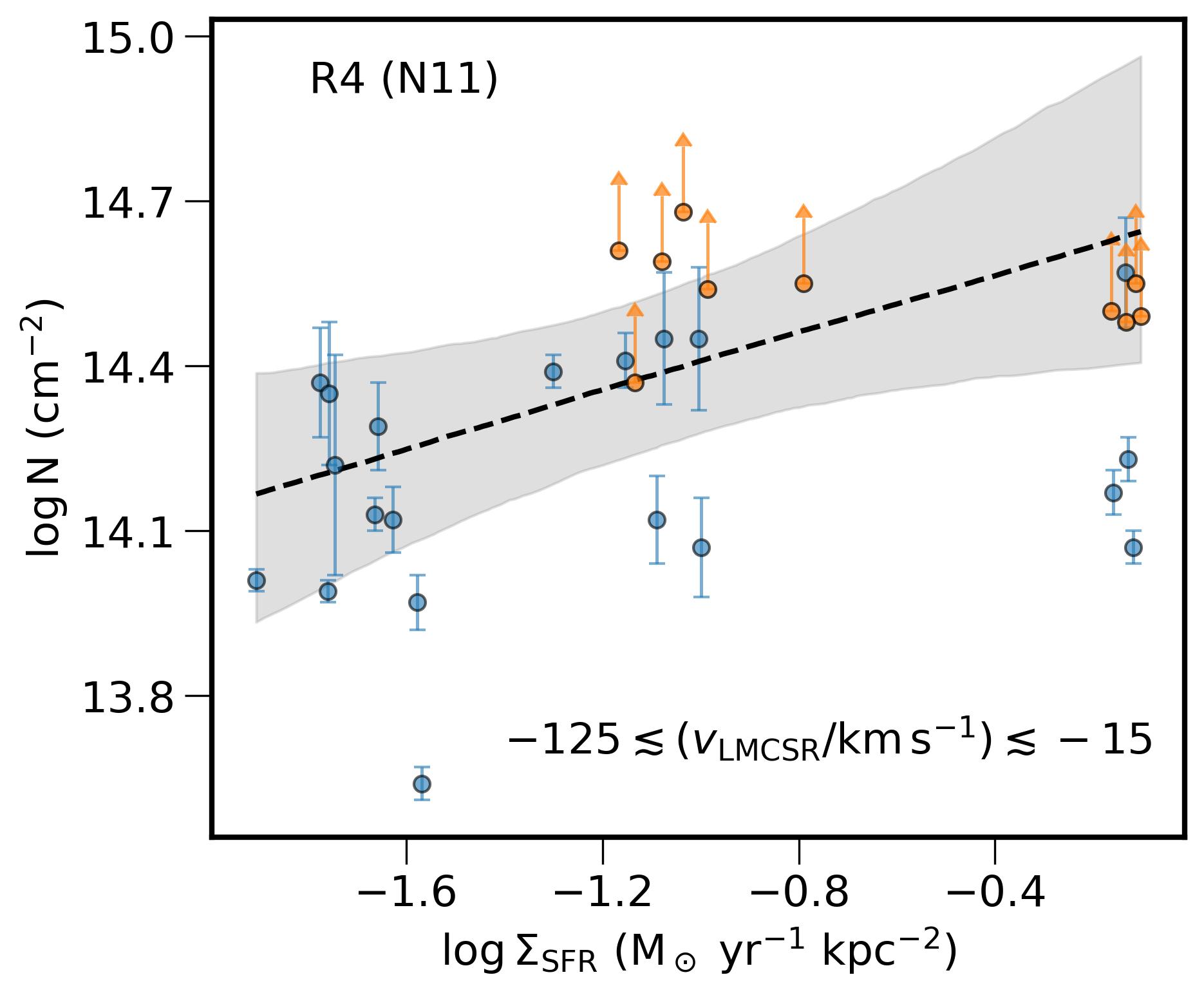}
  \includegraphics[scale=0.40, trim=30 0 0 0, clip]
  {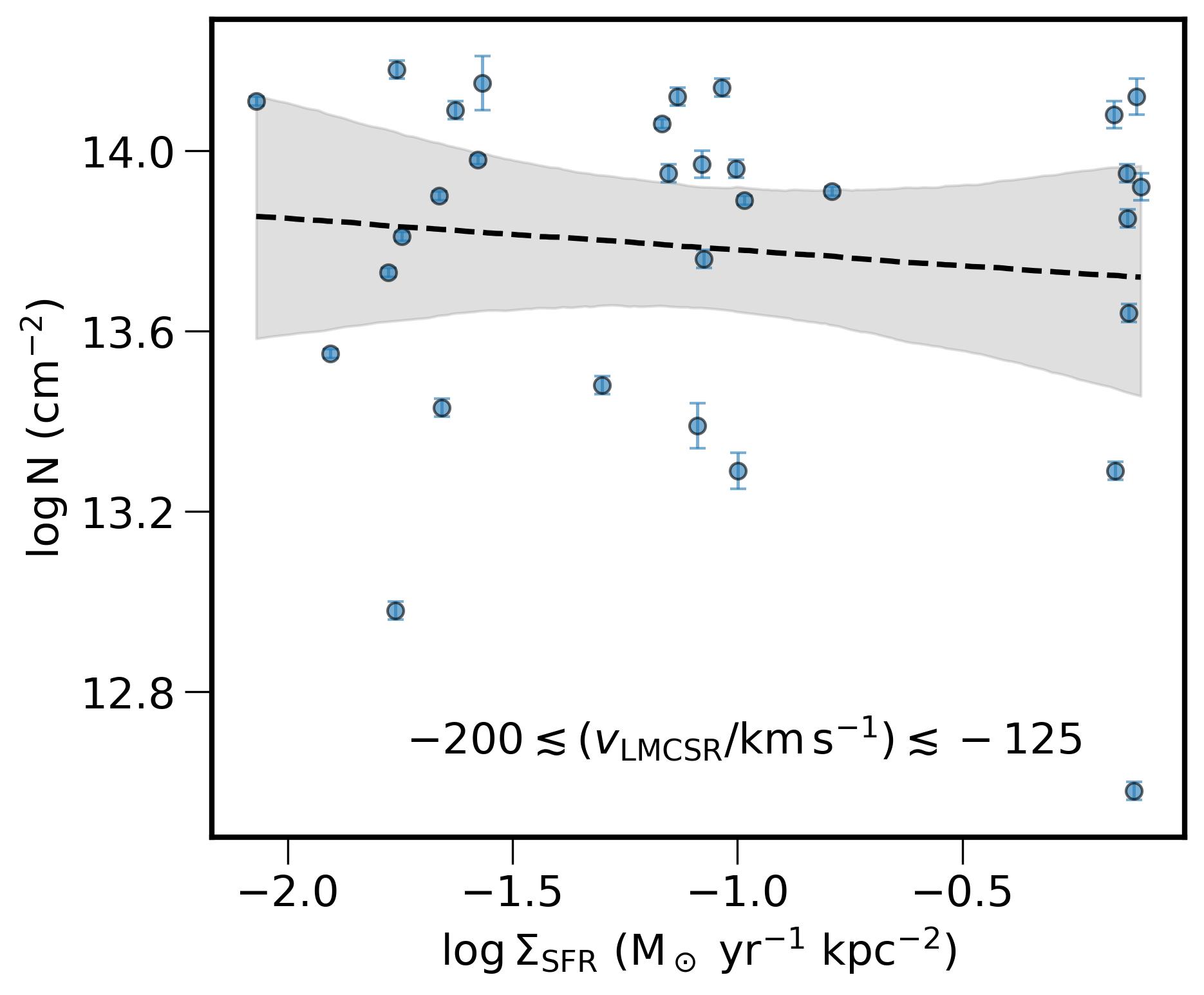}
  
  \caption{Bayesian linear regression between log~$\Sigma_{\mathrm{SFR}}$ and log~$N_{\rm Si\textsc{~ii}}$, for 4 spatial regions (R1--R4) and 2 kinematic windows (\textbf{Left:} $-125 < v_{\rm LMCSR} < v_{\rm DISK}~\kms$ and \textbf{Right:} $-200 < v_{\rm LMCSR} < -125~\kms$). 
    Blue circles indicate observed data points with error bars, while orange circles with arrows denote saturated log~$N_{\rm Si\textsc{~ii}}$ values treated as lower limits in the regression. 
    Each panel shows the median posterior regression line (black) and the 95\% credible interval (gray band) derived from the \texttt{PyMC} model. 
    The fits account for measurement uncertainties and censored (saturated) data using a survival analysis.}
  \label{fig:sfrsd}
\end{figure*}

However, we emphasize that the LMC disk boundary is not uniform across the galaxy. By integrating from a single global velocity boundary, we may systematically overestimate wind column densities along some sightlines by including residual disk gas, while underestimating them along others where the disk boundary occurs at more negative velocities. To mitigate this effect and also to explore the spatial variations, we analyze the LMC disk by dividing it into four spatial regions (R1–R4) and adopting locally determined mean disk boundaries for each region. In Figure~\ref{fig:sfrsd}, we plot the $\Sigma_{\rm SFR}$ verses log~$N_{\rm Si\textsc{~ii}}$ for both the slow-moving and ambiguous HVCs regions to assess whether or not blueshifted absorbers on the foreground the LMC's disk correlate with the LMC's stellar activity. In the left panels, which trace the slow-moving gas relative to the LMC disk, all four regions exhibit positive correlations, albeit with varying strengths and scatter. This consistent behavior across the disk suggests that the slower component of the outflow is pervasive throughout the LMC and remains physically coupled to local star-formation activity. In contrast, the right panels, corresponding to the ambiguous HVCs region, show region-dependent and markedly different behaviors. In R1, centered on the 30~Doradus star-forming complex, we observe a negative correlation similar to that seen in the global analysis in Figure~\ref{fig:ha_logN_cor}. Given that R1 contains the largest number of sightlines, this result possibly reflect the same superposition effect, in which intrinsically low-column LMC wind material is blended with higher-column foreground or non-wind absorption, producing an apparent anti-correlation between $\Sigma_{\rm SFR}$ and log~$N_{\rm Si\textsc{~ii}}$.

On the leading side of the LMC, R2 instead exhibits a statistically significant positive correlation in the ambiguous HVCs regime. This is particularly noteworthy given that this region does not host star formation as intense as 30~Doradus and is not located on the trailing side where stripped material might naturally accumulate. This behavior suggests that the observed trend is unlikely to be driven solely by local star-formation intensity and instead might reflect geometric effects unique to the leading side of the LMC that allows high-velocity outflowing gas to be observed with minimal foreground contamination. Additionally, the interaction between the LMC and the ambient halo medium may lead to compression or confinement of outflowing gas on the leading side, allowing higher-velocity material to persist at relatively higher column densities. The behavior observed on the leading side of the LMC may also reflect the dynamical interaction between the LMC and the MW halo. As the LMC moves through the Galactic halo medium, ram pressure can compress gas on the leading side of the disk and potentially confine outflowing material. Such confinement may allow higher-velocity clouds to remain detectable at larger column densities than would otherwise be expected in a freely expanding wind. This environmental effect could therefore contribute to the positive correlation observed in R2, even though the local star formation rate is lower than in the 30~Doradus region. Finally, R3 (trailing side) and R4 (N11) do not exhibit a clear correlation in the ambiguous HVCs regime and have substantial scatter, consistent with a mixed possible origins including LMC wind, Magellanic CGM, and foreground Milky Way HVC gas.

\subsection{Distribution of Doppler $b$ values}\label{subsec:kinematics}
\begin{figure*}[ht!]
  \centering
  \includegraphics[width=0.85\textwidth]{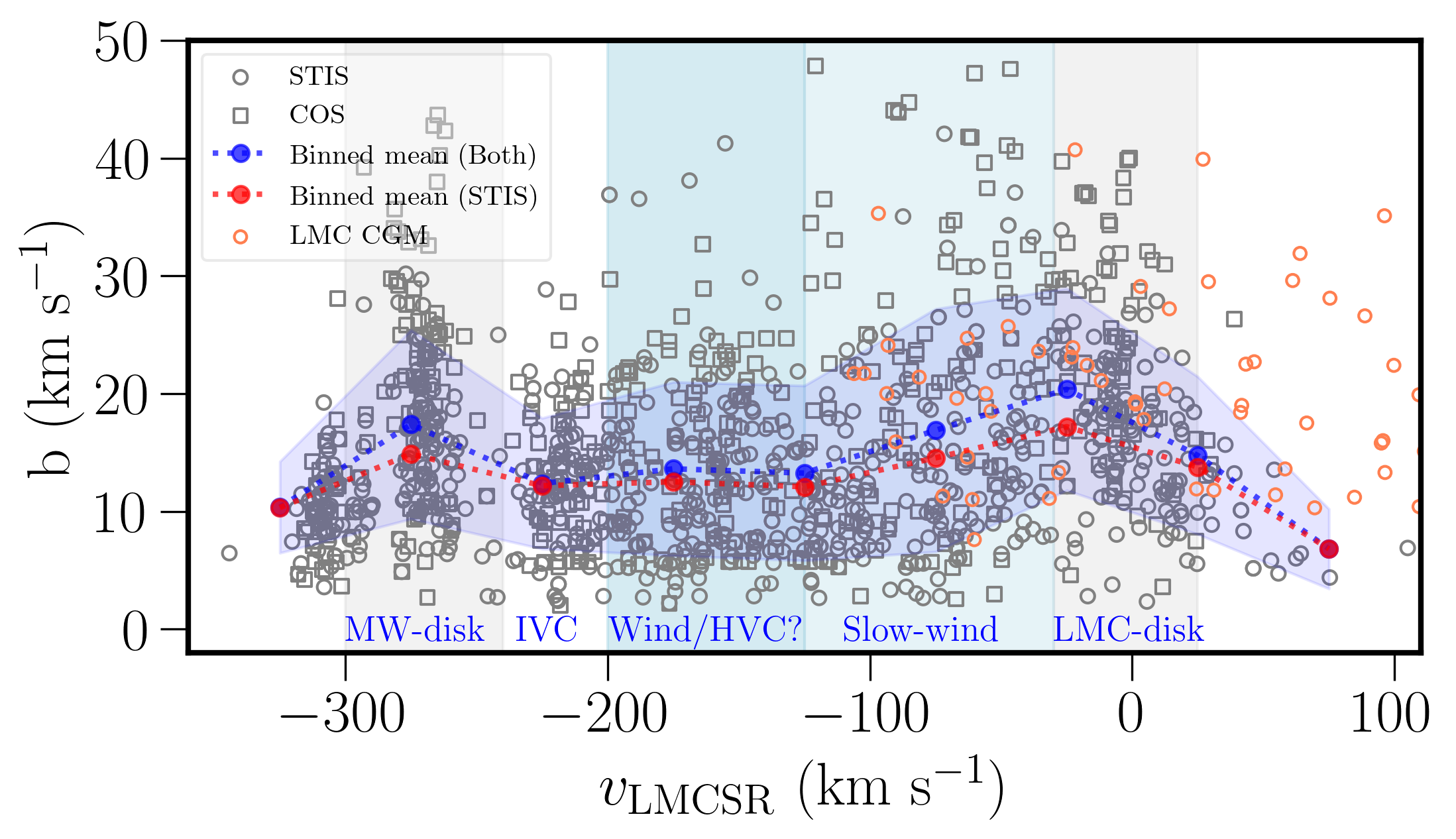}
  \caption{Doppler $b$-value distributions of individual Voigt-fitted components for Si\textsc{~ii} as a function of velocity with respect to the LMC. The gray circles and squares represent individual $b$ values observed with STIS and COS, respectively. The binned averages, computed in $50\, \kms$ intervals, are shown as red dots for STIS and as blue dots for STIS+COS; the blue envelope around those mean values represents their $1\, \sigma$ scatter. Approximate kinematic widths for various regions of interest are also shaded and labeled in the Figure. For comparison, $b$ values for LMC CGM from \citet{2024ApJ...976L..28M} are plotted as orange circles.}
  \label{fig:b_SiII}
\end{figure*}

The Doppler $b$ parameter measures the effective line broadening produced by both thermal and non-thermal motions of ions along the line of sight, making it a valuable diagnostic of the physical conditions of the absorbing gas. Larger $b$ values may arise from higher temperatures, turbulence, bulk motions, or unresolved velocity structure, whereas smaller values are generally associated with cooler and more kinematically quiescent gas. Here we use Si\textsc{~ii} absorption to examine the distribution of $b$-values across a broad kinematic range, because Si\textsc{~ii} lines are typically less saturated than O\textsc{~i} in dense LMC disk regions and are stronger than S\textsc{~ii} in the low–column-density, high-velocity wind components. The overall $b$-value distributions across all kinematic range between MW disk and LMC disk are broadly consistent between STIS and COS measurements (see Figure~\ref{fig:b_SiII}). Some COS measurements exhibit slightly larger $b$-values, which likely reflect the lower spectral resolution of COS compared to STIS. 

\begin{figure*}[ht!]
  \centering
  \includegraphics[scale=0.53, trim=0 30 0 0, clip]{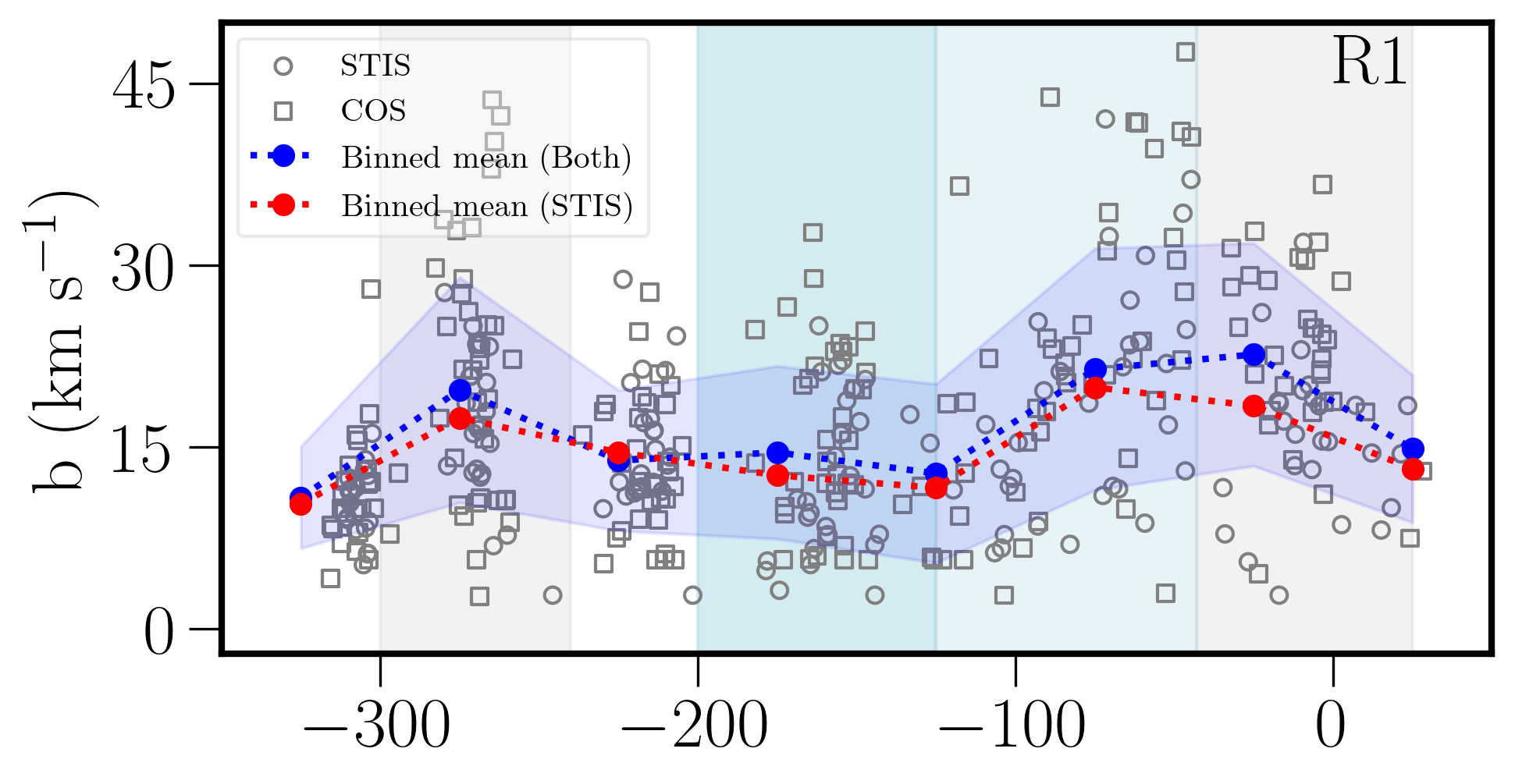}
  \includegraphics[scale=0.53, trim=32 30 0 0, clip]
  {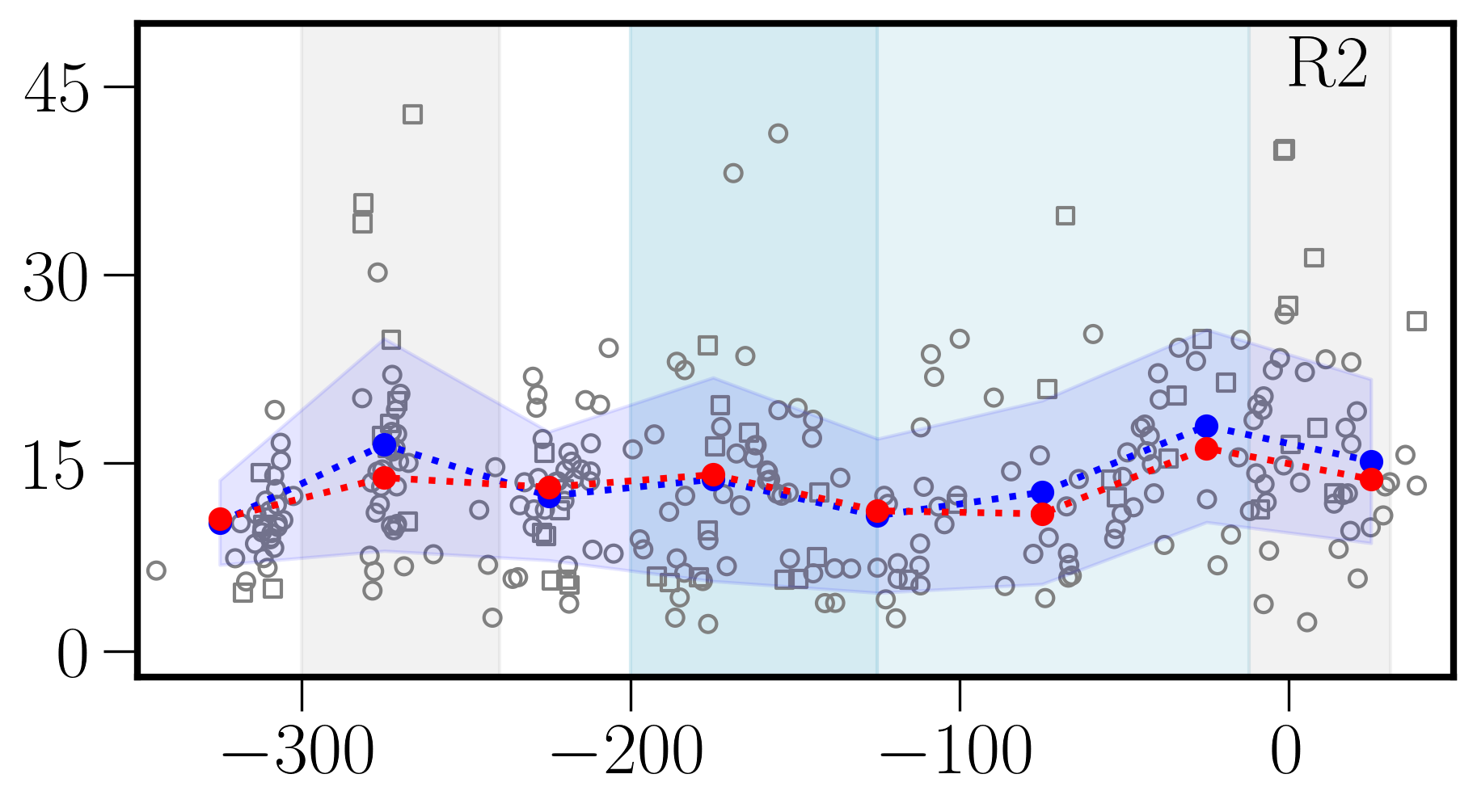}
  \includegraphics[scale=0.53, trim=0 0 0 0, clip]{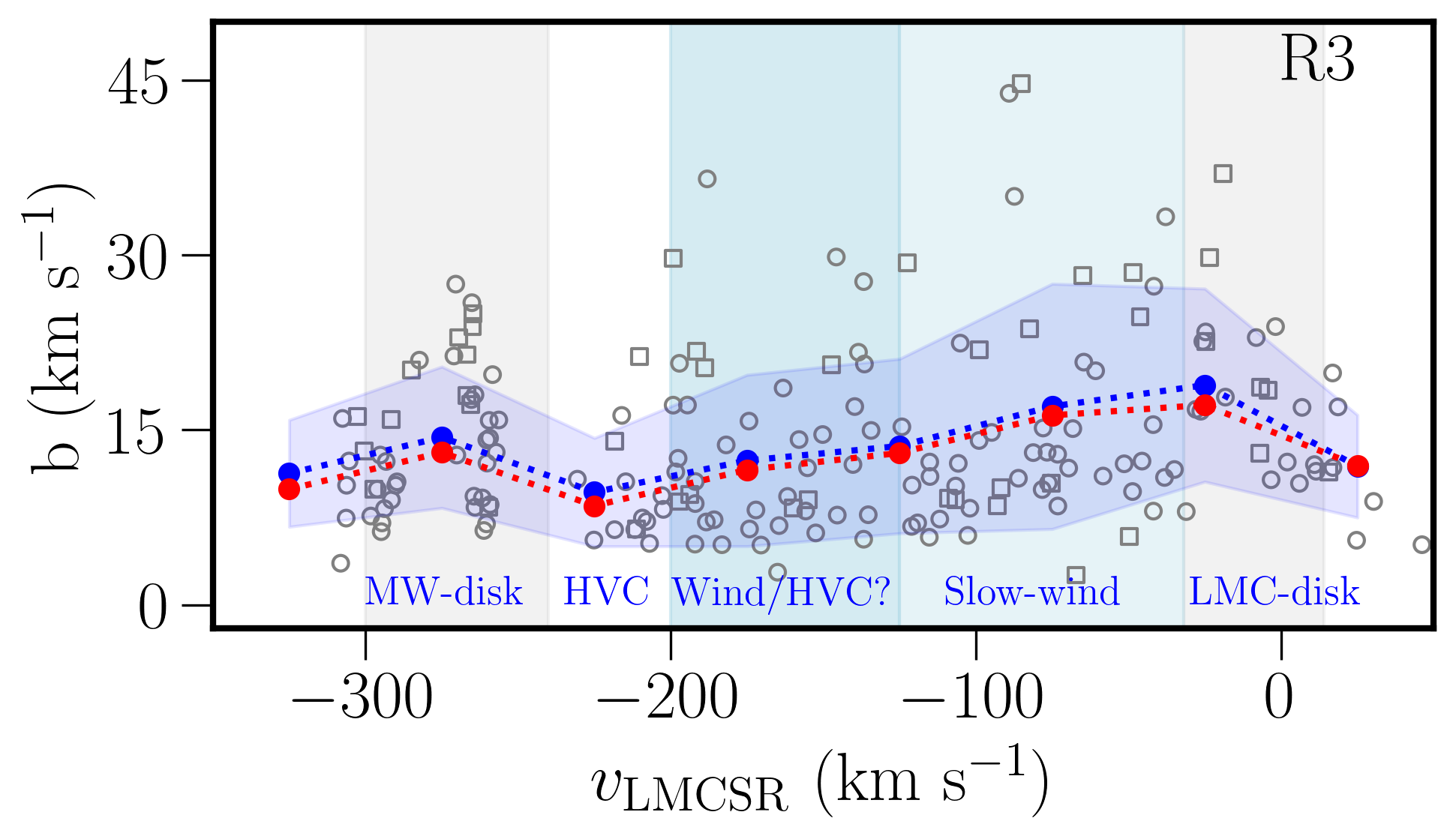}
  \includegraphics[scale=0.53, trim=32 0 0 0, clip]
  {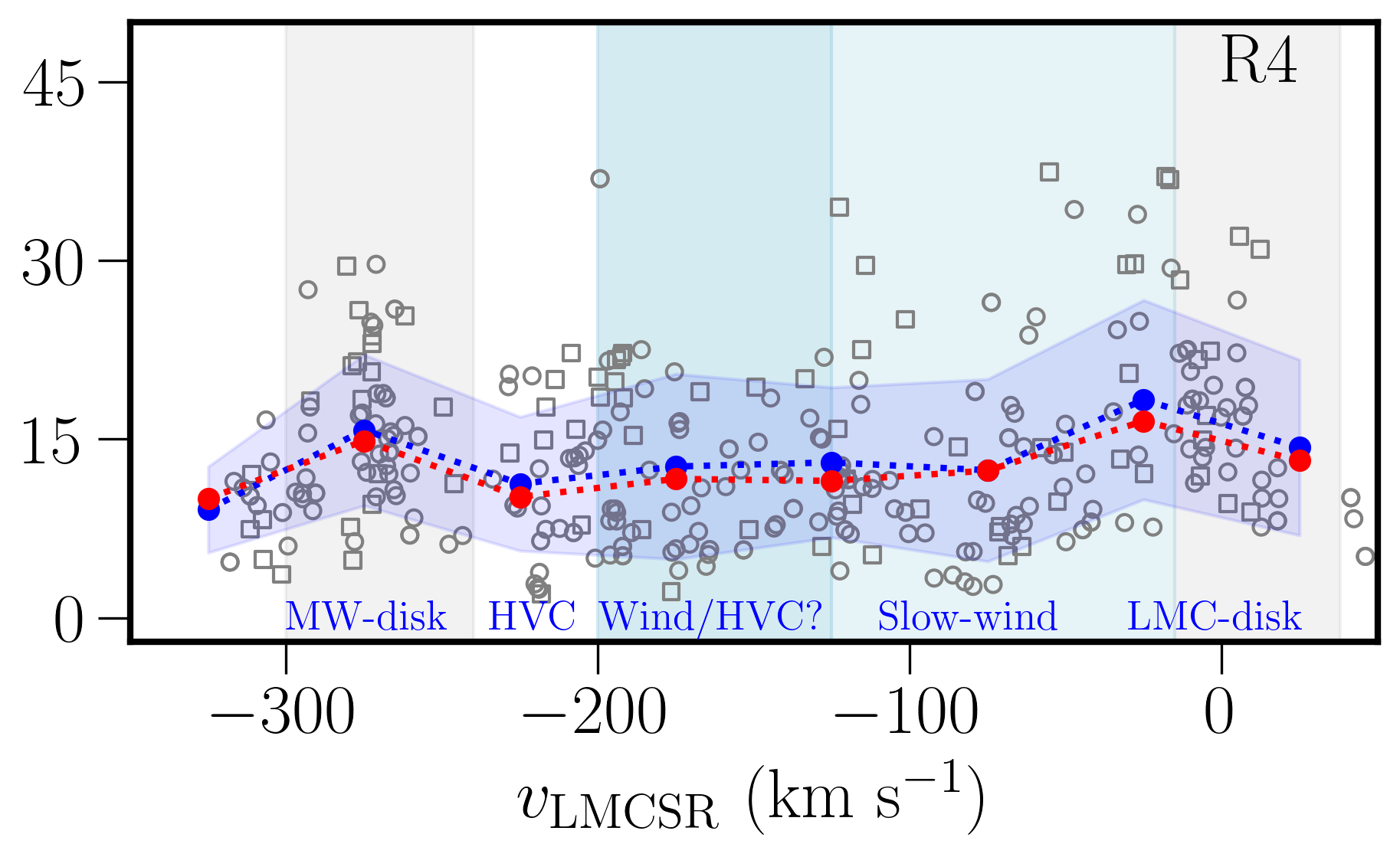}
  \caption{Doppler-$b$ value distributions of individual Voigt-fitted components for Si\textsc{~ii} as a function of velocity with respect to the LMC for 4 different regions defined in Figure~\ref{fig:hi_region_map}. The gray circles and squares again represent individual $b$ values observed with STIS and COS, respectively. The binned averages, computed in $50\, \kms$ intervals, are shown as red dots for STIS and as blue dots for STIS+COS. The light shade in blue represents the $1\, \sigma$ scatter from the binned means for STIS+COS.}
  \label{fig:b_all}
\end{figure*}

Across all kinematic windows---from Milky Way disk to LMC disk---the Doppler $b$-values have substantial scatter, ranging from as low as ${\sim}3\, \kms$ up to ${\sim}40\, \kms$. The $50\,\kms$ binned averages reveal only mild variations, with slightly higher mean $b$-values near the MW and LMC disk velocities than in the intermediate regions, but the differences are not statistically significant. The wide spread likely reflects a combination of varying physical conditions and unresolved velocity structure among the absorbing clouds. The overall distribution of $b$-values remains comparable between the slow-wind and fast-moving components, indicating that the thermal and non-thermal broadening mechanisms do not change dramatically across these regimes. Furthermore, the Doppler $b$-value distributions are broadly similar across all four spatial regions of the LMC (see Figure~\ref{fig:b_all}). Even in regions dominated by intense star formation, such as 30~Doradus, the $b$-values of fast-moving absorbers are not systematically larger than those observed elsewhere. This uniformity suggests that enhanced local star formation does not strongly influence the observed linewidths of the fast-moving absorbers, although the effects of unresolved component blending cannot be excluded.

The observed linewidths cannot be explained purely by thermal broadening. For example, thermal Doppler widths of $b\sim20~\kms$ for Si~\textsc{ii} would correspond to temperatures of order $10^6$ K, well above the temperature at which low-ionization species such as Si~\textsc{ii} can survive. Therefore, additional broadening mechanisms must contribute to the measured linewidths. These may include unresolved velocity substructure, turbulence, and bulk motions. In particular, higher-resolution STIS/E140M observations of several sightlines demonstrate that features appearing as single broad components in COS spectra can resolve into multiple narrower absorbers. Consequently, the measured Doppler parameters should generally be regarded as upper limits to the intrinsic linewidths of individual clouds.

For comparison, we also plot the Doppler $b$-values for LMC CGM (see orange circles in Figure~\ref{fig:b_SiII}) from \citet{2024ApJ...976L..28M}. These Si\textsc{~ii} absorbers tracing the LMC CGM have characteristic Doppler parameters of $\langle b \rangle \approx 20.4~\kms$ with a dispersion of $\sigma_b \approx 8~\kms$ within $\sim$17~kpc, and slightly larger values of $\langle b \rangle \approx 24~\kms$ with a similar dispersion beyond $\sim$17~kpc. These mean $b$-values are systematically larger than the mean Doppler parameters measured for the bulk of our low-ion components and lie near the upper $1\sigma$ envelope of our observed distribution (see Figure~\ref{fig:b_all}). Consequently, the subset of components in our sample with relatively large Doppler $b$-values may be partially probing gas associated with the compact LMC CGM and/or stripped CGM material, rather than purely entrained disk gas.

\subsection{Observed ion ratios}\label{subsection:ion_ratio}
\begin{figure*}[ht!]
  \centering
  \includegraphics[scale=0.40, trim=0 0 0 0, clip]{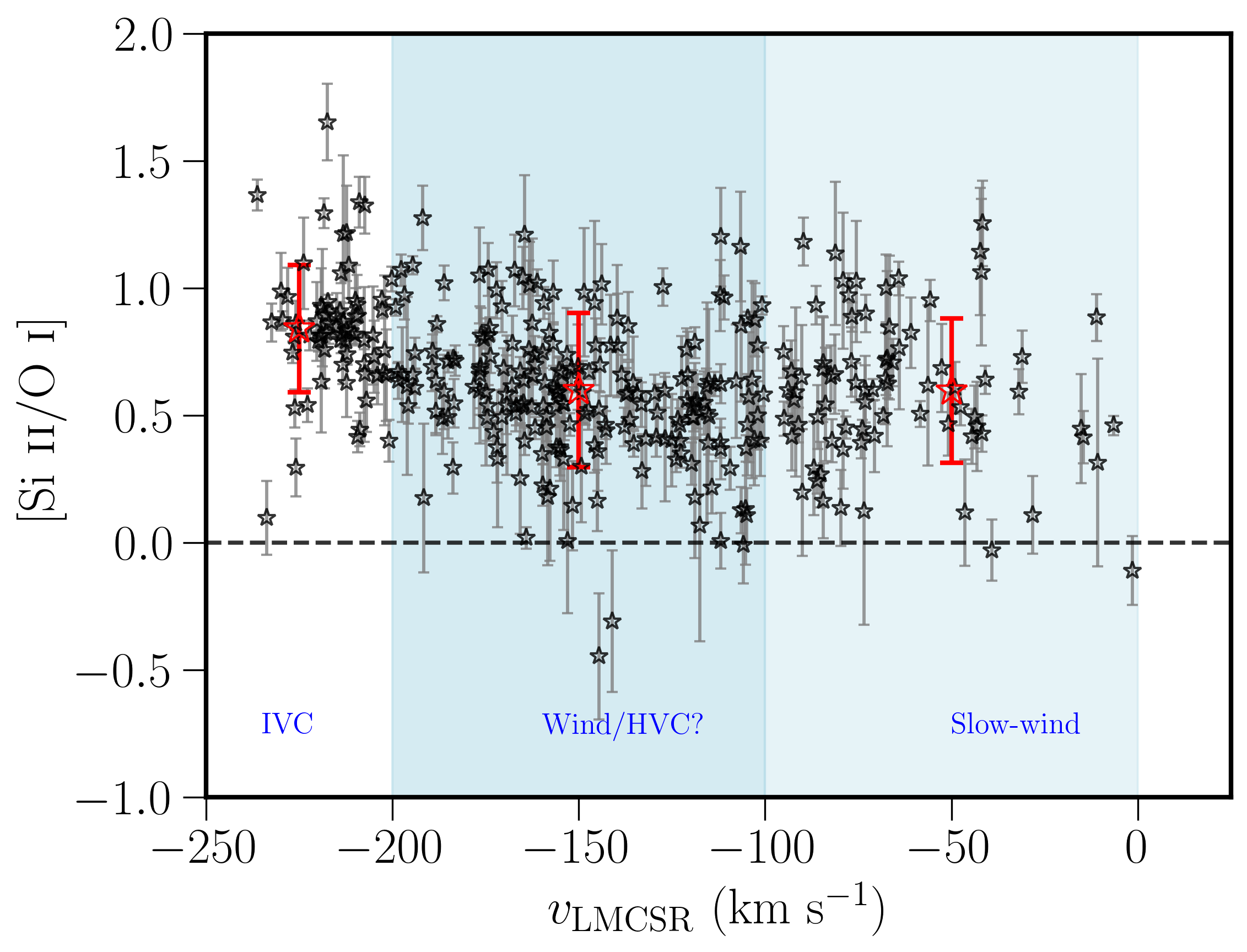}
  \includegraphics[scale=0.40, trim=0 0 0 0, clip]
  {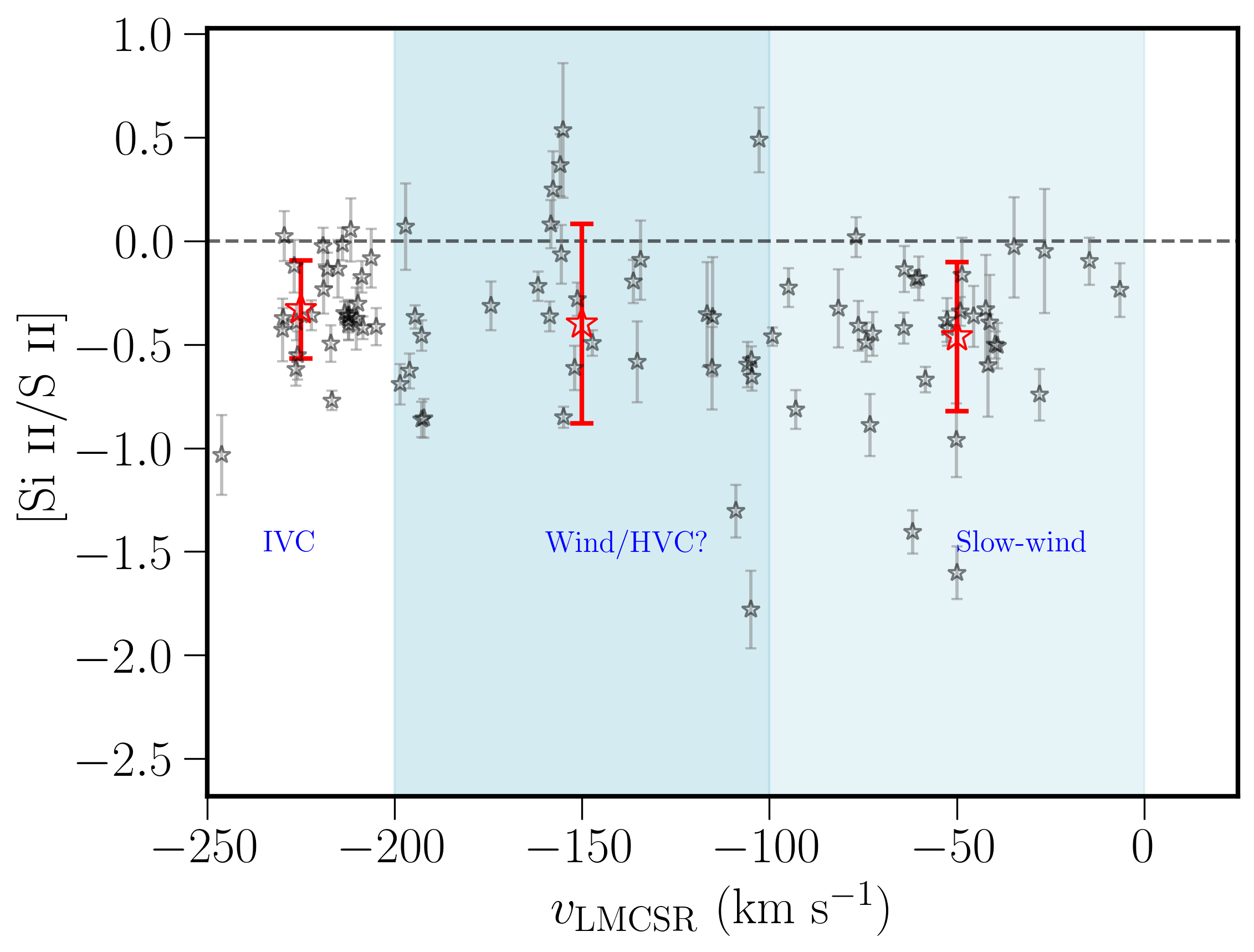}
  \caption{Ion ratios (\textbf{Left:} [Si\textsc{~ii}/O\textsc{~i}] and \textbf{Right:} [Si\textsc{~ii}/S\textsc{~ii}]) as a function of velocity with respect to the LMC disk ($v_{\rm LMCSR}$) for the sightlines across the entire LMC disk. Ion ratios are computed using closest-neighbour velocity matching between corresponding absorption components, and only components at velocities more negative than the left boundary of the LMC disk (as defined from H\textsc{~i} emission; see Section~4) are included.}
  \label{fig:ion_ratios}
\end{figure*}

To explore ionization and depletion patterns across the LMC disk, we analyze the observed low-ion ratios [Si\textsc{~ii}/O\textsc{~i}] and [Si\textsc{~ii}/S\textsc{~ii}]. We adopt solar abundances from the \citet{2009ARAA..47..481A} catalog. These non-ionization corrected ion ratios are computed by matching closest absorption components in velocity space, where we carefully excluded components associated with the LMC disk using the method in which we compared the UV absorption and H\textsc{~i} emission data (see Section~\ref{section:Prob_wind}). This selection isolates gas that is kinematically distinct from the disk and likely associated with outflowing or foreground material even when the LMCSR velocity is near zero, as is the case for some sightlines.  We present these ratios for all LMC disk sightlines in Figure~\ref{fig:ion_ratios}. It is important to note that the S\textsc{~ii} $\lambda1250$ and S\textsc{~ii} $\lambda1253$ transitions used to measure S\textsc{~ii} column densities are relatively weak. As a result, these lines are not widely detected in the spectrum and most often limited to the kinematic disk of the LMC. Consequently, there are significantly fewer sightlines with measurable [Si\textsc{~ii}/S\textsc{~ii}] ratios compared to those with [Si\textsc{~ii}/O\textsc{~i}].

Accurate interpretation of these ion ratios requires consideration of the nucleosynthetic origin of the atomic species, their depletion strengths, and their susceptibility to ionization. All of the atomic species used here are $\alpha$-elements, which are primarily produced by type-II supernovae \citep{2006NuPhA.777..424N}. Therefore, variations in the observed ratios are unlikely to be driven by chemical evolution effects. Instead, the ratios are expected to be shaped predominantly by the competing influences of dust depletion and ionization. The level of dust depletion depends on the local density environment and also correlate with metallicity (\citealt{2024A&A...683A.216D} and references therein). In addition, depletion patterns differ between refractory elements such as silicon and volatile elements such as sulfur and oxygen \citep{2017ApJ...838...85J,2009ApJ...700.1299J,1996ApJ...470..893S}. In the MW's ISM, silicon is subject to substantially stronger depletion than oxygen and sulfur \citep{2009ApJ...700.1299J}.

Both ion ratios exhibit substantial dispersion across all kinematic windows. The spread is particularly pronounced in the `slow-wind' and `Wind/HVC?' regions, whereas the region associated with MW's IVCs appears comparatively tighter. In the ambiguous `Wind/HVC?' range, the mean value of [Si\textsc{~ii}/O\textsc{~i}] is $0.60 \pm 0.30$, while the slow-wind region shows a nearly identical average of $0.60 \pm 0.28$. This is broadly consistent with the 
mean value reported by \citet{2009ApJ...702..940L}, who found 
$\langle$[Si\,\textsc{ii}/O\,\textsc{i}]$\rangle = +0.48^{+0.15}_{-0.25}$ for the LMC 
HVC complex at $+90 \lesssim v_{\rm LSR} \lesssim +175~{\rm km~s^{-1}}$. In contrast, the IVC components display a higher mean of $0.84 \pm 0.25$. Despite the large dispersions in all three regimes, the IVC population therefore tends to occupy systematically elevated [Si\textsc{~ii}/O\textsc{~i}] values relative to the other two groups, whose averages are similar within the uncertainties. For [Si\textsc{~ii}/S\textsc{~ii}], the mean ratio is roughly $-0.4$ in each of the three velocity intervals. However, the IVC region again shows a smaller spread, suggestive of a more uniform cloud population. By contrast, the `Wind/HVC?' and `slow-wind' regimes display considerably broader scatter, indicating a mixture of physical conditions and possibly multiple origins.

\begin{figure*}[ht!]
  \centering
  \includegraphics[scale=0.38, trim=0 54 0 0, clip]{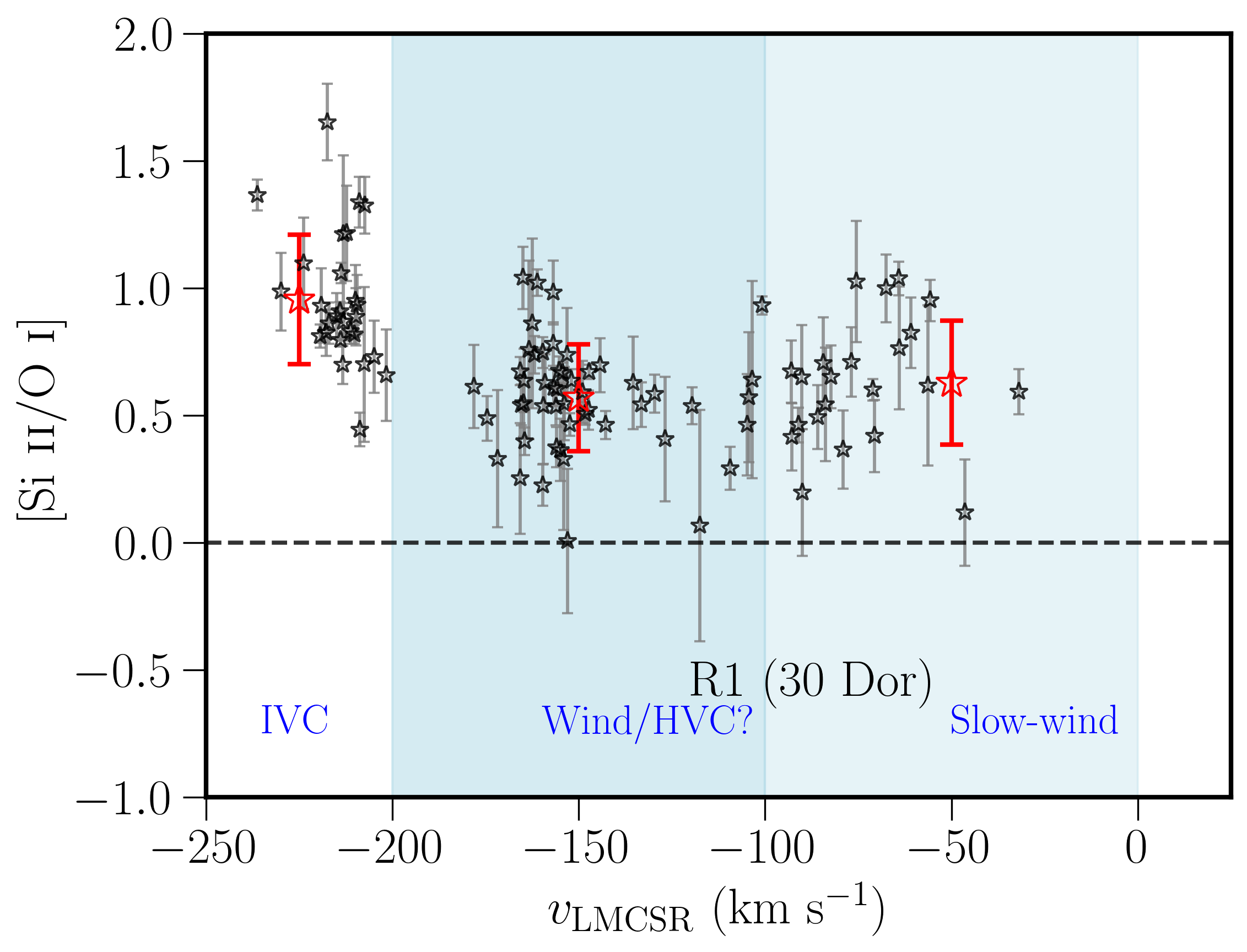}
  \includegraphics[scale=0.38, trim=40 54 0 0, clip]
  {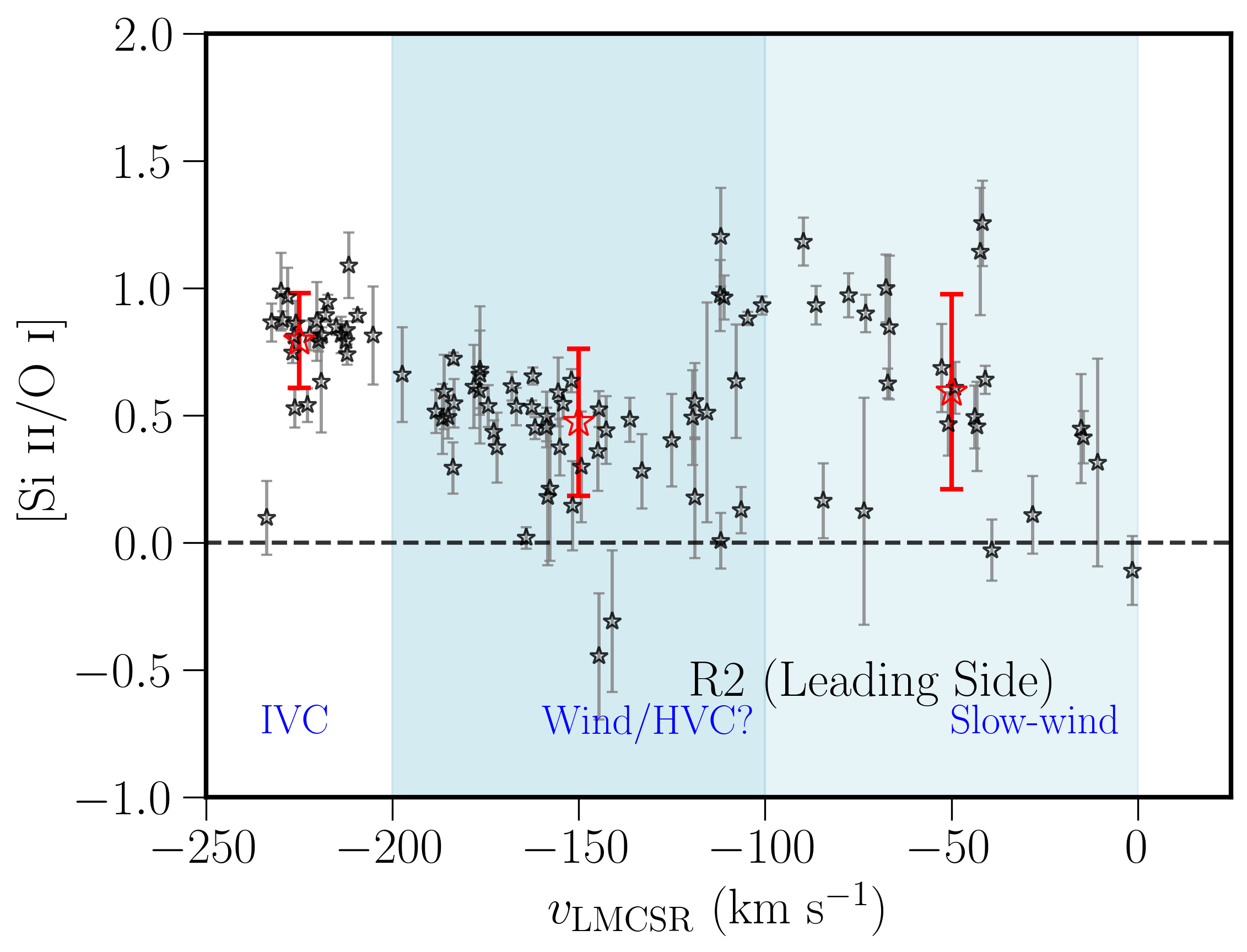}
  \includegraphics[scale=0.38, trim=0 0 0 0, clip]{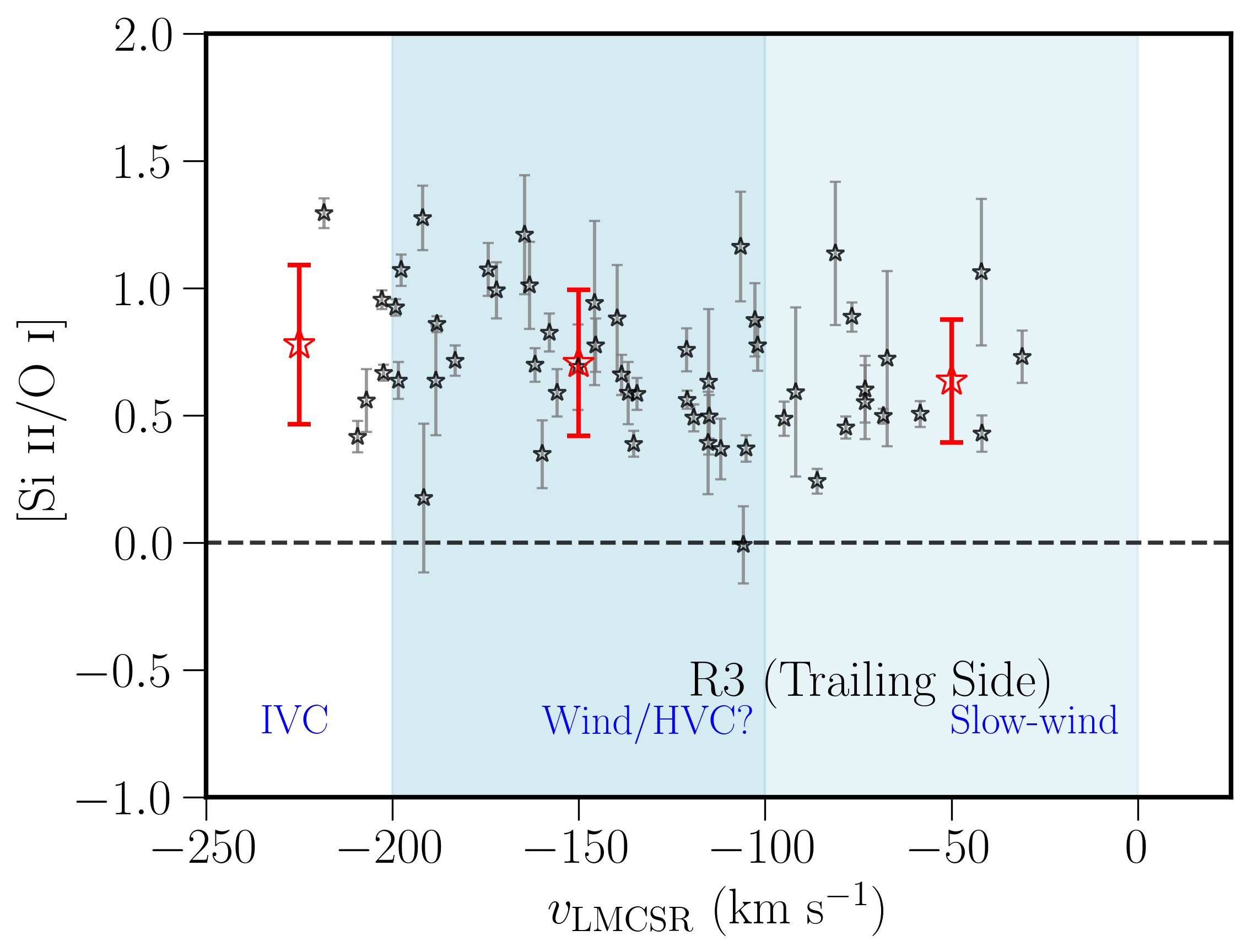}
  \includegraphics[scale=0.38, trim=40 0 0 0, clip]
  {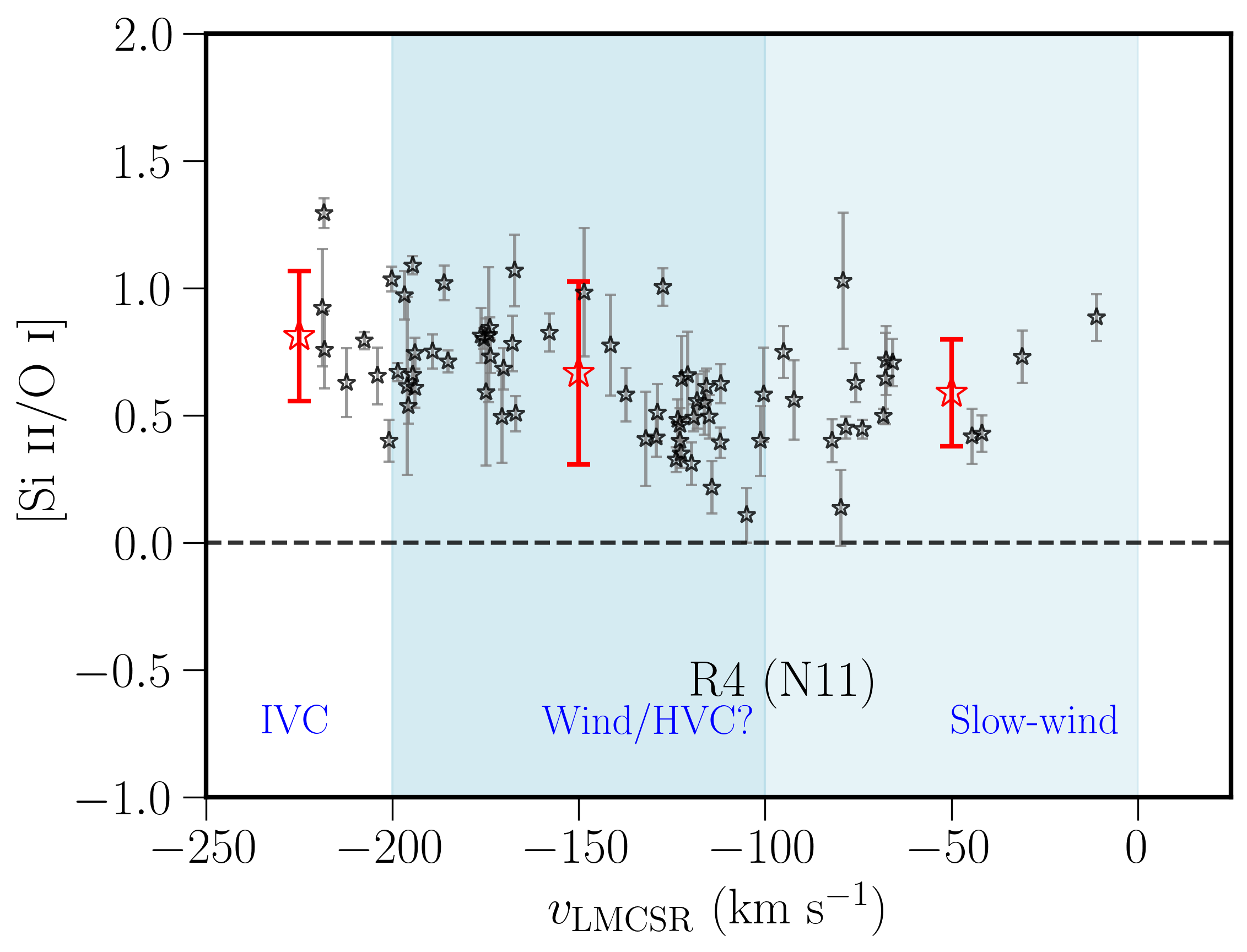}
  \caption{Similar to left panel in Figure~\ref{fig:ion_ratios} but showing the spatial subdivision of the [Si\textsc{~ii}/O\textsc{~i}] ratio for four regions of the LMC: R1 (30~Doradus), R2 (leading side), R3 (trailing side), and R4 (N11).} 
  \label{fig:ion_ratios_R1-R4}
\end{figure*}

We further subdivide the LMC into four regions (R1: 30~Dor, R2: leading side, R3: trailing side, and R4: N11). The overall [Si\textsc{~ii}/O\textsc{~i}] ratio trends in the global sample remain present within each region, but there is significant dispersion at nearly all velocities (see Figure~\ref{fig:ion_ratios_R1-R4}). However, we note that there are fewer components in the IVC velocity range for R3 and R4 compared to R1 and R2. The subdivision further reveals subtle environmental differences. For instance, the kinematic `Wind/HVC?' region in the foreground of 30~Dor (R1) has a large spread in [Si\textsc{~ii}/O\textsc{~i}] values over a narrow velocity range, which is consistent with the complex, multiphase nature of gas surrounding the main starburst. Despite these regional variations, no single spatial (i.e., R1, R2, R3, \& R4) or kinematic region (i.e, ``Wind/HVC?'' \& ``Slow-wind'') has a concentration of ratios that dramatically deviates from the rest. 

\begin{figure*}[ht!]
\centering

\includegraphics[scale=0.60, trim=0 0 0 0, clip]
{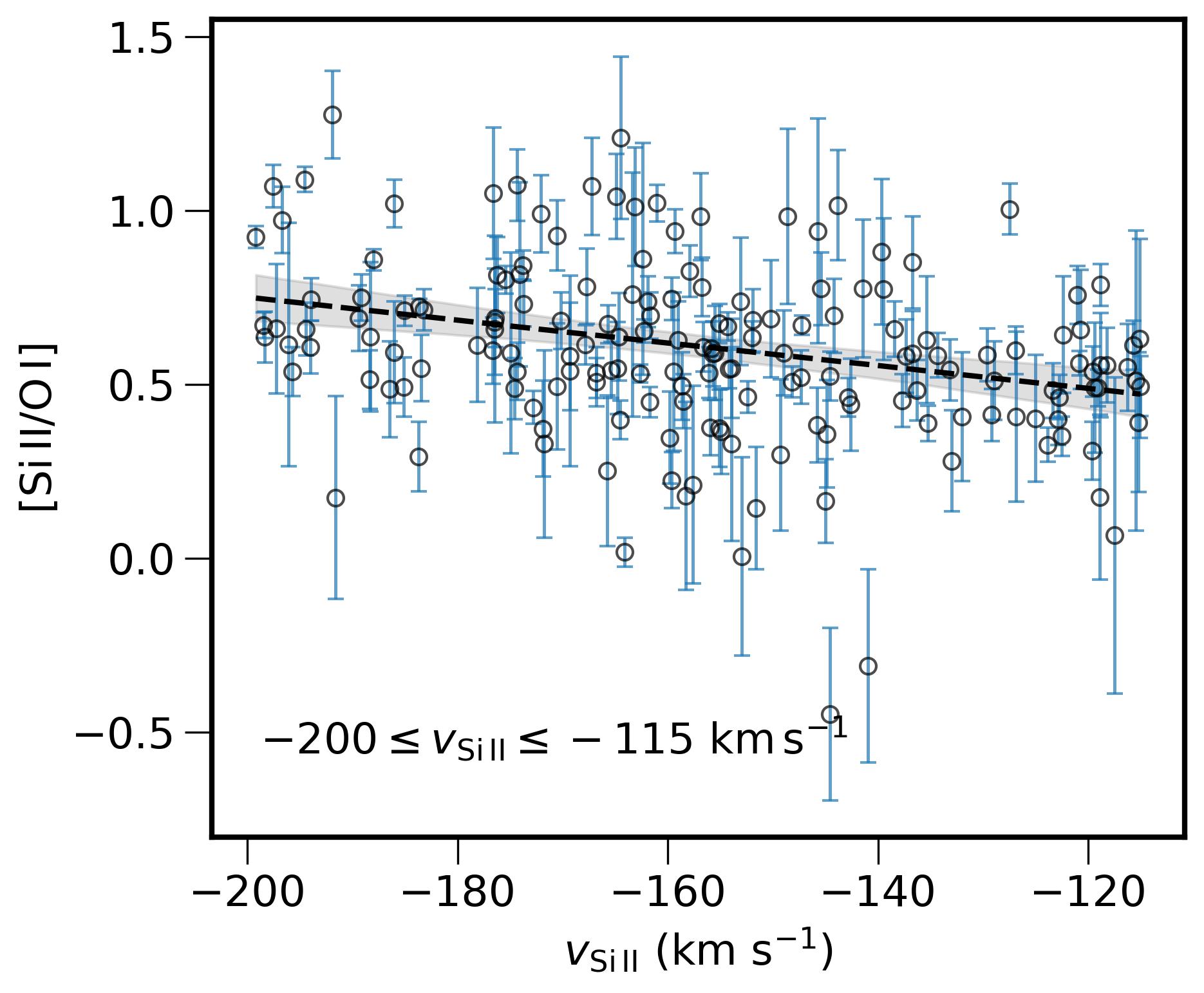}

\vspace{0.4cm}

\includegraphics[scale=0.45, trim=0 30 0 0, clip]
{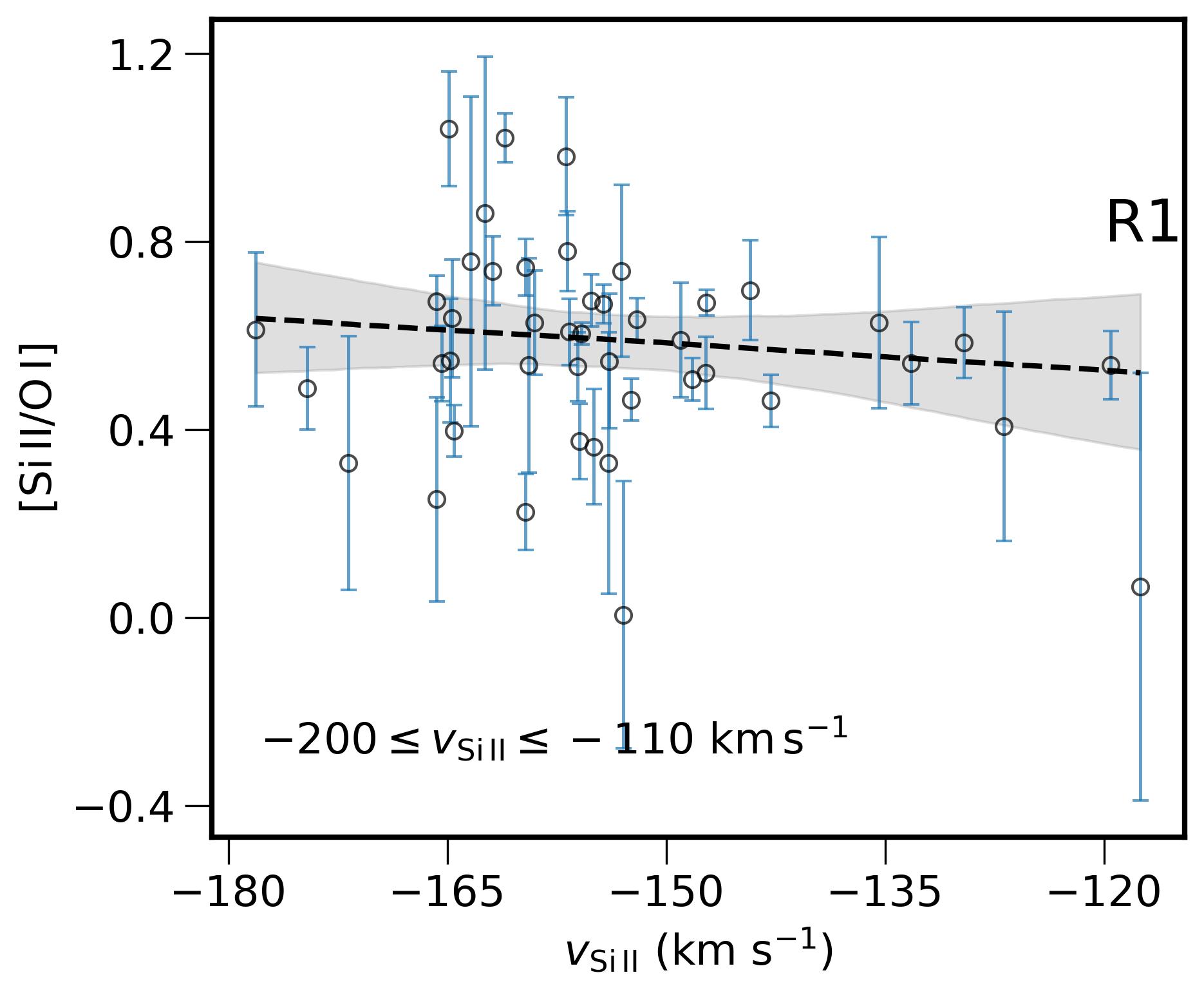}
\includegraphics[scale=0.45, trim=28 30 0 0, clip]
{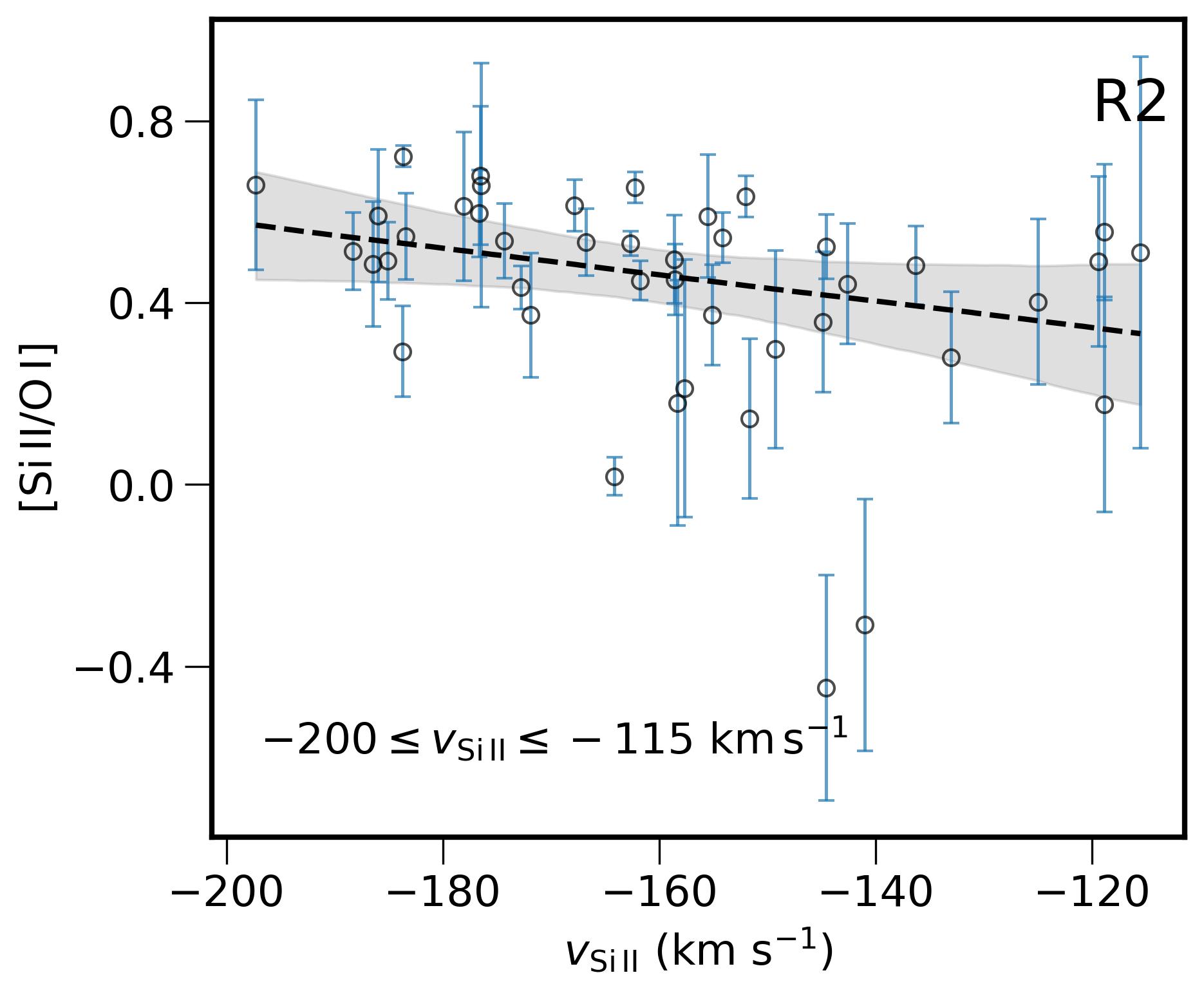}

\includegraphics[scale=0.45, trim=0 0 0 0, clip]
{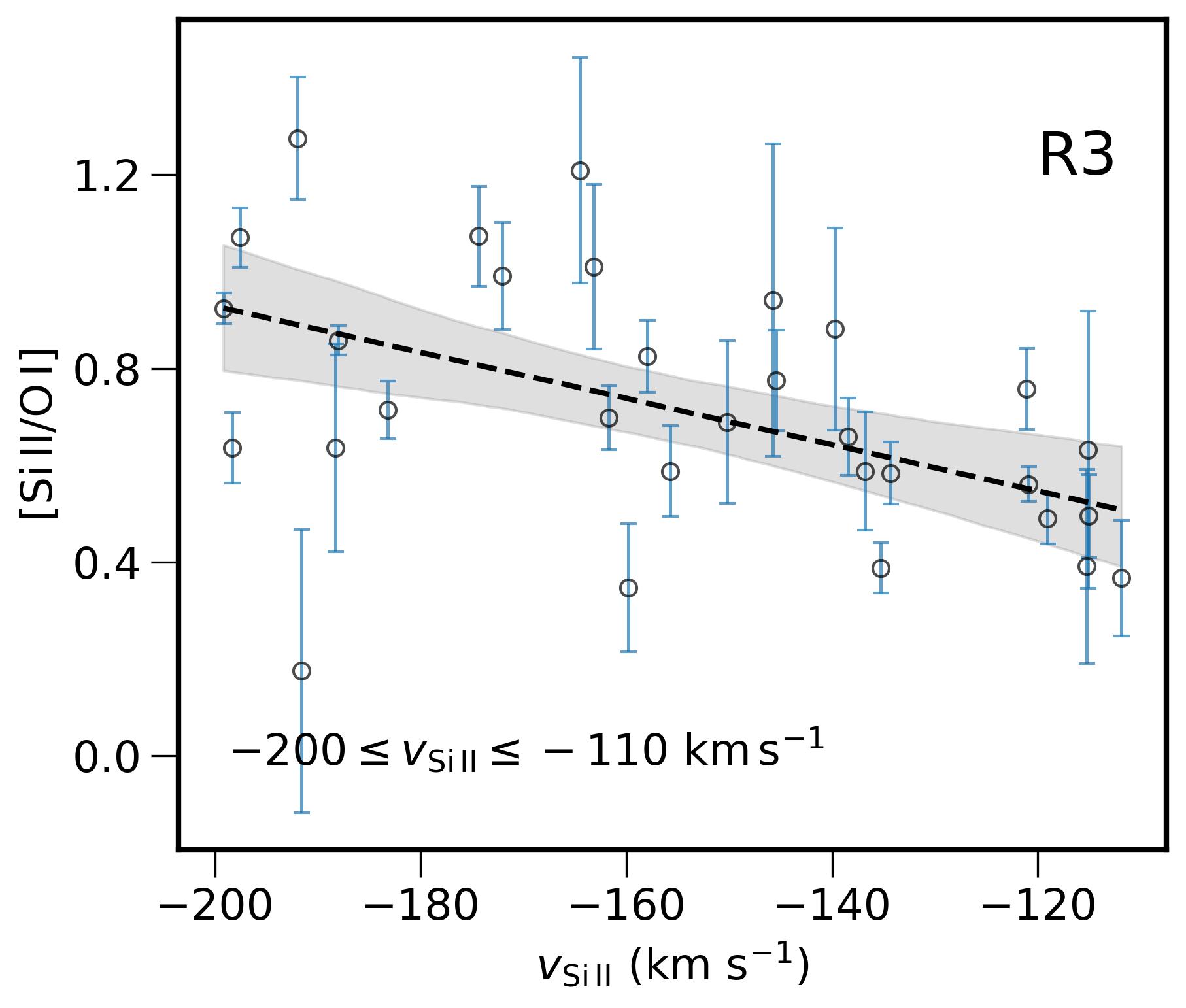}
\includegraphics[scale=0.45, trim=28 0 0 0, clip]
{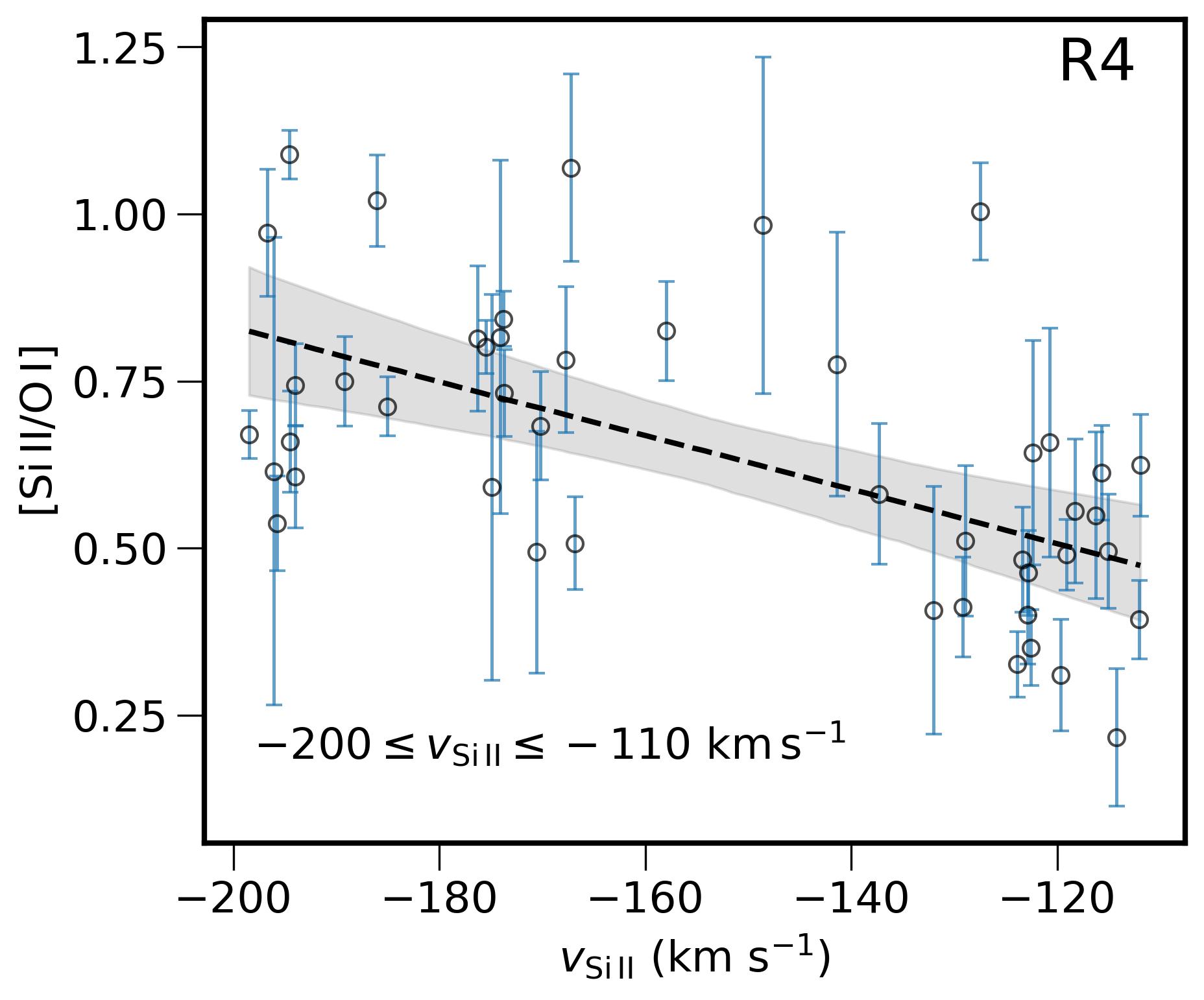}

\caption{
Bayesian regression fit of [Si\textsc{~ii}/O\textsc{~i}] versus velocity with respect to the LMC disk in the ambiguous `Wind/HVC?' regime.
\textbf{Top:} global sample.
\textbf{Bottom:} spatial subdivision into R1 (30~Doradus), R2 (leading side), R3 (trailing side), and R4 (N11).
}
\label{fig:ion_ratios_fit}
\end{figure*}

At the same time, the regression analysis in the ambiguous `HVC/Wind' velocity regime for the global sample hints at a weak but suggestive trend: components at more negative velocities tend to have slightly higher levels of ionization (see Figure~\ref{fig:ion_ratios_fit}). A similar tendency can be observed in R2, R3, and R4, although the significance is limited by the substantial scatter. In contrast, R1 does not have a clear monotonic behavior in which the ratio decreases toward the LMC disk velocities. This is consistent with the broad scatter discussed in Section~\ref{subsection:kinematic_dist}, which suggests that the ambiguous-velocity gas may include components with multiple origins or physical conditions. While these regression fits should be interpreted with caution, they are broadly consistent with more strongly accelerated material may, on average, have an elevated ionization within the ``HVC/Wind' velocity range.

\section{Photoionization modeling of the ambiguous HVCs}
\label{section:Cloudy}

\subsection{Motivation}
A key goal of this work is to assess the origin of the absorbers at $+90 \lesssim v_{\rm LSR} \lesssim +175~\kms$ as there is an ongoing debate as whether they are associated with high-velocity material expelled from the LMC or foreground gas associated with the MW HVCs population. In our \citet{2025ApJ...984..161P} study, we identified multiple absorbers in this velocity range toward the LMC's 30~Doradus region using \textit{HST} ULLYSES observations. We conducted photoionization modeling of two absorbers at $v_{\rm LSR}\approx +115\,\kms$ and found that one of the clouds is relatively neutral, dust-free, and low-metallicity, which is consistent with many of the MW's HVCs and Magellanic tidal debris. However, another absorber had a metallicity and depletion patterns that are consistent with the present-day LMC. These results suggest that there might be significant LMC wind that is able to reach speed up to $|v_{\rm outflow}|\approx150\,\kms$ and that a population of either foreground MW HVC population or Magellanic tidal gas also resides in the same kinematic space. 

In this work, we extend our radiative modeling efforts to additional sightlines that probe the entire angular extent of the LMC's disk to assess the fraction of these clouds that are associated with the LMC's wind and with foreground contamination such as the MW HVCs, tidally stripped materials, and/or the Magellanic CGM/coronal gas. To investigate the physical conditions, elemental abundances, and dust depletion patterns of the cool ($T\approx10^4\,\rm K$) gas in the kinematic range of $+100 \lesssim v_{\rm LSR} \lesssim +150\,\kms$, we performed photoionization modeling toward 13 sightlines (see Table~\ref{tab:cloudy_results}) using the radiative transfer code \textsc{Cloudy} (version 25.00; \citealt{Gunasekera2025}). The selected sightlines met the following criteria: (1) projected toward range of star-forming conditions in the LMC, (2) availability of \hi\, column density measurements from emission-line observations, (3) the detection of multiple volatile and refractory species that included at least one unsaturated O\textsc{~i} or S\textsc{~ii} transition, and (4) their inclusion in the ULLYSES DR7 catalog. These criteria are met primarily by the 13 sightlines modeled here. A small number of additional
sightlines also satisfy similar observational requirements; however, these are
located very close on the sky to one of the modeled sightlines and therefore
probe nearly the same local environment. To avoid
over-weighting individual regions with multiple adjacent sightlines, we selected
a non-redundant set of 13 absorbers that samples the available environments
across the LMC disk. The spatial distribution of these sightlines across the LMC disk is indicated by the black stars in Figures~\ref{fig:ha_map} and \ref{fig:hi_region_map}. Three of these sightlines probe the 30~Doradus region, two lie near N11, and the remaining are distributed across smaller discrete star-forming regions. We previously analyzed two of these sightlines that probe absorbers in the foreground of the  30~Doradus starburst region in  \citet{2025ApJ...984..161P}. And, we analyze the two sightlines that probe the N11 star-forming region in our \citet{Horton_2026} study. A detailed description of the component selection for each modeled absorber, including the O\textsc{~i} components associated with the H\textsc{~i} emission (see Table~\ref{tab:cloudy_results}) and the additional ionic constraints used in the modeling, is provided in Appendix~\ref{sec:cloudy_more}.

\subsection{Methodology}

The modeling approach we used closely follows the methodology described in \citet{2025ApJ...984..161P}, where that study provides full details of the adopted radiation fields, parameter grid, and optimization procedure. Briefly, each absorber is modeled as a plane-parallel slab of uniform density illuminated by a composite radiation field that includes contributions from the Milky Way disk, the Magellanic Clouds, and the extragalactic UV background (\citealt{2005ApJ...630..332F, bland2019, 2013ApJ...771..132B, khaire2019}; see \textsc{galrad} software). Each $(ra,\,dec,\,d_\odot)$ position in this combined radiation-field model corresponds to a predicted radiation-field strength and spectral shape. However, the distances to the absorbing clouds are not known a priori; we only know that they lie between us and the LMC, where the embedded stars serve as the background sources. Therefore, we performed simulations over a wide range of possible cloud distances, corresponding to incident hydrogen-ionizing photon fluxes of $4.65\lesssim\log{\left(\Phi_{\rm H}/{\rm photons\,\cm^{-2}\,\s^{-1}}\right)}\lesssim7.05$ and hydrogen number densities within
$-3\le \log{\left(n_{\rm{H}}/\cm^{-3}\right)} \le 3$. The \hi\, column density serves as the model stopping criterion. We adopted the \hi\, column density values from \citet{2009ApJ...702..940L}, which are in range of $18.22 \lesssim \rm log(N_{\hi}/cm^{-2}) \lesssim 19.23$ (see Table~\ref{tab:cloudy_results}). As in \citet{2025ApJ...984..161P}, we use \textsc{Cloudy}’s \texttt{optimize} command to determine the best-fit combination of $\Phi_{\rm H}$ and $n_{\rm H}$ that reproduces the column densities of the volatile species that are measured from \textit{HST} and \textit{FUSE} observations (see Table~\ref{tab:Voigt_results}).

To distinguish which \textsc{Cloudy} radiative transfer model best reproduces the properties of the absorbers, we compared how well the observed and modeled ionic column densities (and limits) of volatile species agree. In particular, we use the ionic column densities of O\textsc{~i}, S\textsc{~ii}, and P\textsc{~ii} as these atomic species are less affected by dust depletion (see Figure~\ref{fig:obsvsprd_N} in Appendix). We use the best matched model to assess the ionization state of the gas and to compute the ionization corrections (ICs) that we used to determine the gas-phase abundance. The ionization correction is expressed as a difference between the true and observed ion abundance, i.e.,
\begin{equation}
    \mathrm{IC}(\mathrm{X}^i) = [\mathrm{X/H}] - [\mathrm{X}^i/\mathrm{H\;I}],
\end{equation}
where the superscript $i$ is indicating a different ionization species of element~X. 

We also used these radiative transfer models to determine the depletion of a refractory element by comparing its ionization--corrected metallicity with the ionization--corrected oxygen abundance. The depletion, $\delta_\mathrm{O}$(X), of a refractory element X is measured as a difference between ionization-corrected abundances of X and oxygen:
\begin{equation}
    \delta_\mathrm{O}(\mathrm{X}) \equiv [\mathrm{X/H}] - [\mathrm{O/H}].
\end{equation}
The best-fit values for $\Phi_{\rm H}$, $n_{\rm H}$, ionization--corrected metallicities, level of dust depletion, and hydrogen ionization fraction ($\chi_{\mathrm{H\textsc{~ii}}}$) for the modeled absorbers along all 13~sightlines are listed in Table~\ref{tab:cloudy_results}. For representative sightline BI\,272, we graphically illustrate these results and ionization correction values for each ion in Figure~\ref{fig:BI272_cloudy}. In the Appendix (Figures \ref{fig:cloudy_rest1} and \ref{fig:cloudy_rest2}), we include photoionization result figures for 8~more sightlines, where the 2~sightlines that probe 30~Doradus are found in \citet{2025ApJ...984..161P} and the 2 that probe N11~ are in \citet{Horton_2026}. The error bars on the ionization corrections in the top panel represent the highest and lowest possible correction to the ion abundance if the model column densities (i.e., O\textsc{~i}, Si\textsc{~ii}, or Fe\textsc{~ii}) were 10\% higher or lower. We include this range to demonstrate the low sensitivity this change has on the ionization correction. Note that the observed column density errors are typically on the order of $0.1\, \rm dex$. This range of sensitivity is also propagated through to the calculated element abundance and the associated error. 

The photoionization modeling also helps us determine the characteristic thickness and temperature of these absorbing clouds. Furthermore, since these absorbers lie somewhere along the line of sight between the star in the LMC and the Sun, their radial distances from either galaxy means that the clouds will experience a different exposure to this ionizing photon radiation field. This enables us to place constraints on the distances to the absorbers based on their ionization conditions. We stress that there is degeneracy in the best matched $\phi_{\rm H}$ values that can provide the necessarily illumination to reproduce the observed properties of the absorbers due to there being a minima in the radiation field strength between these galaxies. We have listed the physical properties of these clouds, incident ionizing field strength, and the inferred distances in Table~\ref{tab:cloudy_results_2}.

\subsection{Results}

\begin{figure*}[ht!]
  \centering
  \includegraphics[scale=0.42, trim=0 0 0 0, clip]{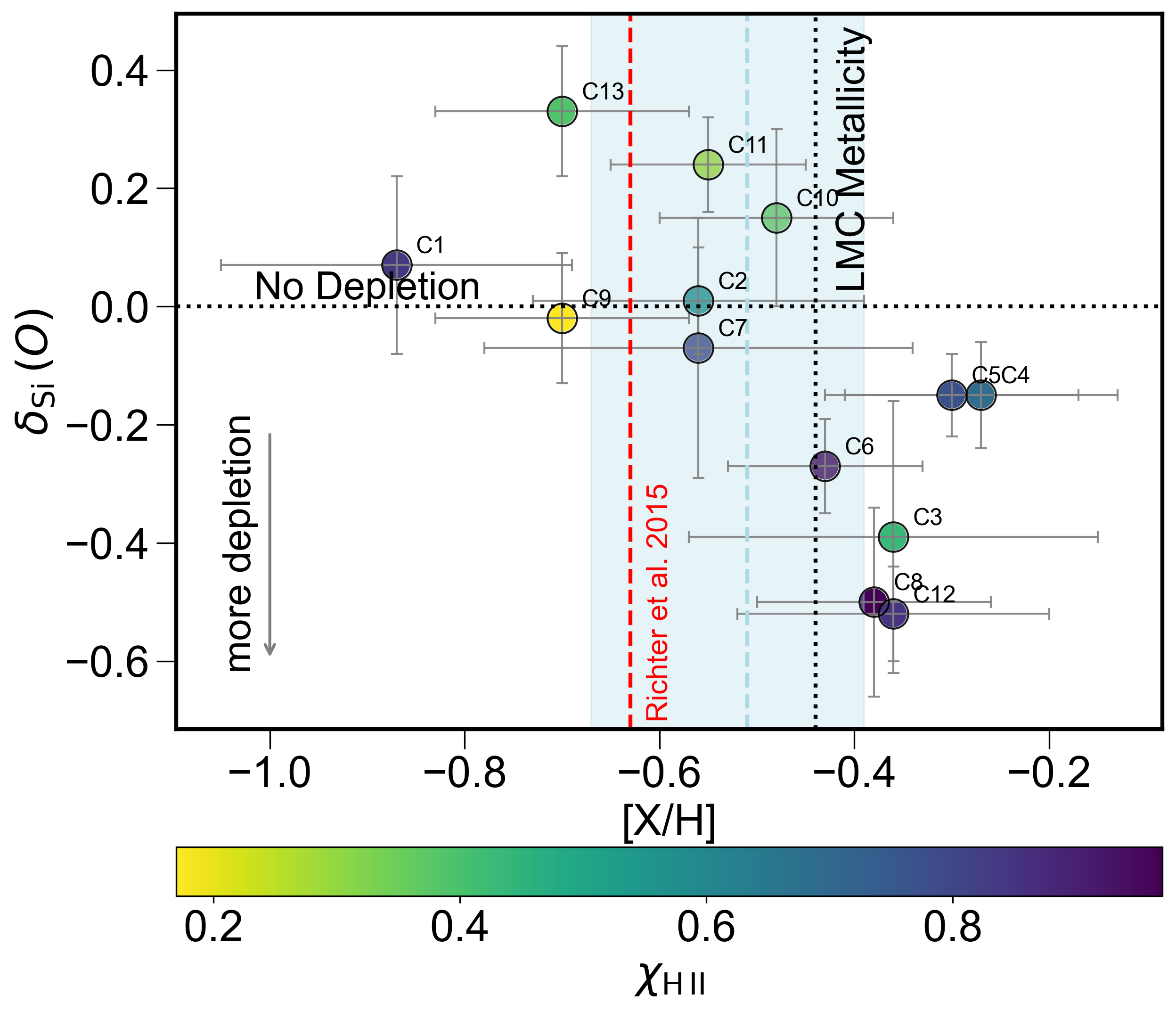}
  \includegraphics[scale=0.42, trim=0 0 0 0, clip]
  {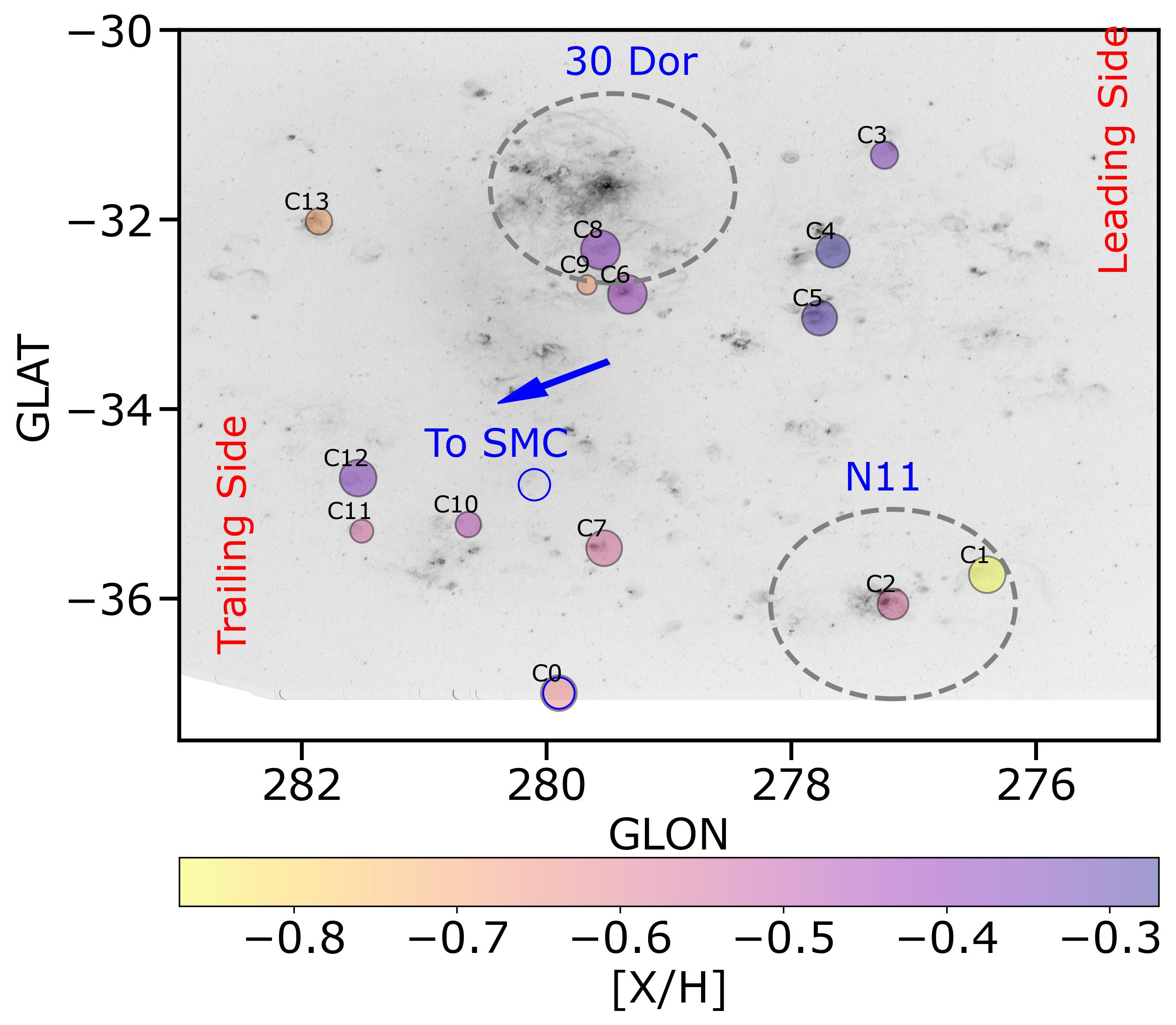}
  \caption{\textbf{Left:} Dust depletion versus metallicity for the fast-moving components relative to the LMC, estimated from \textsc{cloudy} photoionization modeling for the sightlines listed in Table~\ref{tab:cloudy_results}. The color scale indicates the hydrogen ionization fraction ($\chi_{\rm H\textsc{~ii}}$) of each component. The horizontal dashed line at $\delta_{\rm Si}(\rm O)=0$ marks the level of zero depletion, while the vertical dotted line shows the present-day metallicity of the LMC from \citet{russell1992}. For comparison, the mean metallicity ($\rm [O\textsc{~i}/H\textsc{~i}] = -0.51^{+0.12}_{-0.16}$; not ionization corrected) of the HVCs toward LMC at $+90 < v_{\rm LSR} < +175~{\rm km~s^{-1}}$ from \citet{2009ApJ...702..940L} is shown by the vertical dashed line in blue and the scatter by the shaded region. \textbf{Right:} Spatial distribution of the fast-moving absorption components across the LMC, overlaid on the H$\alpha$ emission map from MCELS. Colored circles mark individual sightlines (C1--C13), with colors indicating the gas-phase metallicity $[\mathrm{X/H}]$ derived from \textsc{Cloudy} photoionization modeling. The size of each marker scales with the hydrogen ionization fraction, such that larger symbols correspond to more highly ionized gas. Two blue circles indicate directions where MW's HVCs at comparable velocities have been reported by \citet{2015A&A...584L...6R}. The filled blue circle (C0) denotes the sightline for which a metallicity measurement is available, while the open blue circle marks a direction without metallicity constraints. For these two sightlines, the marker size is not scaled with ionization fraction, as such information is unavailable. Dashed circles denote the locations of the major star-forming regions 30~Doradus and N11. The arrow indicates the direction toward the SMC, and the leading and trailing sides of the LMC disk are labeled.}
  \label{fig:cloudy_summary}
\end{figure*}

Our \textsc{Cloudy} photoionization modeling result reveals substantial chemical and physical diversity, indicating that the $+90 \lesssim v_{\rm LSR} \lesssim +175~\kms$ absorption does not arise from a single population of clouds. The inferred oxygen abundances span nearly an order of magnitude,
with $[{\rm O/H}] \approx -0.9$ to $-0.2$ and the relative abundances of refractory (Si or Fe) vs. volatile (O or S)
exhibit a wide range of dust depletion levels that span from $\delta_{\rm O}({\rm Si})=-0.5$ to $+0.3$ (see Table~\ref{tab:cloudy_results}). Several absorbers along multiple sightlines (C3, C4, C5, C6, C8, and C12) cluster near the present-day LMC metallicity value from \citet{russell1992} (see left panel in Figure~\ref{fig:cloudy_summary}) and have substantial depletion of refractory elements onto dust grains as indicated by the significantly negative values of
$\delta_{\rm O}({\rm Si})$ and $\delta_{\rm O}({\rm Fe})$. The C6 and C8 sightlines probe absorbers directly in the foreground of 30~Doradus and the C3, C4, and C5 sightlines lie along the periphery of the 30~Doradus quadrant (see right panel in Figure~\ref{fig:cloudy_summary}). These dust-rich absorbers are projected toward the higher Galactic latitude side of the LMC, which contains the 30~Doradus starburst region; their LMC-like chemical composition suggests that they trace cool, entrained gas associated with the LMC outflow driven by this active starburst region. This is further supported by previous evidence that LMC-driven winds can reach comparable speeds in regions of strong feedback, particularly near 30~Doradus (e.g., \citealt{Ciampa2020, 2025ApJ...984..161P}).

While a second subset of absorbers (sightlines C1, C2, C7, C9, C10, C11, C13) exhibit metallicities lower than the first group we discussed, about half of them (C2, C7, and C10) have values which may be consistent with the present-day LMC metallicity from \citet{russell1992} within $1\,\sigma$. While C1 has the lowest oxygen abundance value of $[{\rm O/H}] \approx -0.9$, most of the absorbers in this second subset actually have metallicities overlapping with the mean metallicity found by \citet{2009ApJ...702..940L} of $\rm [O\textsc{~i}/H\textsc{~i}] = -0.51^{+0.12}_{-0.16}$ (not ionization corrected) for the HVCs toward LMC at $+90 < v_{\rm LSR} < +175~{\rm km~s^{-1}}$. We found that some of these absorbers (i.e., along sightlines C2, C7, and C9) exhibit little to no dust depletion, with $\delta_{\rm O}({\rm Si}) \approx 0$. The lack of dust tracers for these absorbers are difficult to reconcile with a purely LMC wind origin and suggests that either this material is mixed with clouds from other origins or that this material is not associated with the LMC's wind. 

Finally, we found positive values of
$\delta_{\rm O}({\rm Si}) \equiv [{\rm Si/H}] - [{\rm O/H}]$ for a small number of absorbers that lie along sightlines C10, C11, and C13, meaning that their silicon appears enhanced relative to oxygen. 
This behavior is not expected from standard dust depletion patterns, which preferentially remove refractory elements such as silicon from the gas phase and therefore produce negative values of $\delta_{\rm O}({\rm Si})$.
Instead, positive $\delta_{\rm O}({\rm Si})$ values likely point to additional complexity, such as differential dust processing (e.g., partial grain destruction that selectively returns silicon to the gas phase \citep{1996ApJ...469..740J, 2004ApJ...614..796S} or residual uncertainties in ionization corrections.
Regardless of the specific cause, these components further emphasize that the gas in this velocity range is chemically heterogeneous and cannot be explained by a single, uniform origin.

We also compare the \textsc{Cloudy}-derived metallicities and depletion patterns with the empirical Si\textsc{~ii} offset from the extrapolated slow-wind relation defined in Section~\ref{subsection:kinematic_dist}. Because several modeled sightlines contain more than one Si\textsc{~ii} component, we perform this comparison at the component level rather than assigning a single classification to each sightline. Using the best-fit relation given in equation~\ref{eq:bestfitsii} and the residual scatter of 0.58~dex as the threshold for consistency with the slow-wind sequence, 12 of the 19 modeled Si~\textsc{ii} components lie within the near-trend region, whereas 7 are offset to higher column densities (see Figure~\ref{fig:population_test} in the Appendix).

The comparison with the \textsc{Cloudy} results does not reveal a clean separation between near-trend and high-offset components. Several chemically LMC-like, dust-bearing absorbers fall close to the extrapolated slow-wind trend,
including components associated with C3, C4, and C12, consistent with the possibility that some ambiguous-velocity gas is a high-velocity extension of the LMC wind. However, other LMC-like or dust-bearing absorbers, including components associated with C5, C6, and C8, are offset to higher Si~\textsc{ii} column densities. In addition, some sightlines, such as C4, C10, and C13, contain both near-trend and high-offset components. Therefore, the metallicity and depletion patterns do not map one-to-one onto the empirical near-trend/high-offset classification. This result reinforces the conclusion that the Si\textsc{~ii} offset is useful as an empirical diagnostic, but should not be treated as a definitive population classifier.

\begin{deluxetable*}{
    l  
    l  
    c  
    c  
    c  c  c  
    c  
    c  c    
    c  
    c  
}
\tablecaption{Sightlines selected for \textsc{Cloudy} photoionization modeling and the results on abundances and dust depletion patterns for absorbers at $+100 \lesssim v_{\rm LSR} \lesssim +175~\kms$ absorbers}
\label{tab:cloudy_results}
\scriptsize
\tablehead{
\multicolumn{4}{c}{} &
\multicolumn{3}{c}{Metallicity} &
\colhead{} &
\multicolumn{2}{c}{Dust Depletion} & \\
\cline{5-7} \cline{9-10}
ID & Sightline &
$\vlsr$\tablenotemark{\scriptsize \rm \textcolor{blue} a} &
$\log N_{\rm \hi}$\tablenotemark{\scriptsize \rm \textcolor{blue} a} &
{[X/H]\tablenotemark{\scriptsize \rm \textcolor{blue} b}} &
{[Si/H]} & {[Fe/H]} &&
$\delta_{\rm Si}(\rm O)$ &
$\delta_{\rm Fe}(\rm O)$ & \colhead{$\log  N_{\rm Si\textsc{~iv}}$\tablenotemark{\scriptsize \rm \textcolor{blue} m}} &  \colhead{$\log \ N_{\rm Si\textsc{~iv}}$\tablenotemark{\scriptsize \rm \textcolor{blue} o}}\\
 && (km\,s$^{-1}$) & ($\cm^{-2}$) &
 (dex) & (dex) & (dex) &&
 (dex) & (dex) &
 (cm$^{-2}$) & (cm$^{-2}$)
}
\startdata
C1 & Sk$-$65$^{\circ}$22\tablenotemark{\scriptsize \rm \textcolor{blue} h} & $+137$ &$18.45\pm0.17$ & $-0.87\pm0.18$ & $-0.80\pm0.21$ & $-0.87\pm0.18$ && $+0.07\pm0.15$ & $-0.00\pm0.09$ & $10.45$ & $12.70\pm0.12$ \\
C2 & N11$-$ELS$-$032\tablenotemark{\scriptsize \rm \textcolor{blue} h} & $+146$ &$18.62\pm0.15$ & $-0.56\pm0.17$ & $-0.55\pm0.16$ & $-0.64\pm0.18$ && $+0.01\pm0.09$ & $-0.08\pm0.13$ & $9.31$ & \ldots \\
C3 & BI$-$272 & $+141$ &$18.64\pm0.13$ & $-0.36\pm0.21$ & $-0.75\pm0.20$ & $-0.69\pm0.14$ && $-0.39\pm0.23$ & $-0.34\pm0.18$ & $9.33$ & \ldots \\
C4 & Sk$-$67$^{\circ}$191 & $+132$ &$18.77\pm0.12$ & $-0.27\pm0.14$ & $-0.43\pm0.14$ & $-0.64\pm0.13$ && $-0.15\pm0.09$ & $-0.36\pm0.08$ & $10.70$ & \ldots \\
C5 & Sk$-$67$^{\circ}$104 & $+115$ &$18.79\pm0.12$ & $-0.30\pm0.13$ & $>-0.36$ & $-0.46\pm0.13$ && $>-0.06$ & $-0.15\pm0.07$ & $11.65$ & $12.48\pm0.10$ \\
C6 & Sk$-$68$^{\circ}$80 & $+126$ &$19.07\pm0.07$ & $-0.43\pm0.10$ & $>-0.99$ & $-0.89\pm0.09$ && $>-0.38$ & $-0.27\pm0.08$ & $13.95$ & $12.99\pm0.03$ \\
C7 & Sk$-$68$^{\circ}$15 & $+144$ &$18.22\pm0.15$ & $-0.56\pm0.22$ & $-0.64\pm0.21$ & $-0.52\pm0.23$ && $-0.07\pm0.22$ & $-0.04\pm0.24$ & $9.66$ & $12.16\pm0.12$ \\
C8 & Sk$-$69$^{\circ}$175\tablenotemark{\scriptsize \rm \textcolor{blue} p} & $+116$ &$19.18\pm0.10$ & $-0.38\pm0.12$ & $-0.88\pm0.18$ & $-0.80\pm0.15$ && $-0.50\pm0.16$ & $-0.42\pm0.13$ & $14.24$ & \ldots \\
C9 & BI\,173\tablenotemark{\scriptsize \rm \textcolor{blue} p} & $+118$ &$18.99\pm0.06$ & $-0.70\pm0.13$ & $\geq -0.81$ & $-0.68\pm0.12$ && $\geq -0.11$ & $-0.02\pm0.11$ & $8.86$ & \ldots \\
C10 & Sk$-$69$^{\circ}$50 & $+147$ &$19.23\pm0.10$ & $-0.48\pm0.12$ & $-0.33\pm0.17$ & $-0.49\pm0.11$ && $+0.15\pm0.15$ & $-0.00\pm0.08$ & $10.24$ & \ldots \\
C11 & Sk$-$70$^{\circ}$13 & $+146$ &$18.96\pm0.07$ & $-0.55\pm0.10$ & $-0.32\pm0.08$ & $-0.69\pm0.10$ && $+0.24\pm0.08$ & $-0.14\pm0.10$ & $9.13$ & $13.01\pm0.04$ \\
C12 & Sk$-$70$^{\circ}$32 & $+143$ &$19.00\pm0.15$ & $-0.36\pm0.16$ & $>-0.90$ & $-0.89\pm0.16$ && $>-0.54$ & $-0.52\pm0.08$ & $12.69$ & $12.28\pm0.74$ \\
C13 & Sk$-$71$^{\circ}$45 & $+118$ &$18.93\pm0.08$ & $-0.70\pm0.13$ & $-0.36\pm0.10$ & $-0.77\pm0.09$ && $+0.33\pm0.11$ & $-0.07\pm0.11$ & $9.30$ & $13.01\pm0.09$
\enddata
\tablenotetext{}{
\textsuperscript{\color{blue}a}~The velocity centroid $\vlsr$ is measured in the local standard of rest (LSR) frame for the \hi\ emission line, and $\log N_{\rm \hi}$ is the neutral hydrogen column density derived from the GASS H\,{\sc i} survey \citep{2009ApJ...702..940L}. 
\textsuperscript{\color{blue}b}~Metallicity [X/H] is measured relative to solar, where X = O (oxygen) for all sightlines except C6 and C8, for which X = S (sulfur). 
\textsuperscript{\color{blue}h}~\textsc{Cloudy} photoionization modeling results for these sightlines are adopted from \citet{Horton_2026}. 
\textsuperscript{\color{blue}p}~\textsc{Cloudy} photoionization modeling results for these sightlines are adopted from \citet{2025ApJ...984..161P}.
\textsuperscript{\color{blue}m}~\textsc{Cloudy} photoionization modeling predicted column density.
\textsuperscript{\color{blue}o}~Observed column density.}
\end{deluxetable*}

\begin{deluxetable*}{
    l  
    l  
    c  
    c  
    c  
    c  
    c  c    
}
\scriptsize 
\tablecaption{Physical conditions of the absorbers at $+100 \lesssim v_{\rm LSR} \lesssim +175~\kms$\label{tab:cloudy_results_2}}
\tablehead{
\\[1pt]
  \multicolumn{6}{c}{} & \multicolumn{2}{c}{Location\tablenotemark{\scriptsize \rm \textcolor{blue} b}} \\
  \cline{7-8}
  \colhead{ID} & \colhead{Sightline} & \colhead{$\log n_{\mathrm{H}}$} & \colhead{Thickness\tablenotemark{\scriptsize \rm \textcolor{blue} a}} & \colhead{$\chi_{\rm H\textsc{~ii}}$} & \colhead{$\log \phi$} &
  \colhead{$d_{\rm LMC}$} & \colhead{$d_{\rm MW}$} \\
  & & \colhead{($\cm^{-3}$)} &\colhead{(pc)} & & \colhead{($\cm^{-2}\,\s^{-1}$)} & \colhead{(kpc)} & \colhead{(kpc)}
}

\startdata
C1 & Sk$-$65$^{\circ}$22\tablenotemark{\scriptsize \rm \textcolor{blue} h} & $-0.41$ & 2.4 & 0.84 & 6.2 & 3.1 & 5.2 \\
C2 & N11$-$ELS$-$032\tablenotemark{\scriptsize \rm \textcolor{blue} h} & $+0.05$ & 1.2 & 0.57 & 6.2 & 2.7 & 4.9 \\
C3 & BI$-$272 & $-0.34$ & 3.1 & 0.43 & 5.5 & 7.6 & 12.2  \\
C4 & Sk$-$67$^{\circ}$191 & $-0.32$ & 4.1 & 0.69 & 6.3 & 3.2 & 3.7  \\
C5 & Sk$-$67$^{\circ}$104 & $-1.31$ & 41 & 0.77 & 5.5 & 8.3 & 13.5 \\
C6 & Sk$-$68$^{\circ}$80 & $-1.62$ & 160 & 0.96 & 6.2 & 3.4 & 4.5\\
C7& Sk$-$68$^{\circ}$15 & $-0.08$ & 0.7 & 0.80 & 6.2 & 2.7 & 4.9  \\
C8 & Sk$-$69$^{\circ}$175\tablenotemark{\scriptsize \rm \textcolor{blue} p} & $-2.13$ & 67 & 0.97 & 5.8 & 3.6 & 6.6  \\
C9 & BI\,173\tablenotemark{\scriptsize \rm \textcolor{blue} p} & $-0.29$ & 6.5 & 0.17 & 5.2 & 7.5 & 4.3  \\
C10 & Sk$-$69$^{\circ}$50 & $-0.49$ & 17.2 & 0.36 & 5.9 & 4.4 & 8.3  \\
C11 & Sk$-$70$^{\circ}$13 & $-0.20$ &  4.7 & 0.28 & 5.7 & 6.0 & 11.5 \\
C12 & Sk$-$70$^{\circ}$32 & $-1.28$ &  62 & 0.85 & 6.0 & 3.9 & 7.4 \\
C13 & Sk$-$71$^{\circ}$45 & $-0.22$ & 4.6 & 0.39 & 5.9 & 4.8 & 8.5 \\
\enddata
\tablenotetext{}{
\textsuperscript{\color{blue}a}~The characteristic thickness of the absorbing neutral gas layer is estimated from the hydrogen number density and the neutral hydrogen column density as $r = N_{\rm H\textsc{i}}/n_{\rm H}$. 
\textsuperscript{\color{blue}b}~Distances of the absorbing clouds are predicted either from the center of the Large Magellanic Cloud ($d_{\rm LMC}$) or from the observer’s position in the Milky Way disk ($d_{\rm MW}$). 
\textsuperscript{\color{blue}h}~\textsc{Cloudy} photoionization modeling results for these sightlines are adopted from \citet{Horton_2026}. 
\textsuperscript{\color{blue}p}~\textsc{Cloudy} photoionization modeling results for these sightlines are adopted from \citet{2025ApJ...984..161P}.
}

\end{deluxetable*}

\begin{figure}[ht!]
    \centering
    \includegraphics[width=\columnwidth]{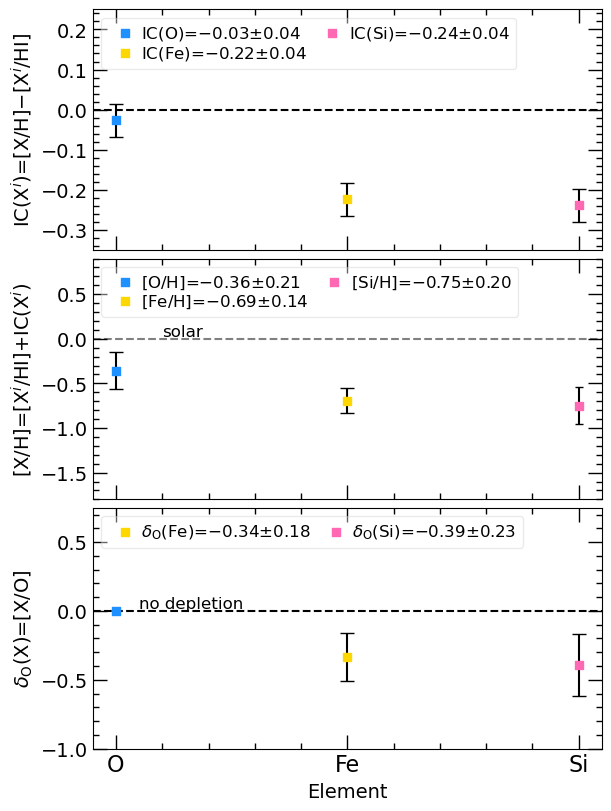}
    \caption{Chemical abundance analysis for the $ v_\mathrm{LSR} \approx +145\,\kms$ component toward BI\,272 with optimal hydrogen number density $\log{\left(n_{\rm H}/\cm^{-3}\right)} = -0.34$ and hydrogen ionizing photon flux of $\log{\left(\Phi_{\rm H}/{\rm photons\,\cm^{-2}\,\s^{-1}}\right)} = 5.55$. Top panel: ionization corrections for the low ions O\textsc{~i}, Si\textsc{~ii}, and Fe\textsc{~ii}. The dashed gray horizontal line at zero indicates no ionization correction. Middle panel: comparison of the elemental abundances after correcting for ionization. The dashed gray horizontal line at zero marks the solar abundance.  Bottom: comparison of the levels of depletion among different elements $\delta_\mathrm{O}$(X) = [X/O]. The dashed gray horizontal line at zero indicates no depletion relative to oxygen.}
    \label{fig:BI272_cloudy}
\end{figure}

\section{High ions}
\label{section:high-ions}

\begin{figure*}
  \centering
  \includegraphics[scale=0.50, trim=0 0 0 0, clip]{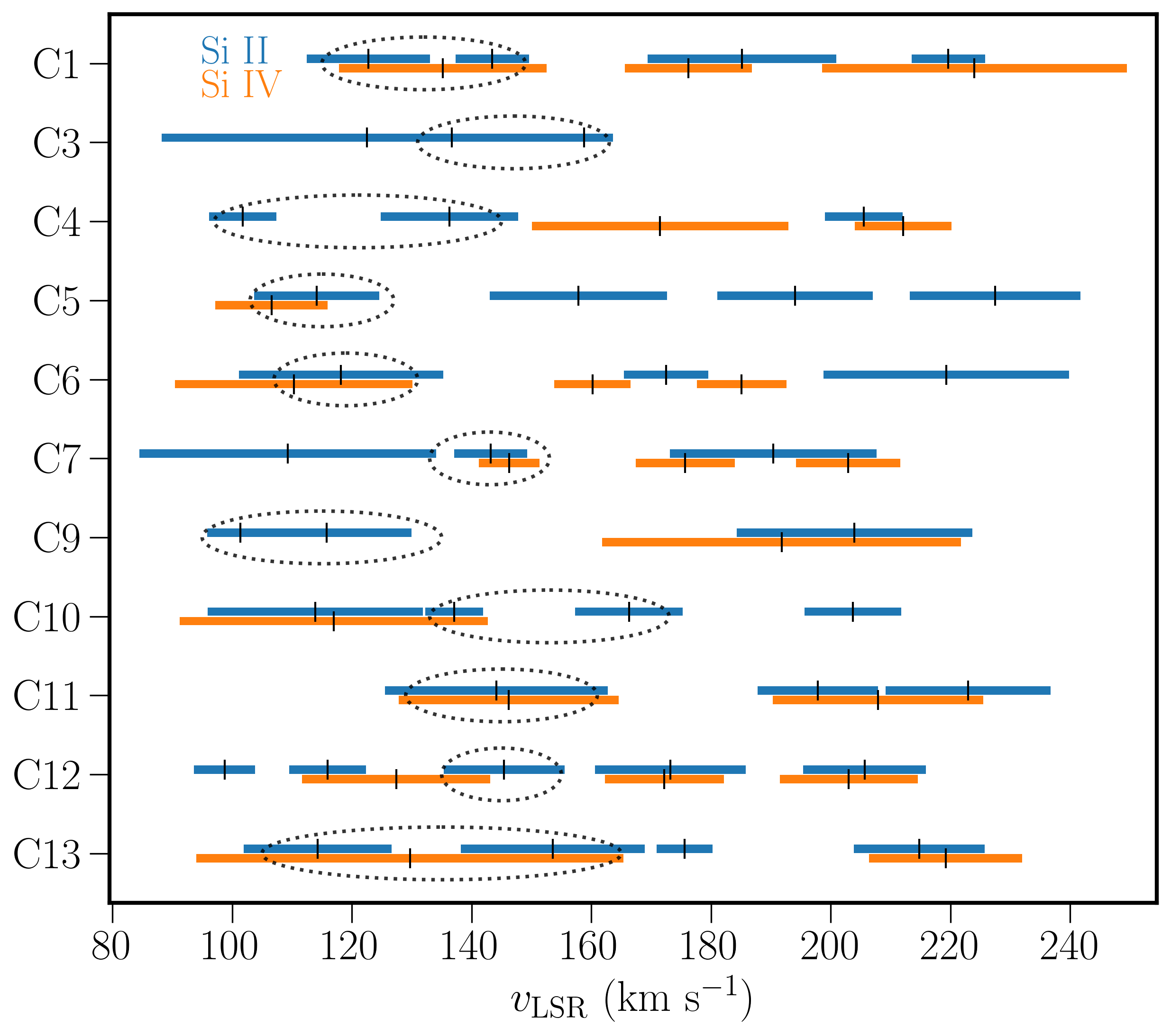}
  \caption{Comparison of full width at half maximum (FWHM) distributions for Si\textsc{~ii} (as blue bars) and Si\textsc{~iv} (as orange bars) toward the sightlines with available \textsc{Cloudy} photoionization modeling for the ambiguous HVC components in the velocity range $+90 < v_{\rm LSR} < +175~{\rm km~s^{-1}}$. Out of the 13 sightlines (C1–C13) listed in Table~\ref{tab:cloudy_results}, we exclude C2 and C8 for which high-ion measurements are unavailable due to complex continuum structure. The velocity centroids for each component are marked by black ticks. The dotted ellipses highlight the Si\textsc{~ii} components used in the photoionization modeling.}
  \label{fig:high_ions}
\end{figure*}

We compare the low-ion and high-ion phases for a subset of sightlines with available \textsc{Cloudy} photoionization modeling by examining their Si\textsc{~ii} and Si\textsc{~iv} absorption. Although the photoionization analysis was performed for 13~sightlines (C1–C13, see Table~\ref{tab:cloudy_results}), we were not able to conduct Voigt profile fitting for C2 and C8. Therefore, we exclude these two sightlines from our analysis. We compare the velocity centroids and FWHM of all detected Si\textsc{~iv} and Si\textsc{~ii} components in Figure~\ref{fig:high_ions} and indicate components that were used in the \textsc{Cloudy} models with dotted ellipses.

In the ambiguous HVCs window ($+90 < v_{\rm LSR} < +175~{\rm km~s^{-1}}$), the Si\textsc{~iv} components have $\langle b\rangle = 20.2 \pm 10.9~{\rm km~s^{-1}}$ for all modeled sightlines, which are scattered across the LMC's disk. The Si\textsc{~iv} absorption in this velocity range is noticeably broader than the Si\textsc{~ii} at $13.9 \pm 8.6~{\rm km~s^{-1}}$; although this is only for a small subset of the total sample (13/170), the global kinematic width of Si\textsc{~ii} at $13.6 \pm 7.4~{\rm km~s^{-1}}$ is in relative agreement (see Figure~\ref{fig:b_SiII}). In the slow LMC wind velocity regime ($+175 < v_{\rm LSR} < +230~{\rm km~s^{-1}}$), the Si\textsc{~iv} widths decrease slightly ($\langle b\rangle = 16.9 \pm 9.5~{\rm km~s^{-1}}$) and become more comparable to Si\textsc{~ii} at $14.7 \pm 6.0~{\rm km~s^{-1}}$ (or $16.9 \pm 10.3~{\rm km~s^{-1}}$ for the global sample). Overall, this pattern---systematically larger $b$ for Si\textsc{~iv} than for Si\textsc{~ii} in the ambiguous HVCs regime---indicates that the high ions preferentially trace warmer and/or more turbulent gas. 

Next, we focus only on the components for which we have conducted \textsc{Cloudy} modeling (enclosed by the ellipses in Figure~\ref{fig:high_ions}). In cases where an ellipse encloses two Si\textsc{~ii} components (i.e., two blue bars), the modeling was performed using the summed low-ion column densities of those components to remain consistent with broader \hi\ emission. Along several sightlines (notably C3, C4, C9, and C10), we find no obvious Si\textsc{~iv} counterpart to the Si\textsc{~ii} components; this is consistent with the \textsc{Cloudy} predictions for these cases, which yield negligible photoionized Si\textsc{~iv} column densities ($\log{N_{\rm Si\textsc{~iv}}}\approx 9.3–10.7$, see Table~\ref{tab:cloudy_results}). In contrast, C1 and C13 exhibit a single, comparatively broad Si\textsc{~iv} component spanning the velocity range occupied by two closely separated Si\textsc{~ii} components. For these two sightlines, the modeled photoionized Si\textsc{~iv} column densities ($\log{N_{\rm Si\textsc{~iv}}}\approx 10.45$ for C1 and $9.30$ for C13) fall far below the observed values ($12.70\pm0.12$ and $13.01\pm0.09$, respectively), indicating that only photoionization cannot account for the high-ion absorption and that an additional collisional contribution is required. Several other cases (C5, C6, C7, and C11), the Si\textsc{~iv} components broadly track the Si\textsc{~ii} velocities, but with modest centroid offsets. Here, the \textsc{Cloudy} predictions suggest a range of photoionized contributions, from partial (e.g., C5: predicted $\log{N_{\rm Si\textsc{~iv}}}=11.65$ vs. observed $12.48\pm0.10$) to potentially dominant (e.g., C6: predicted $\log{N_{\rm Si\textsc{~iv}}}=13.95$ vs. observed $12.99\pm0.03$). Finally, C12 stands out because the Si\textsc{~iv} component is broader and more offset from Si\textsc{~ii}, yet the modeled column density ($\log N_{\rm Si\textsc{~ii}}=12.69$) is consistent with the observed $\log N_{\rm Si\textsc{~ii}}=12.28\pm0.74$ within the large measurement uncertainty, suggesting that photoionized gas alone may not explain the large offset Si\textsc{~iv} absorption. 

For the sightlines C3, C4, C5, C6, C8, and C12---each of which shows clear LMC-like metallicity and dust depletion patterns (see left panel of Figure~\ref{fig:cloudy_summary})---the Si\textsc{~iv} behavior further constrains the nature of the high-speed outflows. For the C3 and C4 sightlines, no Si\textsc{~iv} absorption is detected, indicating that these high-velocity outflows are traced primarily by Si\textsc{~ii}. For the C5 and C6 sightlines, the photoionized contribution to Si\textsc{~iv} appears significant or dominant. Although the Si\textsc{~iv} column density for the C8 sightline is not measurable due to continuum issues, the relatively high \textsc{Cloudy}-predicted column density suggests that photoionization likely dominates there as well. For the C12 sightline, the photoionized contribution cannot be neglected despite the broader and offset Si\textsc{~iv} profile. Taken together, these sightlines with LMC-like metallicity and significant dust depletion pattern are generally consistent with predominantly photoionized gas.

\section{Discussion} \label{section:Discussion}

\subsection{Outflow mass, outflow rate, and mass loading factor}
We estimate the global mass of the low-ion outflow traced by Si\textsc{~ii} by adopting a characteristic \hi\, disk radius of $R \approx 3.7$~kpc \citep{1998ApJ...503..674K} as a representative spatial scale for the LMC. While the radial extent of the outflow is not directly measured, this value provides a reasonable ballpark estimate for the characteristic size of the system. To determine the column density contributing to the bulk outflow, we integrate the mean column-density--velocity curve presented in Figure~\ref{fig:SiII_logN} as a blue line between the median disk boundary ($v_{\rm LMCSR} = -34~{\rm km~s^{-1}}$) and $v_{\rm LMCSR} = -100~{\rm km~s^{-1}}$, which is the interval that encompasses the majority of the slow wind components. The column density derived from this approach is consistent with the typical values obtained from the AOD measurements over the same range. The total mass in the outflow is proportional to the hydrogen column density and the projected surface area subtended by the flow,
\begin{equation}
M_{\rm outflow} = \mu\, \langle N_{\rm H} \rangle\, A\, \cos i ,
\end{equation}
where $i = 23.4^\circ$ is the inclination of the LMC \citep{2022ApJ...927..153C} and $\mu = 1.3\,m_{\rm p}$ accounts for the mean particle mass including helium. We parameterize the area as $A = c_f \pi R^2$, adopting a covering fraction of $c_f = 0.8$ based on the fraction of sightlines with column densities higher than global average trend. To convert the observed Si\textsc{~ii} column densities into hydrogen, we assume a characteristic LMC metallicity of $Z = 0.5\,Z_\odot$ \citep{1992ApJ...384..508R}. The resulting outflow mass becomes
\begin{equation}
M_{\rm outflow} =
\frac{1.3\, m_{\rm p}\, \pi R^{2}\, c_f\, \cos i}
{(Z/Z_{\odot})\,({\rm Si/H})_{\odot}}
\int_{v_1}^{v_2} N_{\rm Si\textsc{~ii}}(v)\, dv ,
\end{equation}
where $v_1$ corresponds to the adopted disk boundary and $v_2 = -100~{\rm km~s^{-1}}$. Using this approach, we obtain a Si\textsc{~ii } traced baryonic outflow mass for low ionization species of
\begin{equation}
M_{\rm baryon} \approx 1.8 \times 10^{7}\ M_{\odot} \left(\frac{R}{3.7\,\rm kpc}\right)^2\left(\frac{c_f}{0.8}\right).
\end{equation}
Including the higher-velocity material in the range $-100 \ga v_{\rm LMCSR} \ga -200~{\rm km~s^{-1}}$ would increase the total column density only minimally; this is because the Si\textsc{~ii} column densities in this regime are typically small compared to those of the dominant slow component.

To estimate the mass outflow rate, we adopt a characteristic distance that the LMC-like clouds have reached from the disk that is based on our radiative transfer modeling (see Table~\ref{tab:cloudy_results_2}) results. These models imply the LMC's winds span outward up to a galactocentric distance of $d \approx 3.7-8.3$~kpc. Assuming a representative outflow velocity of $v_{\rm LMCSR} = -67~{\rm km~s^{-1}}$ (see Figure~\ref{fig:SiII_logN}), which is the average velocity centroid for the slow part of the wind, the corresponding flow timescale is
\begin{equation}
t_{\rm flow} = \frac{d}{|v|} \approx 54\text{--}121~{\rm Myr}.
\end{equation}
Dividing the total outflow mass derived above by this travel time yields a nearside mass outflow rate of
\begin{equation}
\dot{M}_{\rm out} \approx 0.15\text{--}0.33~M_{\odot}~{\rm yr^{-1}}.
\end{equation}
Adopting a global LMC star formation rate of ${\rm SFR} \approx 0.25~M_{\odot}~{\rm yr^{-1}}$ over the time period considered from \citep{2009AJ....138.1243H}, we obtain a mass loading factor
\begin{equation}
\eta \equiv \frac{\dot{M}_{\rm out}}{{\rm SFR}} \approx 0.6\text{--}1.3,
\end{equation}
again referring only to the nearside component. If the outflow is approximately symmetric about the disk, inclusion of material on the farside would increase the outflow mass, $\dot{M}_{\rm out}$ and $\eta$ by roughly a factor of two. 

We emphasize that these estimates are subject to systematic uncertainties associated with several assumptions in the calculation, including the adopted metallicity, the ionization state of the gas, and the covering fraction of the outflow. Variations in these parameters could change the inferred mass by factors of a few. A simple symmetric extension to both sides would imply a total cool outflow rate of $\dot{M}_{\rm out,tot}\approx 2\dot{M}_{\rm out,near}
\approx 0.3$--$0.7~M_\odot~{\rm yr^{-1}}$. This value is slightly above the
lower limit of $\dot{M}_{\rm out}\gtrsim 0.4~M_\odot~{\rm yr^{-1}}$ inferred by
\citet{2016ApJ...817...91B}, despite that their estimate was based on a more quiescent LMC region, which may not represent the average trend across the LMC. However, part of this agreement may reflect the different assumptions used to estimate the flow timescale. \citet{2016ApJ...817...91B} adopted $R_{\rm out}=3.7~{\rm kpc}$, corresponding to the approximate
radius of the LMC H\textsc{~i} disk, as the characteristic distance reached by
the outflowing gas. This value lies toward the lower end of the cloud-distance
range considered in our analysis based on the photoionization modeling. These estimates are also consistent with the \ha\, study of \citet{Ciampa2020}, particularly when both IVC and HVC material are included. Further, our inferred symmetric mass-loading factor ($\eta_{sym} \approx 1.2$--2.6) is broadly consistent with observational constraints for nearby dwarf galaxies. For example, \citet{2019ApJ...886...74M} derived $\eta \approx 0.2$--7 for a sample of low-mass starburst dwarfs with stellar masses of $M_\ast \approx 10^{7}$--$10^{9.3}\,M_\odot$ based on deep \ha\, imaging. 

Our cool-gas outflow rate is substantially larger than the conservative high-ion outflow rate inferred by \citet{2024ApJ...974...22Z}, who found $\dot{M}_{\rm out}\gtrsim 0.03~M_\odot~{\rm yr^{-1}}$ and $\eta\gtrsim 0.15$ for disk-wide ionized outflows traced primarily by C\textsc{~iv}. This difference is not unexpected because the two studies trace different gas phases and integrate over different velocity ranges. \citet{2024ApJ...974...22Z} focused on warmer, more highly ionized material with bulk velocities of only ${\sim}20$--$60~{\rm km~s^{-1}}$, whereas our analysis isolates cooler, low-ionization gas blueshifted relative to the LMC disk over a broader velocity range. Together, these results suggest that the LMC wind is strongly multiphase with a larger mass flux in the cool low-ionization phase and a smaller, but dynamically important contribution from the warmer ionized phase.

The inferred mass-loading factor, $\eta \approx 0.6$--$1.3$ for the nearside
cool outflow, or roughly $\eta \approx 1.2$--$2.6$ if the outflow is assumed to
be symmetric about the disk, is broadly consistent with expectations for
stellar-feedback-driven winds in low-mass galaxies. Observational and
theoretical studies of dwarf galaxies commonly find mass-loading factors of
order unity or larger, with substantial scatter arising from differences in gas
phase, aperture, assumed geometry, and star-formation history
\citep[e.g.,][]{2015MNRAS.454.2691M, 2017MNRAS.469.4831C, 2019ApJ...886...74M, 2021MNRAS.504.3412L}. In this context, the LMC does not appear exceptional in its mass loading, but it is a particularly valuable benchmark because its proximity allows the outflow to be
resolved across the disk and connected directly to individual star-forming
regions such as 30~Doradus and N11.

We next examine how individual star-forming environments contribute to the total mass budget by estimating regional outflow rates and loading factors. We begin with the 30~Doradus quadrant. As shown in Figure~\ref{fig:ha_map}, the majority of the sightlines in this region lie within ${\sim}1^\circ$ of the center of 30~Doradus. This angular extent corresponds to an in-plane radius of $R \approx 0.87\,\kpc$, which we adopt as the radius of the outflowing material associated with this starburst. Using the same methodology as for the global estimate, we integrate the mean column-density--velocity profile for R1 (Figure~\ref{fig:SiII_logN_regions}, top-left panel) between the adopted median disk boundary of $v_{\rm LMCSR} = -42~{\rm km~s^{-1}}$ for this region and $v_{\rm LMCSR} = -125~{\rm km~s^{-1}}$. Among the sightlines in R1, two of the absorbers that we conducted radiative transfer modeling with \textsc{cloudy} have metallicity and dust depletion that is consistent with the LMC. These models place the absorbing clouds at typical galactocentric distances of $d \approx 3.5~\kpc$ from the LMC's disk and therefore adopt this value as the representative travel distance for these outflows. Using an velocity centroid of $v_{\rm LMCSR} \approx -83~{\rm km~s^{-1}}$ yields a flow timescale of $t_{\rm flow} \approx 41~{\rm Myr}$. This implies a nearside outflow mass of $M_{\rm out} \approx 1.7 \times 10^{6}\,M_{\odot}$ and a corresponding mass outflow rate of $\dot{M}_{\rm out} \approx 0.04\,M_{\odot}\,{\rm yr^{-1}}$. Adopting a regional star formation rate of ${\rm SFR}_{\rm 30~Dor} \approx 0.18\,M_{\odot}\,{\rm yr^{-1}}$ from \citet{2023ApJ...944...26N}, we infer a mass loading factor of $\eta_{\rm 30~Dor} \approx 0.23$. These values are in broad agreement with our earlier estimate of the 30~Doradus outflow in \citet{2025ApJ...984..161P}. However, the present approach benefits from the larger sightline sample (66 compared with 8), velocity-resolved integration, and constraints from photoionization modeling, thereby providing a more robust characterization of the outflow in this region.

The other key region is N11, the second largest star-forming complex in the LMC after 30 Doradus, although its stellar populations indicate a somewhat younger episode of activity. The oldest stellar association (LH9) has an age of ${\sim}3.5$–7 Myr \citep{1999AJ....118.1684W,2007A&A...465.1003M}. The pre-main sequence stars across the complex span ${\sim}2–10\,\rm Myr$ \citep{2010A&A...511A..79V}, with the majority of the stars being older than $6$ Myr. For the two fast moving absorbers that we conducted radiative transfer modeling via \textsc{cloudy}, we did not find that they had chemical or dust depletion patterns consistent with an LMC wind origin. This further suggests that the gas in this region does not reach the high outflow velocities observed for 30~Doradus. Using the same angular aperture of ${\sim}1^\circ$ adopted for 30~Doradus to enable a direct comparison, we derive a nearside outflow mass of $M_{\rm out} \approx 6.1 \times 10^{5}\,M_{\odot}$ for N11. Although this outflow mass is smaller than that of 30~Doradus, the inferred outflow rate is $\dot{M}_{\rm out}\gtrsim 0.06~M_{\odot}\,\mathrm{yr^{-1}}$, with a mass-loading factor of $\eta_{\rm out}\gtrsim 2.5$, both exceeding the values obtained for 30~Doradus. This suggests that, although N11 is less extreme in absolute output, it may be comparatively more efficient at converting recent star formation into large-scale gas removal. 

Finally, both the leading (R2) and trailing (R3) regions have uniformly distributed OB stellar targets in the ULLYSES sample that cover comparable projected areas, as illustrated in Figure~\ref{fig:hi_region_map}. For the leading side, we derive a total nearside outflow mass of $M_{\rm out} \approx 3.6 \times 10^{6}\, M_{\odot}$ using an effective radius of $R \approx 1.7~\mathrm{kpc}$. For the trailing side, we obtain a larger mass of $M_{\rm out} \approx 4.3 \times 10^{6}\,M_{\odot}$ using an effective radius $R \approx 1.6~\mathrm{kpc}$. These two effective radii correspond to the projected radii of the R2 and R3 regions traced in Figure~\ref{fig:hi_region_map}, assuming a distance to the LMC of $50\,\rm kpc$. Thus, despite the comparable coverage and area, the trailing side generally contributes more outflowing material than the leading side, consistent with expectations from gas stripping effects. We have listed summary of the outflow masses, outflow rates, and mass loading factors for all the regions in Table~\ref{tab:outflow_summary}.

\begin{table*}[ht!]
\centering
\begin{tabular}{ccccc}
\hline
Region & Outflow Mass & Outflow Rate & Mass Loading Factor \\
Name & $M_{\odot}$ & $M_{\odot}$ yr$^{-1}$ & $\eta$ \\ \hline
Total (Global) & $1.8 \times 10^{7}\ M_{\odot} \left(\frac{R}{3.7\,\rm kpc}\right)^2\left(\frac{c_f}{0.8}\right)$ & $0.15$--$0.33$ & $0.6$--$1.3$ \\ 
30 Doradus & $1.7 \times 10^6\ M_{\odot} \left(\frac{R}{0.87\,\rm kpc}\right)^2\left(\frac{c_f}{1.0}\right)$ & $0.04$ & $0.23$ \\ 
N11 & $6.1 \times 10^5\ M_{\odot} \left(\frac{R}{0.87\,\rm kpc}\right)^2\left(\frac{c_f}{1}\right)$ & $\gtrsim 0.06$ & $\gtrsim 2.5$ \\ 
Leading Side (R2) & $2.3 \times 10^6\ M_{\odot} \left(\frac{R}{1.7\,\rm kpc}\right)^2\left(\frac{c_f}{0.8}\right)$ & - & - \\ 
Trailing Side (R3) & $3.4 \times 10^6\ M_{\odot} \left(\frac{R}{1.7\,\rm kpc}\right)^2\left(\frac{c_f}{0.8}\right)$ & - & - \\ \hline
\end{tabular}
\caption{Summary of total and regional outflow masses, outflow rates, and mass loading factors.}
\label{tab:outflow_summary}
\end{table*}

\subsection{Origin of the ambiguous clouds}
The gaseous environment surrounding the LMC is intrinsically complex, shaped by the combined influence of internal feedback and external interactions. In addition to large-scale outflows driven by stellar feedback within the LMC disk, the region hosts multiple overlapping gas reservoirs, including the cool CGM of the Magellanic Clouds \citep{2024ApJ...976L..28M}, the warm–hot Magellanic Corona \citep{2020Natur.585..203L, 2022Natur.609..915K}, tidally stripped material produced during past LMC-SMC interactions (e.g., \citealt{1974ApJ...190..291M, 2005A&A...432...45B, 2007ApJ...668..949B, 2016ARA&A..54..363D}), and foreground gas associated with the MW halo and its population of HVCs \citep{2015A&A...584L...6R, 2017A&A...607A..48R}. These components can coexist at similar velocities making it challenging to assign a unique physical origin to individual absorption components based on kinematics alone.

In practice, distinguishing LMC wind material from foreground MW's HVCs requires multiple diagnostics beyond velocity alone. In this work, we rely on three main indicators. First, chemical abundances and dust depletion patterns provide valuable, although not unique, constraints on the origin of the gas. Throughout this work, we compare the derived metallicities with the present-day LMC abundance \citep{russell1992} and the chemical properties of the LMC outflow reported by \citet{2009ApJ...702..940L}. Absorbers with metallicities and depletion patterns comparable to those of the present-day LMC are therefore consistent with an LMC origin. However, recent studies have shown that some unambiguous MW HVCs also exhibit similarly high metallicities (e.g., \citealt{2023ApJ...944...65C,2023ApJ...946L..48F}) and measurable dust depletion (e.g., \citealt{2025ApJ...988..251V}), indicating that these diagnostics alone cannot uniquely distinguish between LMC-associated gas and foreground MW HVCs. Second, the spatial distribution of absorbers relative to major star-forming regions offers an important clue, since gas accelerated by stellar feedback should preferentially appear in directions connected to active star formation. Third, the kinematic continuity between the
slow-wind regime and some of the higher-velocity absorbers suggests that part
of the ambiguous-velocity gas may represent a high-velocity extension of the
LMC outflow sequence. However, the broad scatter in the column-density--velocity relation and the lack of a clean separation in the photoionization results indicate that this diagnostic alone cannot unambiguously determine the gas origin. Together, these diagnostics suggest that the ambiguous velocity regime likely contains a mixture of LMC-driven wind material, Magellanic-associated gas, and foreground MW HVCs.

Historically, several studies have interpreted absorption at $+90 \lesssim v_{\rm LSR} \lesssim +175~\kms$ toward the general direction of the LMC as arising from foreground Milky Way HVCs (e.g., \citealt{1981ApJ...243..460S, 1999Natur.402..386R}). However, these interpretations were largely based on sparse and non-uniform sightline coverage and did not systematically probe lines of sight that intersect the dense \hi\ disk of the LMC itself. In particular, two key observational results motivated a foreground origin. First, high-velocity \hi\ absorption was detected at $\vlsr \approx +140~\kms$ both toward the LMC and several degrees away from the Magellanic Clouds, suggesting a spatially extended structure that may not be uniquely associated with the LMC \citep{1990A&A...233..523D}. Second, UV absorption at $\vlsr \approx +150~\kms$ was identified in the spectrum of a Milky Way halo star at $d_\odot \lesssim 13.3$~kpc projected near the periphery of the LMC’s \hi\ disk \citep{2015A&A...584L...6R}. While these results demonstrate that MW HVCs contribute to absorption in this velocity range, the limited spatial sampling and lack of sightlines directly aligned with the LMC’s \hi\ disk prevented a definitive assessment of whether a distinct LMC-associated component is also present.

However, other studies conclude that the high-velocity gas toward the LMC is associated with the LMC's galactic winds. A comprehensive study by \citet{2009ApJ...702..940L} using 139 stellar sightlines established the chemical fingerprint of the $+90 \lesssim \vlsr \lesssim +175 \,\kms$ absorbers in the foreground of the LMC. They measured a mean value for the oxygen metallicity of $\rm [O\textsc{~i}/H\textsc{~i}] = -0.51^{+0.16}_{-0.12}$ for these absorbers, which aligns closely with the abundance of the LMC (see Figure~\ref{fig:cloudy_summary}). That study also found evidence of dust from depletion patterns, where they determined that the $\rm [Si\textsc{~ii}/S\textsc{~ii}]$ and $\rm [Fe\textsc{~ii}/Si\textsc{~ii}]$ ratios are subsolar. The presence of dust in circumgalactic clouds provides strong support that they have a galactic origin as dust is created by stars, albeit the source could be the LMC, MW, or another satellite galaxy. High-resolution spectroscopy of the 30~Doradus halo using [O\textsc{~iii}] emission from \citet{2003MNRAS.344..741R} reveals faint high-velocity gas produced by supernovae, with velocities overlapping those of the kinematically ambiguous HVCs. Furthermore, a spectroscopically resolved \ha\, emission map by \citet{Ciampa2020} revealed that the wind's morphology is asymmetric and most concentrated around 30~Doradus; they provide compelling evidence that this starburst region is significantly polluting the velocity range that hosts these ``ambiguous HVCs''.

Our recent \citet{2025ApJ...984..161P} study suggests that absorbers in the velocity range $+100 \lesssim v_{\rm LSR} \lesssim 150\, \kms$ might represent a mixed association of these distinct populations in the direction of 30~Doradus. More than half of the sightlines that we modeled cluster around the present-day LMC metallicity within uncertainties as illustrated in Figure~\ref{fig:cloudy_summary}, indicating a strong chemical connection to the LMC rather than to the more metal-poor Milky Way halo cloud population. These LMC-like clouds are distributed across a wide spatial range, spanning both the leading and trailing sides of the LMC disk and extending well beyond the immediate vicinity of 30~Doradus. At the same time, the lower portion of the LMC disk extending to N11 region (Figure~\ref{fig:cloudy_summary}, right) shows a grouping of sightlines with similar metallicities, including C2, C7, C10, and C11. These values are comparable to that measured for the MW HVC along C0 reported by \citet{2015A&A...584L...6R}. This direction, together with another nearby sightline without a metallicity measurement where \citet{2015A&A...584L...6R} also detected MW HVC absorption, is indicated by the blue circles in the figure. The spatial coherence and chemical similarity raise the possibility that at least some of these absorbers trace the same foreground MW HVC structure rather than material physically associated with the LMC wind. Notably, one of the more metal-poor, dust-free and relatively neutral sightlines lies near the 30~Doradus region, while the N11 star-forming region does not exhibit particularly enhanced metallicity and dust depletion, indicating that non--LMC-wind-like gas clouds are also present in regions otherwise dominated by LMC activity.

In summary, the fast-moving gas at $+100 \lesssim v_{\rm LSR} \lesssim 150~\kms$ likely arises from a combination of three origins: (1) LMC-driven winds with higher  metallicity and dust depletion, (2) Magellanic-associated HVCs with lower metallicity and no dust depletion, and (3) foreground HVCs associated with the MW. The gas from all three populations lies on the near side of the LMC and overlap in velocity space. Assessing the extent of foreground clouds associated with the MW can most accurately be accomplished by searching for absorption signatures along MW halo stars that lie at known distances, though this is beyond the efforts of this study.

\section{Summary} \label{section:Summary}
In this work, we used ultraviolet absorption-line spectroscopy of 170 OB stars from the \textit{HST}/ULLYSES program, combined with \hi\, 21-cm emission from GASS and GASKAP and \ha\, emission from MCELS and SHASSA to probe the LMC’s stellar-driven galactic wind. Our main findings are summarized below.

\begin{enumerate}
    \item \textbf{Cloud acceleration:} 
    Using multiple Si\textsc{~ii} transitions, we find a clear and continuous decline in column density with increasing outflow velocity relative to the LMC disk, extending from the slow-wind regime into the kinematically ambiguous velocity range. This monotonic trend is preserved across different regions of the LMC and is consistent with models in which lower–column density clouds are more efficiently accelerated to higher velocities. Within the ambiguous ``HVC/wind?'' regime, the Si\textsc{~ii} absorbers have a broad range of offsets from the extrapolated slow-wind relation, including a high-column-density tail. However, this distribution does not provide strong evidence for two statistically distinct populations, so we interpret the offset only as an empirical diagnostic of possible mixed origins. The bulk of the wind spans velocities of $-30$ to $-100~\kms$ relative to the LMC, while the fastest outflow components reach speeds of up to $-175~\kms$.

    \item \textbf{Origin of the ``ambiguous HVCs'':}
Photoionization modeling of 13 sightlines in the fast-moving, kinematically ambiguous range ($100 \lesssim v_{\rm LSR} \lesssim 150~{\rm km~s^{-1}}$) reveals large diversity in metallicity, dust depletion, and ionization state. More than half of the components in this velocity range exhibit metallicities consistent with the present day value of LMC reported by \citep{russell1992} , have significant dust depletions and are spatially distributed in star-forming regions across the disk, likely suggesting a substantial contribution from LMC-driven outflows. At the same time, the presence of lower-metallicity and more weakly ionized components—some projected near active star-forming regions—indicates that this population is not homogeneous and likely includes contributions from Magellanic-associated and MW halo gas. The ambiguous velocity regime therefore represents a physically mixed interface where outflowing LMC material, stripped halo gas, and foreground clouds coexist, underscoring the limitations of velocity-based classifications alone.

\item \textbf{Outflow mass, outflow rate, and loading factor:}
We find a global Si\textsc{~ii}-traced nearside outflow mass of $\sim1.8\times10^{7}\,M_{\odot}$, implying $\dot{M}_{\rm out}\approx0.15$--$0.33\,M_{\odot}\,\mathrm{yr^{-1}}$ and a mass-loading factor of $\eta\approx0.6$--1.3. On regional scales, 30~Doradus dominates the energetics and contributes nearly ten percent of the total outflow mass, but shows a modest efficiency ($\eta\approx0.2$), whereas N11, despite its smaller outflow mass (about 3 percent of the total outflow mass), exhibits a comparable outflow rate and a substantially higher loading factor ($\eta\gtrsim2.5$). Across the disk outskirts, both the leading and trailing sides are uniformly sampled, yet the trailing side contains more mass, consistent with ram-pressure stripping.

\item \textbf{Linewidth kinematics:}
Across the full velocity range, the Si\textsc{~ii} linewidths span $b\approx3$--$40~\kms$, indicating that broadening beyond the thermal contribution is required. The observed linewidths likely reflect a combination of unresolved velocity substructure, turbulence, and bulk motions, with unresolved blending expected to contribute particularly in the COS spectra. We find no sharp discontinuity in the $b$-value distribution between the slow-wind and fast-moving regimes, suggesting that the kinematic properties of the absorbing gas remain broadly similar across these velocity ranges. The broadest components overlap the Doppler parameter distribution measured for the LMC CGM, although the present data do not uniquely distinguish intrinsically broad absorbers from unresolved blends.

\item \textbf{Stellar activity:}
The comparison of column density with stellar activity indicates that the relation is not uniform across velocity regimes. The slower-moving components remain broadly consistent with an origin influenced by stellar feedback, but the higher-velocity gas shows substantial scatter and regional variations which suggest the data favor a scenario in which direct outflows, entrainment of ambient material, and unrelated foreground or halo clouds all contribute.

\item \textbf{Ionization:}
The low-ion ratio analysis provides additional insight into the physical evolution of the outflow. In the kinematically ambiguous regime, the global Bayesian regression hints at a weak tendency for components at more negative velocities with respect to the LMC to exhibit higher levels of ionization. Similar behavior is visible in R2, R3, and R4, though with substantial scatter. In contrast, R1 (30~Dor) shows no clear monotonic trend, likely reflecting the presence of multiple overlapping cloud populations, as also suggested by the column-density structure. Taken together, these results suggest at least part of the rapidly moving material represents gas that becomes progressively more ionized as it is accelerated away from the disk, while additional populations of foreground or unrelated clouds contribute to the observed dispersion. Further, the combined observed and \textsc{Cloudy}-predicted Si\textsc{~iv} properties indicate that the high-speed outflows associated with the LMC-like metallicity and dust-rich sightlines are largely consistent with predominantly photoionized gas, although a modest collisionally ionized contribution cannot be ruled out in some cases.

\end{enumerate}

\acknowledgments

Support for this program was provided by NASA through the grants HST-AR-16602.001-A and HST-AR-17052.008-A from the Space Telescope Science Institute, which is operated by the Association of Universities for Research in Astronomy, Inc. under NASA contract NAS5-26555. Additional support for Horton was provided by NSF grant 2334434. This study used ULLYSES observations obtained with the NASA/ESA Hubble Space Telescope \citep{2025ApJ...985..109R}, retrieved from the Mikulski Archive for Space Telescopes (MAST: http://archive.stsci.edu listed under \href{https://doi.org/10.17909/t9-jzeh-xy14}{DOI: 10.17909/T9-JZEH-XY14}). STScI is operated by the Association of Universities for Research in Astronomy, Inc. under NASA contract NAS 5-26555. This work has made use of the Vienna Atomic Line Data Base (VALD) database, operated at Uppsala University, the Institute of Astronomy RAS in Moscow, and the University of Vienna. This scientific work uses data obtained from Inyarrimanha Ilgari Bundara / the Murchison Radio-astronomy Observatory. We acknowledge the Wajarri Yamaji People as the Traditional Owners and native title holders of the Observatory site. CSIRO’s ASKAP radio telescope is part of the Australia Telescope National Facility (\url{https://ror.org/05qajvd42}). Operation of ASKAP is funded by the Australian Government with support from the National Collaborative Research Infrastructure Strategy. ASKAP uses the resources of the Pawsey Supercomputing Research Centre. Establishment of ASKAP, Inyarrimanha Ilgari Bundara, the CSIRO Murchison Radio-astronomy Observatory and the Pawsey Supercomputing Research Centre are initiatives of the Australian Government, with support from the Government of Western Australia and the Science and Industry Endowment Fund. GASKAP-HI is partially funded by the Australian Government through an Australian Research Council Australian Laureate Fellowship (project number FL210100039 awarded to NMc-G). This study used archived \hi\ LMC data obtained through \url{https://www.astro.uni-bonn.de/hisurvey/AllSky_gauss/}.

\software{Astropy \citep{astropy:2013,astropy:2018,astropy:2022}, Cloudy \citep{ferland2017}, GIZMO \citep{gizmo,gadget},
VoigtFit \citep{2018arXiv180301187K},
galrad (\href{https://github.com/Deech08/galrad}{https://github.com/Deech08/galrad}).
}



\clearpage
\bibliographystyle{aasjournal} 
\bibliography{References} 

\appendix \label{section:appendix}

\section{Bayesian Regression Analysis}
\label{sec:bayesian}
We modeled the relationship between the logarithm of the SFR surface density, 
$\log (\Sigma_{\mathrm{SFR}})$, and the logarithm of the ionic column density, 
$\log N$, using a Bayesian linear regression framework implemented in \texttt{PyMC}. 
This approach allows us to incorporate both measurement uncertainties and lower-limit 
constraints (arising from saturated absorption lines) in a statistically consistent manner.

We assumed a linear relation of the form
\begin{equation}
    \log N = \alpha + \beta \, \log (\Sigma_{\mathrm{SFR}}) + \epsilon ,
\end{equation}
where $\alpha$ and $\beta$ are the intercept and slope, respectively, and 
$\epsilon$ represents the intrinsic scatter in $\log N$ at fixed 
$\log (\Sigma_{\mathrm{SFR}})$. 
The intrinsic scatter $\epsilon$ was modeled as a Gaussian random variable with a 
standard deviation $\sigma$, such that each sightline satisfies
\begin{equation}
    \log N_i \sim \mathcal{N}(\mu_i, \sqrt{\sigma^2 + \delta_i^2}),
\end{equation}
where $\mu_i = \alpha + \beta \, \log (\Sigma_{\mathrm{SFR},i})$ and $\delta_i$ 
is the measurement uncertainty in $\log N_i$.

We adopted broad, uninformative priors on all model parameters:
\begin{align}
    \beta &\sim \mathcal{N}(0,\,5), \\
    \alpha &\sim \mathcal{N}(14,\,5), \\
    \sigma &\sim \mathrm{HalfNormal}(1).
\end{align}
These choices reflect weak prior knowledge about the normalization and slope of 
the relation while constraining $\sigma$ to be positive.

For the detected (uncensored) data points, we modeled the observed $\log N$ values directly as normally distributed about the regression mean.

\begin{equation}
\log N_{\mathrm{obs}} \sim \mathcal{N}\left( \mu_i,\, \sqrt{\sigma^2 + \delta_i^2} \right),
\end{equation}

For sightlines where $\log N$ represents a lower limit due to line saturation, 
we used a censored likelihood formulation.  
If $\log N_{\mathrm{lim},i}$ is a lower limit for the $i$-th sightline, the 
likelihood contribution is
\begin{equation}
    P(\log N_{\mathrm{true},i} > \log N_{\mathrm{lim},i}) 
    = 1 - \mathrm{CDF}\!\left(\log N_{\mathrm{lim},i}\,\middle|\, 
    \mu_i,\, \sqrt{\sigma^2 + \delta_i^2}\right),
\end{equation}
where $\mathrm{CDF}$ is the cumulative distribution function of the normal distribution.
This ensures that censored data points inform the fit by constraining the posterior 
probability of parameters to favor models consistent with the lower-limit nature 
of those measurements.

The total posterior probability of the model is then obtained by combining the likelihoods of both detected and censored data. The posterior distributions of $\alpha$, $\beta$, and $\sigma$ were obtained 
using the No-U-Turn Sampler (NUTS) implemented in \texttt{PyMC}.  
We ran four independent Markov chains with 2000 tuning steps and 2000 posterior 
samples per chain, yielding a total of 8000 posterior draws.  

To visualize the inferred relation, we drew 1000 random samples from the 
posterior distribution of $\alpha$ and $\beta$ and computed the corresponding 
predicted values of $\log N$ across the range of observed 
$\log (\Sigma_{\mathrm{SFR}})$.  
The median and 95\% credible intervals of these predictions define the best-fit 
relation and its associated uncertainty band.

\section{Photoionization Modeling}
\label{sec:cloudy_more}
As described in Section~5.1, the absorption components used for the photoionization modeling were selected based on their kinematic association with the H\textsc{~i} 21-cm emission. Because the H\textsc{~i} emission has substantially lower spectral resolution than the UV absorption data, one or more UV absorption components were combined where necessary to match a single broad H\textsc{~i} emission feature. Specifically, the O\textsc{~i} components at $\vlsr=+136$ and $+156~\kms$ (BI\,272), $+115$ and $+138~\kms$ (Sk$-$67$^\circ$191), $+113$ and $+143~\kms$ (Sk$-$68$^\circ$15), $+136$ and $+165~\kms$ (Sk$-$69$^\circ$50), $+110$ and $+145~\kms$ (Sk$-$70$^\circ$13), and $+112$ and $+133~\kms$ (Sk$-$71$^\circ$45) were combined, while single O\textsc{~i} components at $\vlsr=\mathrm{115}~\kms$ (Sk$-$67$^\circ$104), $\vlsr=\mathrm{121}~\kms$ (Sk$-$68$^\circ$80), and $\vlsr=\mathrm{147}~\kms$ (Sk$-$70$^\circ$32) were adopted. Corresponding Si\textsc{~ii} and Fe\textsc{~ii} components were included in all cases, while S\textsc{~ii} was additionally used for Sk$-$67$^\circ$191, Sk$-$68$^\circ$80, and Sk$-$70$^\circ$32. For all sightlines, the available upper limits on S\textsc{~ii} and P\textsc{~ii} were also included as constraints in the photoionization modeling where applicable.

\begin{figure}
  \centering
  \includegraphics[scale=0.35, trim=0 0 0 0, clip]{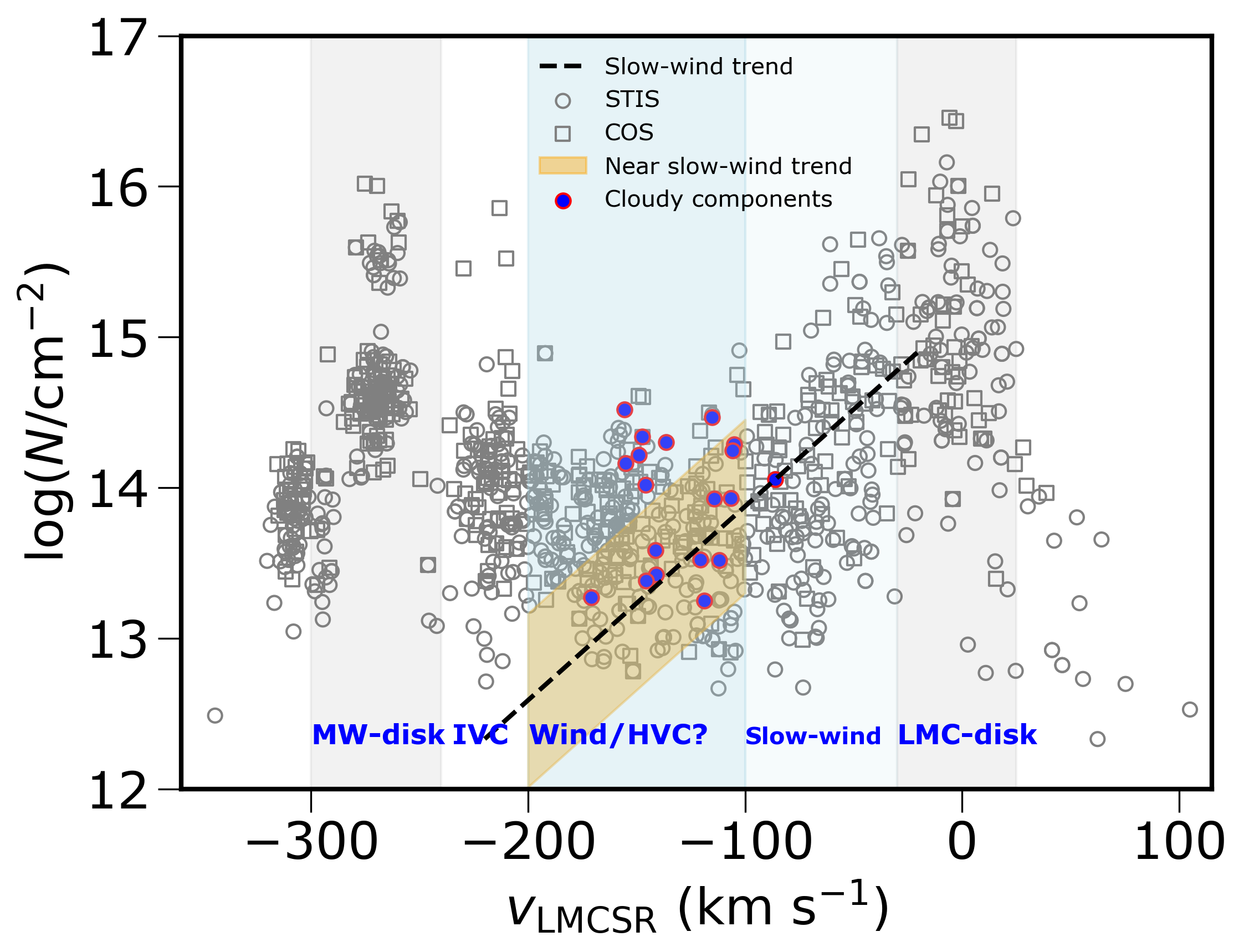}
  \caption{Similar to Figure~\ref{fig:SiII_logN}, this figure shows the Si\textsc{~ii} column density as a function of velocity, with blue points marking the individual components from the 13 sightlines used for the \textsc{Cloudy} photoionization modeling. The dashed black line shows the empirical column-density--velocity relation fitted to the slow-wind components and extrapolated into the ambiguous ``Wind/HVC?'' velocity interval. The shaded region in orange marks components lying within the residual scatter of the extrapolated slow-wind relation, while components above this band have positive offsets larger than the scatter.}
  \label{fig:population_test}
\end{figure}

\begin{figure}
  \centering
  \includegraphics[scale=0.35, trim=0 0 0 0, clip]{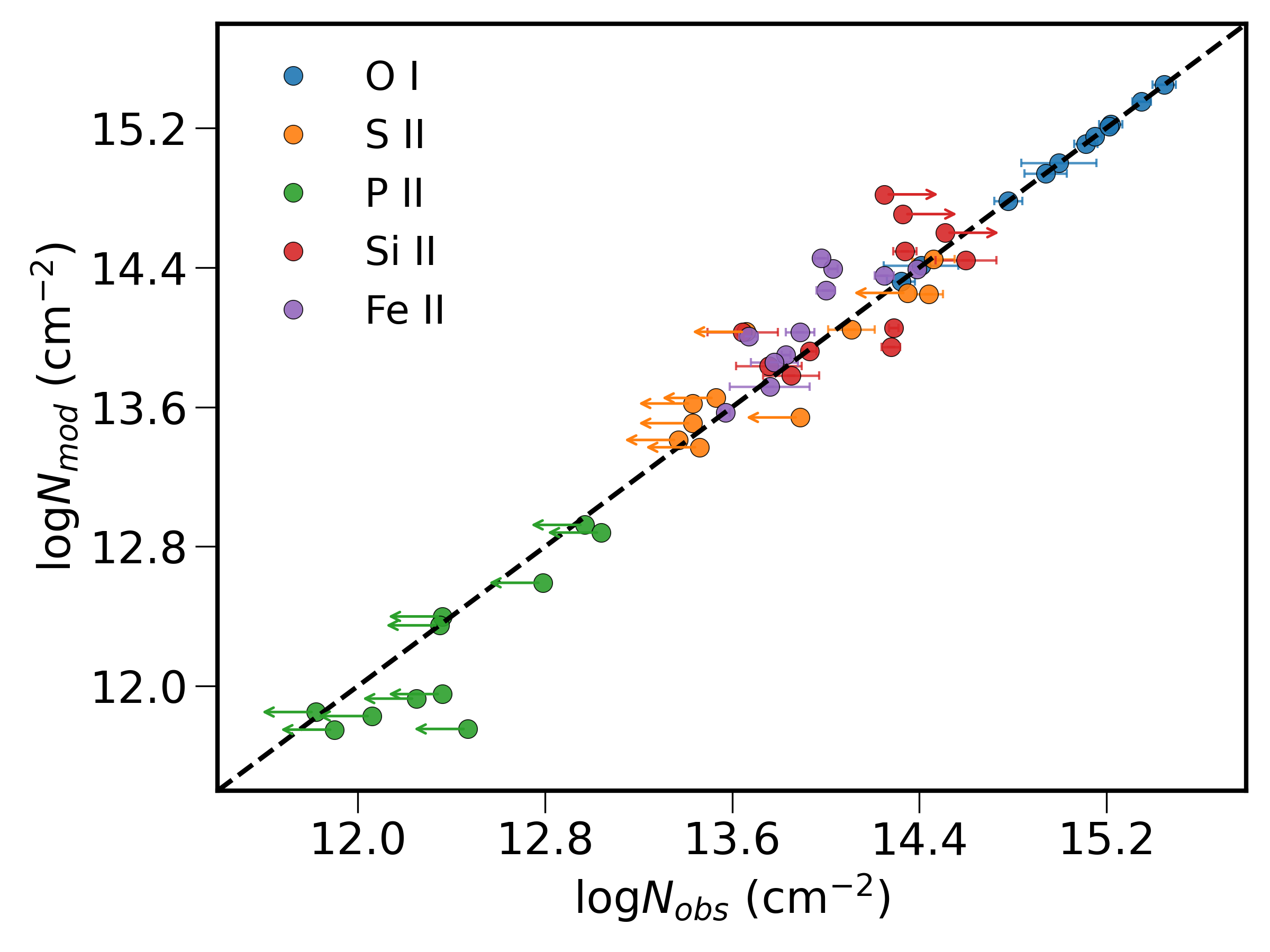}
  \caption{Comparison between the observed ionic column densities and the best-matched
\textsc{Cloudy} model predictions for the absorbers included in the
photoionization analysis. The dashed black line shows the one-to-one relation. Different colors represent different ions used as model constraints, including O\textsc{~i}, S\textsc{~ii}, P\textsc{~ii}, Si\textsc{~ii}, and Fe\textsc{~ii}. For detections, horizontal error bars show the observational uncertainties. For upper or lower limits, arrows indicate the allowed direction of the observed
column density.}
  \label{fig:obsvsprd_N}
\end{figure}

As discussed in Sections~\ref{subsection:kinematic_dist} and \ref{section:Cloudy}, we compare the \textsc{Cloudy}-modeled components with the empirical Si\textsc{~ii} offset from the extrapolated slow-wind relation. Figure~\ref{fig:population_test} shows this comparison at the component level, illustrating that the modeled absorbers occupy both the near-trend and high-offset regions. We also present the comparison between the observed ionic column densities and the best-matched \textsc{Cloudy} model predictions for the absorbers included in the
photoionization analysis in Figure~\ref{fig:obsvsprd_N}. Furthermore, we present the full set of photoionization modeling outputs for eight additional sightlines analyzed in this work. Each figure follows the same layout as Figure~\ref{fig:BI272_cloudy}, showing the ionization corrections, metallicities, and dust estimation for all ions considered. The two sightlines that probe the 30~Doradus region are not shown here, as their photoionization modeling results were previously published in \citet{2025ApJ...984..161P}. Likewise, we omit the two sightlines intersecting the N11 region, whose results are presented in \citet{Horton_2026}.

\begin{figure*}[ht!]
  \centering
  \includegraphics[scale=0.5, trim=0 23 0 0, clip]{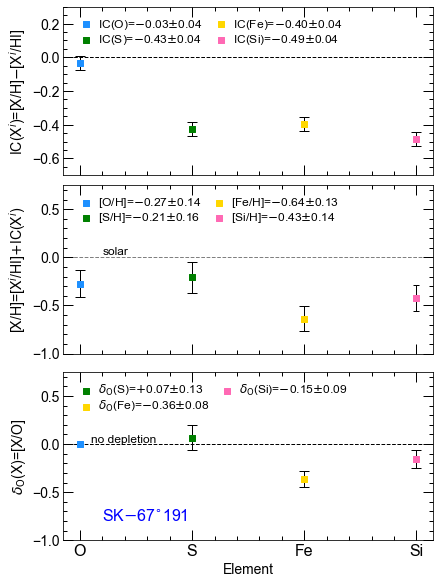}
  \includegraphics[scale=0.5, trim=32 23 0 0, clip]{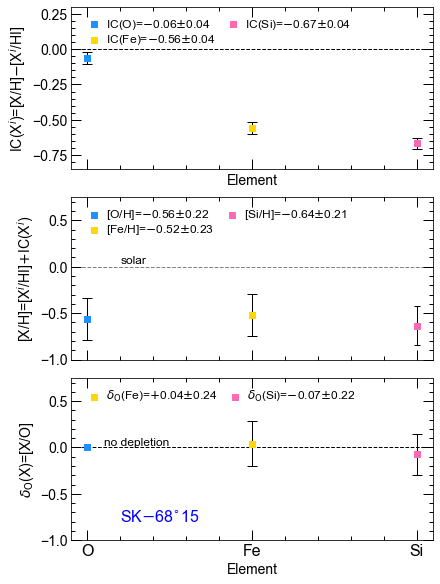}
  \includegraphics[scale=0.5, trim=0 0 0 0, clip]{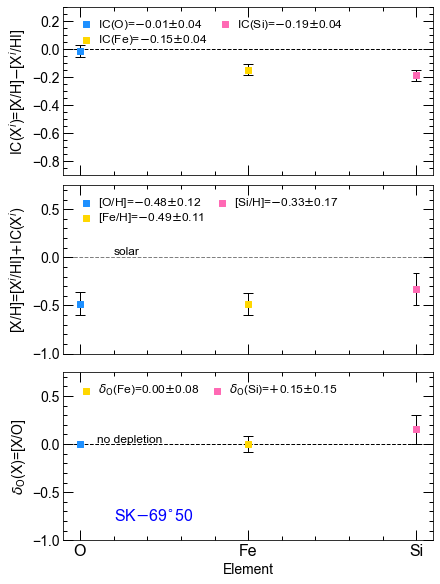}
  \includegraphics[scale=0.5, trim=32 0 0 0, clip]{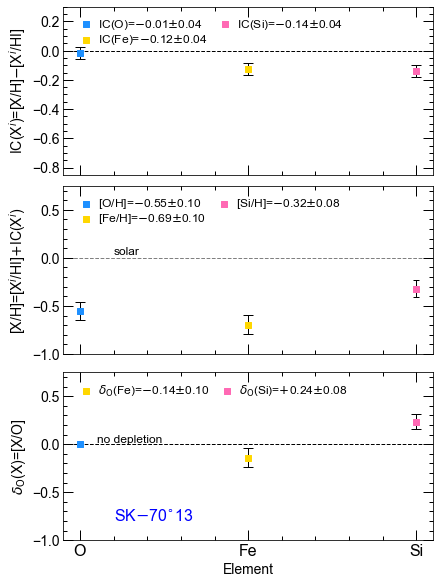}
  \caption{Similar to Figure~\ref{fig:BI272_cloudy} but showing the chemical abundance analysis for the ambiguous HVCs toward sightlines: SK$-$67$^{\circ}$191, SK$-$68$^{\circ}$15, SK$-$69$^{\circ}$50, and SK$-$70$^{\circ}$13.}
  \label{fig:cloudy_rest1}
\end{figure*}

\begin{figure*}[ht!]
  \centering
  \includegraphics[scale=0.5, trim=0 23 0 0, clip]{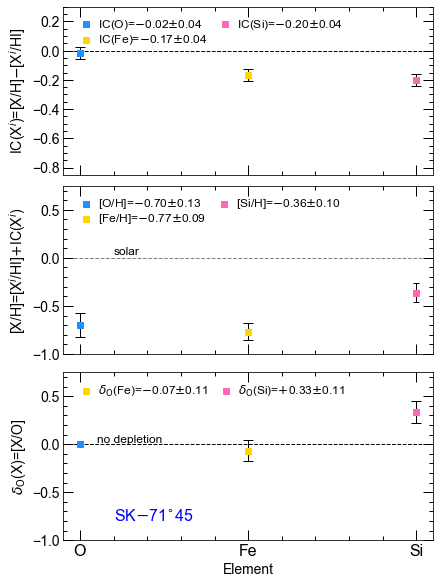}
  \includegraphics[scale=0.5, trim=32 23 0 0, clip]{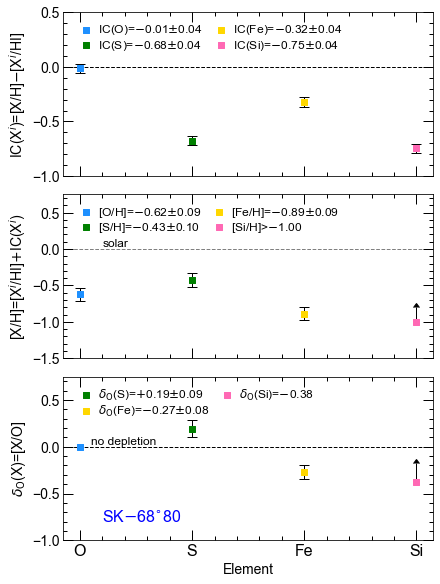}
  \includegraphics[scale=0.5, trim=0 0 0 0, clip]{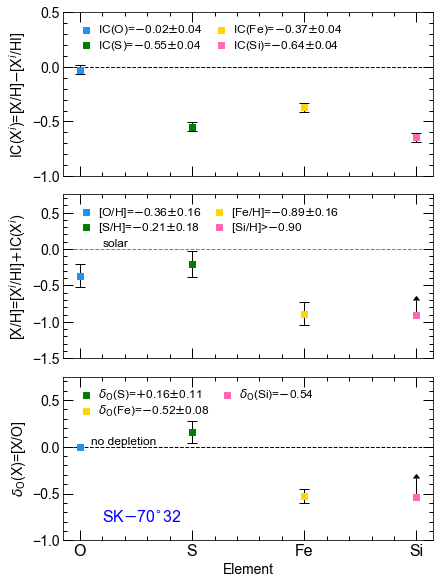}
  \includegraphics[scale=0.5, trim=32 0 0 0, clip]{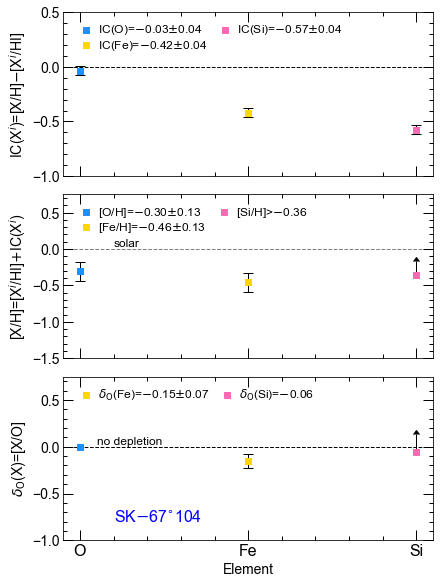}
  \caption{Similar to Figure~\ref{fig:BI272_cloudy} but showing the chemical abundance analysis for the ambiguous HVCs toward sightlines: SK$-$71$^{\circ}$45, SK$-$68$^{\circ}$80, SK$-$70$^{\circ}$32, and SK$-$67$^{\circ}$104.}
  \label{fig:cloudy_rest2}
\end{figure*}

\pagebreak
\newpage
\clearpage

\end{document}